\pdfoutput=1
\documentclass[a4paper,11pt]{article}
\usepackage{amsmath,amssymb,amsfonts,mathrsfs}
\usepackage[multiple,stable]{footmisc}
\usepackage[amsmath,amsthm,thmmarks]{ntheorem}
\allowdisplaybreaks[4]
\usepackage[noconfig]{refstyle}
\usepackage[dvipsnames]{xcolor}
\usepackage[hyperfootnotes=false, linktocpage=true, colorlinks, citecolor=blue, linkcolor=blue, urlcolor=blue]{hyperref}
\usepackage{array,tabularx,booktabs,geometry,subfigure,graphicx,longtable}
\usepackage[bf]{caption}
\usepackage{tikz}
\usepackage[all]{xy}
\usetikzlibrary{shapes,arrows}
\tikzstyle{block} = [rectangle, draw, text width=7em, text centered, rounded corners, minimum height=3em]
\usepackage{graphicx,color}
\usepackage{cite}
\usepackage{tikz}
\usepackage{tikz-cd}
\usetikzlibrary{cd}

\usepackage[affil-it]{authblk}
\usepackage{blindtext}

\usepackage{tikz}
\usetikzlibrary{shapes}
\usetikzlibrary{positioning}
\usetikzlibrary{chains}
\usetikzlibrary{arrows, fit, decorations.pathreplacing, shapes.misc}
\usetikzlibrary{decorations.markings}
\tikzstyle{every picture}+=[remember picture]
\tikzstyle{na} = [baseline=-.5ex]
\usepackage{todonotes}
\usepackage{tikz-3dplot}

\def\bZ{\mathbb{Z}}
\def\bR{\mathbb{R}}

\def\bP{\mathbb{P}}

\def\bF{\mathbb{F}}

\def\cC{\mathcal{C}}
\def\cH{\mathcal{H}}
\def\cL{\mathcal{L}}
\def\cN{\mathcal{N}}

\def\cO{\mathcal{O}}
\def\cC{\mathcal{C}}

\def\cT{\mathcal{T}}

\def\cX{\mathcal{X}}

\def\MCG{\mathrm{MCG}}
\def\Aut{\text{Aut}}
\def\Arf{\text{Arf}}

\makeatletter
\newcommand\xleftrightarrow[2][]{%
  \ext@arrow 9999{\longleftrightarrowfill@}{#1}{#2}}
\newcommand\longleftrightarrowfill@{%
  \arrowfill@\leftarrow\relbar\rightarrow}
\makeatother

\newcommand{\br}{\breve}

\def\fnote#1#2{\begingroup\def\thefootnote{#1}\footnote{#2}
     \addtocounter{footnote}{-1}\endgroup}

\numberwithin{equation}{section} 

\newcommand{\be}{\begin{equation}}
\newcommand{\ee}{\end{equation}}
\newcommand{\ba}{\begin{aligned}}
\newcommand{\ea}{\end{aligned}}

\newcommand{\bea}{\begin{eqnarray}}
\newcommand{\eea}{\end{eqnarray}}

\def\diag{\mathop{\mathrm{diag}}\nolimits}

\def\mb{\mathbb}

\def\mc{\mathcal}
\def\bp{\begin{pmatrix}}
\def\ep{\end{pmatrix}}

\begin{document}

\title{
       \vskip 40pt
       {\huge \bf SymTFTs and Duality Defects from 6d SCFTs on 4-manifolds}}
       
\vspace{10cm}

\author[1]{Jin Chen}
\author[2,3]{Wei Cui}
\author[3,2]{Babak Haghighat}
\author[4,5]{Yi-Nan Wang}

\vspace{10cm}

\affil[1]{Department of Physics, Xiamen University, Xiamen, 361005, China}
\affil[2]{Yanqi Lake Beijing Institute of Mathematical Sciences and Applications (BIMSA), Huairou District, Beijing 101408, P. R. China}
\affil[3]{Yau Mathematical Sciences Center, Tsinghua University, Beijing, 100084, China}
\affil[4]{School of Physics,
Peking University, Beijing 100871, China}
\affil[5]{Center for High Energy Physics, Peking University,
Beijing 100871, China}

\date{}
\maketitle

\fnote{}{Emails: \href{zenofox@gmail.com}{zenofox@gmail.com}, \;\href{cwei@vt.edu}{cwei@vt.edu}, \;\href{babakhaghighat@tsinghua.edu.cn}{babakhaghighat@tsinghua.edu.cn}, \;\href{ynwang@pku.edu.cn}{ynwang@pku.edu.cn} }

\begin{abstract}

In this work we study particular TQFTs in three dimensions, known as Symmetry Topological Field Theories (or SymTFTs), to identify line defects of two-dimensional CFTs arising from the compactification of 6d $(2,0)$ SCFTs on 4-manifolds $M_4$. 
The mapping class group of $M_4$ and the automorphism group of the SymTFT switch between different absolute 2d theories or global variants. 
Using the combined symmetries, we realize the topological defects in these global variants.  
Our main example is $\bP^1 \times \bP^1$.  
For $N$ M5-branes the corresponding 2d theory inherits $\bZ_N$ $0$-form symmetries from the SymTFT. 
We reproduce the orbifold groupoid for theories with $\bZ_N$ $0$-form symmetries and realize the duality defects at fixed points of the coupling constant under elements of the mapping class group. 
We also study other Hirzebruch surfaces, del Pezzo surfaces, as well as the connected sum of $\bP^1 \times \bP^1$. 
We find a rich network of global variants connected via automorphisms and realize more interesting topological defects. 
Finally, we derive the SymTFT on more general 4-manifolds and provide two examples.

\end{abstract}

\newpage

\tableofcontents

\section{Introduction}

Six-dimensional superconformal field theories (6d SCFTs) are UV complete theories with conformal symmetry and supersymmetry in the highest possible dimension \cite{Witten:1995ex, Witten:1995zh, Witten:1995gx, Seiberg:1996vs, Morrison:1996pp, Seiberg:1996qx, Hanany:1997gh, Morrison:2012np, Heckman:2013pva, DelZotto:2014hpa, Heckman:2014qba, Heckman:2015bfa, Heckman:2018jxk}. 
Compactification of 6d SCFTs on a $d$-dimensional manifold $M_d$ leads to many interesting $(6-d)$-dimensional  SCFTs \cite{Gaiotto:2009we, Gaiotto:2009hg, Argyres:2007cn, Argyres:2007tq, Alday:2009aq, Benini:2009mz, Dimofte:2010tz, Dimofte:2011jd, Dimofte:2011ju,  Gukov:2016gkn, Zafrir:2015rga, Jefferson:2018irk, Bhardwaj:2018yhy, Bhardwaj:2018vuu, Apruzzi:2018nre, Apruzzi:2019vpe,  Apruzzi:2019opn, Apruzzi:2019enx, Bhardwaj:2019fzv, Bhardwaj:2020gyu, Bhardwaj:2020kim, Duan:2021ges, Braun:2021lzt, Tian:2021cif, Gaiotto:2015una, Razamat:2016dpl,Assel:2016lad, Bah:2017gph, Kim:2017toz, Kim:2018bpg, Kim:2018lfo, Razamat:2018gro, Chen:2019njf, Razamat:2019mdt, Cecotti:2011iy, Cordova:2013cea, Yagi:2013fda, Lee:2013ida, Gukov:2015sna, Cordova:2016cmu, Gukov:2017kmk, Cho:2020ljj, Gadde:2013sca,Dedushenko:2017tdw,Feigin:2018bkf,  Pasquetti:2019hxf, Assel:2022row,Apruzzi:2016nfr, Gukov:2018iiq, Chen:2022vvd}. 
In particular, for 6d $\cN=(2,0)$ SCFTs realized on the worldvolume of $N$ M5-branes, the twisted compactification over 4-manifolds $M_4$ give a large class of two-dimensional theories denoted by $T_N[M_4]$ \cite{Gadde:2013sca,Dedushenko:2017tdw,Feigin:2018bkf}, and more recently such reductions have been extended to 6d $\cN=(1,0)$ SCFTs \cite{Apruzzi:2016nfr, Gukov:2018iiq, Chen:2022vvd}. The theories $T_N[M_4]$, defined through the 4-manifold $M_4$, are in general difficult to study since in most cases the Lagrangian description is unknown. It is believed that, in interesting cases, they will eventually flow to non-trivial interacting 2d SCFTs.

Global symmetry is one of the most important tools in the study of quantum field theories. Recently, through the association of symmetries with topological defects \cite{Kapustin:2013uxa, Gaiotto:2014kfa}, the concept of symmetry has been greatly generalized to include higher-form symmetries \cite{Kapustin:2013uxa, Gaiotto:2014kfa, GarciaEtxebarria:2019caf, Morrison:2020ool,Albertini:2020mdx, Apruzzi:2020zot,Braun:2021sex, Closset:2020scj, Bhardwaj:2020phs, Closset:2020afy, Apruzzi:2021phx,Hosseini:2021ged,  Bhardwaj:2021mzl,Buican:2021xhs, Cvetic:2021maf,Cvetic:2021sxm, Cvetic:2021sjm, Closset:2021lwy,Tian:2021cif, Genolini:2022mpi,DelZotto:2022fnw, Cvetic:2022uuu}, higher-group symmetries \cite{Sharpe:2015mja, Tachikawa:2017gyf, Cordova:2018cvg, Benini:2018reh, Cordova:2020tij, Lee:2021crt,  DelZotto:2020sop, Apruzzi:2021mlh, Apruzzi:2021vcu, Bhardwaj:2021wif}, and non-invertible symmetries \cite{Verlinde:1988sn, Bhardwaj:2017xup, Chang:2018iay, Thorngren:2019iar, Thorngren:2021yso, Komargodski:2020mxz, Choi:2021kmx, Kaidi:2021xfk, Koide:2021zxj,Choi:2022zal,Arias-Tamargo:2022nlf,Hayashi:2022fkw,Roumpedakis:2022aik,Kaidi:2022uux,Choi:2022jqy,Cordova:2022ieu,Antinucci:2022eat,Bashmakov:2022jtl,Damia:2022rxw,Damia:2022bcd,Choi:2022rfe,Lu:2022ver,Bhardwaj:2022lsg,Lin:2022xod,Apruzzi:2022rei,GarciaEtxebarria:2022vzq, Benini:2022hzx, Wang:2021vki, Chen:2021xuc, DelZotto:2022ras,Bhardwaj:2022dyt,Brennan:2022tyl,Delmastro:2022pfo, Heckman:2022muc,Freed:2022qnc,Freed:2022iao,Niro:2022ctq,Mekareeya:2022spm,Antinucci:2022vyk,Chen:2022cyw,Karasik:2022kkq,Cordova:2022fhg,Decoppet:2022dnz,GarciaEtxebarria:2022jky,Choi:2022fgx,Yokokura:2022alv,Bhardwaj:2022kot,Bhardwaj:2022maz, Hsin:2022heo,Heckman:2022xgu,Antinucci:2022cdi,Apte:2022xtu,Garcia-Valdecasas:2023mis, Delcamp:2023kew, Bhardwaj:2023zix}. 
The fusion of non-invertible defects does not obey the group law and they are described by (higher)fusion categories \cite{Bhardwaj:2022yxj, Bartsch:2022mpm, Bartsch:2022ytj, Bhardwaj:2023wzd}. 
Non-invertible symmetries were first found in 2d CFTs,
% \cite{Verlinde:1988sn, Bhardwaj:2017xup, Chang:2018iay, Thorngren:2019iar, Thorngren:2021yso}
and recently constructed in many higher-dimensional theories.
% \cite{Koide:2021zxj,Choi:2022zal,Arias-Tamargo:2022nlf,Hayashi:2022fkw,Roumpedakis:2022aik,Kaidi:2022uux,Choi:2022jqy,Cordova:2022ieu,Antinucci:2022eat,Bashmakov:2022jtl,Damia:2022rxw,Damia:2022bcd,Choi:2022rfe,Lu:2022ver,Bhardwaj:2022lsg,Lin:2022xod,Apruzzi:2022rei,GarciaEtxebarria:2022vzq, Benini:2022hzx, Wang:2021vki, Chen:2021xuc, DelZotto:2022ras,Bhardwaj:2022dyt,Brennan:2022tyl,Delmastro:2022pfo, Heckman:2022muc,Freed:2022qnc,Freed:2022iao,Niro:2022ctq,Mekareeya:2022spm,Antinucci:2022vyk,Chen:2022cyw,Karasik:2022kkq,Cordova:2022fhg,Decoppet:2022dnz,GarciaEtxebarria:2022jky,Choi:2022fgx,Yokokura:2022alv,Bhardwaj:2022kot,Bhardwaj:2022maz, Hsin:2022heo,Heckman:2022xgu,Antinucci:2022cdi,Apte:2022xtu,Garcia-Valdecasas:2023mis, Delcamp:2023kew, Bhardwaj:2023zix} 
In this paper, we will study the global symmetries and in particular the non-invertible ones, in the theory $T_N[M_4]$, using the concept of SymTFTs.

Symmetry topological field theory (SymTFT) is a topological field theory on a compact $(d+1)$-dimensional space $X_{d+1}$, which encodes symmetries, anomalies, and global structures for theories on the boundary $X_d=\partial X_{d+1}$ \cite{Apruzzi:2021nmk, Apruzzi:2022dlm, vanBeest:2022fss, Kaidi:2022cpf, Bashmakov:2022uek, Kaidi:2023maf}. 
% For quantum field theories constructed from the string theories on d-dimensional manifold $M_d$, the corresponding SymTFT can be obtained from the dimensional reduction on the $\partial M_d$. 
% or more mathematically a lagrangian algebra of the corresponding Drinfield center \cite{Zhang:2023wlu}. 
6d SCFTs are relative theories living on the boundary of a non-invertible 7d TQFT \cite{Witten:1998wy, Witten:2009at}. 
Taking into account this relative nature, it becomes apparent that one has to study the compactification of the 6d/7d coupled system on $M_4$. 
The 2d theory $T_N[M_4]$ is in general also relative and coupled to a non-invertible 3d TQFT obtained from the reduction of the 7d TQFT. This 3d TQFT is the SymTFT for $T_N[M_4]$. 
To make $T_N[M_4]$ absolute, one needs to choose a maximal isotropic sublattice in $H_2(M_4,\bZ_N)$ \cite{Tachikawa:2013hya} or polarization on $M_4$ \cite{Gukov:2020btk}. Combining with other discrete choices of data, one can find all global variants of $T_N[M_4]$.

From the perspective of the 3d SymTFT, the maximal isotropic sublattice is equivalent to a choice of topological boundary condition \cite{Kapustin:2010hk} rendering the fields corresponding to this subset non-propagating background fields. The 0-form symmetry of the 3d SymTFT that transforms between these boundaries gives rise to different 2d topological manipulations among global variants of  $T_N[M_4]$ \cite{Gaiotto:2020iye}. In our setup, this symmetry can be obtained from the automorphism group of $H_2(M_4, \bZ_N)$ denoted by $\Aut_{\bZ_N}(Q)$. 
Thus, by employing the SymTFT, one can obtain the global variants of  $T_N[M_4]$ and study how they transform under topological manipulations. 
% which are untractable without the knowledge of the dimensional reduction of the 7d TQFT. 

Similar to the class S theory, the automorphism group of $H_2(M_4, \bZ)$ or mapping class group $\MCG(M_4)$ leads to Montonen–Olive-like dualities that also transform different global variants of $T_N[M_4]$ into each other. 
Also, there are coupling constants of $T_N[M_4]$ corresponding to geometric parameters of $M_4$ which transform non-trivially under such dualities. 
At fixed points of such transformations, combinations of the duality transformations and topological manipulations, captured by $\MCG(M_4)$, give rise to  topological defects known as \textit{duality defects}. When the combination involves a gauging operation, the defects are non-invertible. In this way, one can construct interesting non-invertible defects in $T_N[M_4]$. 
Similar constructions for the class S theory in four dimensions have been studied in \cite{Kaidi:2022uux, Bashmakov:2022uek}.

The plan of the current paper is as follows. In Section \ref{sec:2}, we give a general overview of SymTFTs from dimensional reduction and the construction of absolute theories. We then proceed in Section \ref{sec:3} to study the compactification of 6d $N=(2,0)$ theory of type $A_{N-1}$ on $\bP^1 \times \bP^1$. 
The SymTFT is the standard $\bZ_N$ gauge theory. 
With the help of this SymTFT, we find the orbifold groupoid and global variants of $T_N[\bP^1 \times \bP^1]$ for all prime $N$, as well the cases of $N=4$ and $N=6$. 
An interesting observation is that when $N$ is even, one can identify both bosonic and fermionic absolute theories of $T_N[\bP^1 \times \bP^1]$ and two topological manipulations, one of which is gauging and another one is stacking the Arf invariant. The reason is that the 4-manifold has spin structure, thus the 2d theory $T_N[\bP^1 \times \bP^1]$ admits spin structure and can be a fermionic theory.  
For the cases of $N$ prime, we identify all global variants and possible topological manipulations. 
We also discuss how to generalize the result to the case when $N=pq$ is not prime, but a product of two primes $p$ and $q$, using two examples, namely $N=4$ for $p,q$ not coprime and $N=6$ for  $p,q$ coprime. 
We can reproduce all orbifold groupoids studied in \cite{Gaiotto:2020iye}, and without too much effort, we can study the cases for $N>6$.

With the knowledge of the global variants and how they transform under duality and topological manipulations, we discuss the topological defects in each global variant of $T_N[\bP^1 \times \bP^1]$. It turns out that there exist duality defects for each $N$. In particular, for $N=2$, one can show that the duality defect can be related to invertible symmetry by duality, and thus is non-intrinsic non-invertible, but for all other cases, the duality is intrinsic non-invertible.

We extend our analysis to the connect sum of $\bP^1 \times \bP^1$. The SymTFT in this case is $\bZ_N \times \bZ_N$ gauge theory. For $N=2$ and $N=3$, we compute the maximal isotropic sublattice and obtain the same orbifold groupoid studied in \cite{Gaiotto:2020iye}. Considering the possible SPT phases, the global variants can be obtained from the complement of these sublattices. 
After that, we determine the mapping class group and the coupling constants. Interestingly, at particular value, these couplings are invariant under a set of dualities of a $D_8$ group. We find some topological manipulations that can undo the action of these dualities in some specific global variant. Since these topological manipulations involving different ways of gauging, we can realize non-invertible defects described by $TY(D_8)$ category.

Next, in Section \ref{sec:4}, we study the other Hirzebruch surfaces. As we will show in the main text, it is sufficient to consider the case of $\bF_1$. The SymTFT is the twisted $\bZ_N$ gauge theory. In analogy with the $\bP^1 \times \bP^1$ case, using the SymTFT, we study the global variants of the theory $T_N[\bF_1]$ and possible topological manipulations for each $N$. 
Similar to the $\bP^1 \times \bP^1$ case, one can find two absolute theories for odd $N$. However, they are not related by gauging, but some other topological operation. 
For even $N$, there are significantly fewer global variants compared with the $\bP^1 \times \bP^1$ case. One observation is that there is an anomaly for gauging $\bZ_2$. 
It is interesting to find the physical understanding of these topological operations. Similarly, we find the mapping class group, couplings and identified topological defects for prime $N$, $N=4$ and $N=6$. 
% find the physical argument about this from the 3d TQFT. Similarly, we construct the topological defects. Except for $N=2$, we find intrinsic duality defects. At $N=2$, the defect generates a $\bZ_2$ invertible symmetry. 
% For example, the compactification of SymTFT of 6D $\cN=(1,0)$ rank-1 SCFTs on $S^1$ correctly describes the symmetries of the 5d KK theory and can be considered as a 6d SymTFT for that 5d KK theory \cite{Apruzzi:2022dlm}. 
% Compactification on the Riemann surface leads to the 4d class S theories. The generalized symmetries and even the non-invertible symmetries have been deduced from the 5d SymTFT which itself is obtained from the reduction of the 7d SymTFT on the corresponding Riemann surface \cite{}. 
We compute the 0-form symmetries of $\bZ_N$ gauge theories up to $N=20$ and twisted gauge theory up to $N=11$ from the perspective of geometry. For odd $N$, the result is the same as the ones found in \cite{Delmastro:2019vnj}, while for even $N$, our result gives the 0-form symmetry for spin (twisted) $\bZ_N$ gauge theories.

We then move to study the Del Pezzo surfaces, particularly $dP_2$, in Section \ref{sec:5}. We calculate the mapping class group of $dP_2$ and determine the couplings constants using invariant volumes. 
Under a transformation generated by a $\bZ_2$ subgroup of the class mapping group, these couplings are invariant at a extended loci in the conformal manifold, which implies the exsistence of the duality defect. However, the SymTFT does not have the form of a standard Dijkgraaf-Witten theory. It would be interesting to study it on its own.

Finally, in Section \ref{sec:6}, we consider $T_N[M_4]$ with a general compact 4-manifold $M_4$, which is allowed to have 1-cycles, 3-cycles, as well as torsional cycles. 
From the 7d TQFT of the $A_{N-1}$ $(2,0)$ theory, we derive the 3d SymTFT for $T_N[M_4]$ using differential cohomology \cite{Apruzzi:2021nmk, vanBeest:2022fss}. 
As examples, we consider the 4-manifold $T^2\times S^2$ with non-trivial (1-)3-cycles and the Enriques surface with torsional cycles. 
We compute the intersection numbers in both examples and analyze the symmetry and mixed anomalies.

\section{Compactification of 6d SCFTs on 4-manifolds} \label{sec:2}

In six dimensions, there exist six-dimensional $\cN=(2,0)$ SCFTs that can be understood as relative theories living on the boundary of a non-invertible seven-dimensional topological quantum field theory \cite{Witten:1998wy, Witten:2009at, Tachikawa:2013hya}. However, the partition function of such theories on a six-manifold $M_6$ is not well-defined, and instead, the theory is better understood as a state in the Hilbert space of the bulk TQFT.

In this section, we will first review how to define the partition vector of these 6d theories by choosing a discrete set of data. Next, we will discuss the dimensional reduction of 6d relative theories, which involves coupling the 7d/6d systems on 4-manifolds to obtain absolute theories upon compactification. Finally, we will focus on the compactification of the 7d TQFT of the 6d SCFTs of type $A_{N-1}$ on a 4-manifold with non-trivial 2-cycles. We will study various properties of the resulting three-dimensional SymTFT that will be useful in subsequent sections.

\subsection{6d SCFTs as relative theories}

In a 6d $\cN=(2,0)$ SCFT of type $g$, the defect group $\mathscr{D}$ is given by the center of $g$ \cite{Tachikawa:2013hya,DelZotto:2015isa}. When $\mathscr{D}$ is non-trivial, the theory is relative and the bulk 7d TQFT is non-invertible. The partition function of such a 6d relative theory on a manifold $M_6$ is not a number but a vector in the Hilbert space of the 6+1-dimensional TQFT on $M_6 \times \bR$. To specify it, we need to choose a discrete set of data that will be discussed in the following.

Consider the specific case of the 6d $\cN=(2,0)$ theory of type $A_{N-1}$, denoted by $\cT_N$. The defect group, in this case, is $\mathscr{D}=\bZ_N$, implying that the theory is relative when $N>1$. The 7d bulk TQFT associated with this theory is described by the action \cite{Witten:1998wy}
\begin{equation} \label{eqn:symTFT-7d}
    S_{7d} = \frac{N}{4\pi} \int_{W_7} c \wedge dc~,
\end{equation}
where $c \in H^3(W_7, U(1))$ is a 3-form field. The corresponding Wilson 3-surfaces of this theory are given by
\begin{equation}
    \Phi(M_3) := e^{\oint_{M_3} c}~, \qquad M_3 \in H_3(W_7, \bZ_N)
\end{equation}

On a constant time 6d slice $M_6 \subset W_7$, the Wilson 3-surfaces satisfy the following equal-time commutation relation \cite{Bashmakov:2022uek}
\begin{equation} \label{Eqn:braiding7d}
\Phi(M_3) \Phi(M_3') = e^{  \langle M_3, M_3' \rangle}\Phi(M_3')\Phi(M_3)~. 
\end{equation}
where the intersection pairing is 
% \begin{equation}
%     H_3(M_6,\bZ) \times  H_3(M_6,\bZ) \to \bZ_N
% \end{equation}
\begin{equation}
    \langle M_3, M_3' \rangle = \frac{2\pi i}{N} \int_{M_6} \omega_{M_3} \cup \omega_{M'_3}
\end{equation}
with $\omega_{M_3} $ and $ \omega_{M'_3}$ being the Poincare dual of $M_3$ and $M'_3$.
Besides, Wilson 3-surfaces $\Phi(M_3)$ also satisfy the following quantum torus algebra 
\begin{equation} \label{eq:qtorus}
\Phi(M_3) \Phi(M_3') = e^{\frac{ 1 }{2}\langle M_3, M_3' \rangle} \Phi(M_3+M_3')~.
\end{equation}

The 6d SCFT can be understood as a state in the Hilbert space $\cH(M_6)$ of this 7d TQFT. 
In order to specify this state, one must first fix a basis for $\cH(M_6)$, which is specified by a maximal isotropic sublattice $\cL \subset H_3(M_6, \bZ_N)$
% \footnote{Note that the maximal isotropic sublattice is defined in the $\cL \subset H^3(M_6, \bZ_N)$ in the early literature \cite{Witten:2009at,Tachikawa:2013hya}. The notation used here is the Poincare dual to theirs. In particular, the pairing should be $Q^{-1}$. 
% }
, i.e. a maximal subset such that 
\footnote{Note that the maximal isotropic sublattice $\cL \subset H_3(M_6,\bZ_N)$ is the dual of those defined in $H^3(M_6,\bZ_N)$ \cite{Tachikawa:2013hya, Gukov:2020btk}.}
\begin{equation}
    \langle M_3, M_3'\rangle = 0 ~, \qquad \forall \; M_3,M_3' \in \cL\,.
\end{equation}
With $\cL$, one can find a set of commuting Wilson 3-surfaces. These Wilson 3-surfaces define a basis in  $\cH(M_6)$ given by 
% \footnote{Following the notation in \cite{Bashmakov:2022uek}, we will absorb $q$ by $M_3 = qM'_3$ with $M_3 \in H_3(M_6,\bZ_N)$ and write $\Phi_q(M_3)$ as $\Phi(M_3)$ for simplicity.}
%
\begin{equation}
\Phi(M_3) |\cL,0 \rangle = |\cL ,0 \rangle, \qquad\quad \forall M_3 \in \cL\,,
\end{equation}
where $|\cL,0 \rangle $ is a state in $\cH(M_6)$ invariant under  $\Phi(M_3)$. 
% is defined such that all  can be simultaneously diagonalized.
The rest of the states in $\mathcal H(M_6)$ are obtained from elements in $\cL^\perp = H_3(M_6,\mathbb Z_N)/\cL$ by  
\begin{equation}
\Phi(M'_3)  |\cL ,0 \rangle = |\cL , M'_3 \rangle, \qquad \quad  \forall \, M'_3 \in \cL^\perp\,.
\end{equation}
An element $M_3 \in \cL$ induces an action on the state $|\cL , M'_3 \rangle$ given by \cite{Tachikawa:2013hya}
\begin{equation}
    \Phi(M_3)|\cL , M'_3 \rangle = e^{\langle M_3, M'_3\rangle} |\cL , M'_3 \rangle~.
\end{equation}

Given a choice of $\cL$, the ``partition vector'' of the 6d SCFT is given by \cite{Bashmakov:2022uek}
\begin{equation}
| A_{N-1} \rangle = \sum_{M'_3 \in \cL^\perp} Z_{\cL}(M'_3) |\cL, M'_3\rangle~,
\end{equation}
where the coefficients $ Z_{\cL}(M'_3)$ are the 6d conformal blocks \cite{Witten:2009at}.
To obtain the partition function, one can consider the geometry $W_7 = M_6 \times I$ and the choice of the $\cL$ can be understood as the choice of the topological boundary condition for the 7d TQFT. The partition function of the 6d SCFT on the boundary is then given by 
\begin{equation} \label{Eqn:Z6dL}
    Z_{\cL}(M_6) = \langle \cL,0 |A_{N-1}\rangle~, 
\end{equation}
and 
\begin{equation} \label{Eqn:Z6dLv}
    Z_{\cL}(M_6,M'_3) = \langle \cL,M'_3 |A_{N-1}\rangle~.
\end{equation}
Thus, by the choice of maximal isotropic sublattice $\cL$ and elements in $\cL^{\perp}$, one can obtain an absolute theory from relative 6d SCFTs.

\subsubsection*{Compactification on 4-manifolds}

Consider the 6d SCFT living on $M_6 = \Sigma_2 \times M_4$ and the corresponding 7d bulk theory on $W_7 = W_3 \times M_4$. After compactifying this 7d/6d coupled system on $M_4$, one obtains a 2d theory on $\Sigma_2$ denoted by $T_{N}[M_4]$ and a 3d TQFT on $\partial W_3=\Sigma_2$. 
% It is expected that this 3d TQFT will be the SymTFT of $T_{N}[M_4]$. 
Note that the 2d theory is a relative theory coupled to a non-trivial TQFT in the 3d bulk. As discussed in \cite{Tachikawa:2013hya, Gukov:2020btk}, to obtain 2d absolute theories, one needs to choose a maximal isotropic sublattice in the internal geometry.

Assuming that $M_4$ does not have any 1-cycles or 3-cycles, $H_3(M_6, \bZ_N)$ splits via the K{\"u}nneth formula as 
\begin{equation*}
H_3(M_6, \bZ_N) \cong   H_2(M_4, \bZ_N) \otimes H_1(\Sigma_2, \bZ_N)~. 
\end{equation*}
Thus, any two 3-cycles $M_3,M'_3 \in H_3(M_6,\bZ_N)$, can be decomposed as 
\begin{equation*}
    M_3= M_2 \times \gamma, \qquad M'_3= M'_2 \times \gamma',
\end{equation*}
with $M_2,M'_2 \in H_2(M_4,\bZ_N)$ and $\gamma,\gamma' \in H_1(\Sigma_2,\bZ_N)$. 
The intersection between $M_3$ and $M'_3$ becomes 
\begin{equation} \label{Eqn:decomPairing}
    \langle M_3,M'_3\rangle = (M_2,M'_2) \times \langle \gamma,\gamma' \rangle\,,
\end{equation}
where $(-,-)$ is defined by the intersection form of $M_4$
\begin{equation}
    Q: H_2(M_4,\mathbb{Z}_N) \times H_2(M_4,\mathbb{Z}_N) \rightarrow \mathbb{Z}_N\,.
\end{equation}
and $\langle-,-\rangle$ is the standard anti-symmetric intersection pairing for $\Sigma_2$.

Similar to the theory of class S \cite{Tachikawa:2013hya, Bashmakov:2022jtl, Bashmakov:2022uek}, to obtain absolute theories on $\Sigma_2$, one needs to specify a maximal isotropic sublattice $L \subset H_2(M_4,\bZ_N)$ such that 
\begin{equation} \label{Eqn:isotropyM4}
    ( M_2, M_2') = 0 ~, \hspace{0.5 in}\,\,M_2,M_2'\in L~.
\end{equation}
According to equation (\ref{Eqn:decomPairing}), choosing an $L$ automatically defines a 6d maximal isotropic sublattice $\cL \subset H_3(M_6,\bZ_N)$ given by 
\begin{equation}
    \cL = L \otimes H_1(\Sigma_2,\bZ_N)\,.
\end{equation}
As we will see, different choices of $L$ define different topological boundary conditions for the 3d SymTFT on $W_3$ and thus lead to distinct absolute theories on $\Sigma_2$ denoted by $T_{N, L}[M_4]$.

Besides that, each absolute theory can have different global properties \cite{Aharony:2013hda, GarciaEtxebarria:2019caf}. 
To specify the global variants, one also needs to choose a specific representative of the non-trivial classes of $L^\perp \otimes H_1(\Sigma_2, \bZ_N)$, with $L^\perp:= H_2(M_4, \bZ_N)/L$.
The choice of representatives in  $L^\perp$ determines the possible stacking of the SPT phases and the choice of elements in $H_1(\Sigma_2, \bZ_N)$ determines the background fields for the corresponding zero-form symmetries in the 2d theory. 
After the choice of $B$, the partition function of the 2d theories is \cite{Bashmakov:2022jtl}
\begin{equation}
    Z_{L}(\Sigma_2,B) = \langle \cL,B |A_{N-1}\rangle = \langle \cL,0|\Phi(B) |A_{N-1}\rangle\,,
\end{equation}
with $B\in L^{\perp}$. By taking $B \in L$, one goes back to the definition in equation \eqref{Eqn:Z6dL}. 
Following the notation introduced in \cite{Gukov:2020btk, Bashmakov:2022jtl, Bashmakov:2022uek}, we will denote the representative by $B$ and the corresponding global variant by $T_{N, L}[M_4, B]$.

\begin{figure}
    \centering
    \begin{tikzpicture}[scale=0.75]
        \fill[blue!20] (0.00,0.00) rectangle (6,3);
        \fill[blue!20] (13,0) rectangle (19,3);
        \draw[ thick] (13,0) -- (13,3);
        \draw[ thick] (19,0) -- (19,3);
	\draw[thick] (6,0) -- (6,3);
        \draw[thick,  ->] (6.7,1.5) -- (12.3,1.5);
        \node[above] at (9,1.6) {Shrink $M_4$};
	\node at (16,1.5) {SymTFT($\cC$)};
 \node at (3,1.5) {7d TQFT};
 \node[below] at (6,0) {$|A_{N-1}\rangle $}; 
 \node[below] at (13,0) {$\langle L(B)|$};
 \node[below] at (19,0) {$|\cT_N[M_4]\rangle $}; 
 \node[above] at (19,3) {$x=\epsilon$};
 \node[above] at (13,3) {$x=0$};
	\end{tikzpicture}
    \caption{Compactification of 7d/6d coupled system on $M_4$ with maximal isotropic sublattice $L$ lead to a 2d theory $T_{N}[M_4]$ on $\Sigma_2$ and its SymTFT on $\Sigma_2 \times I_{(0,\epsilon)}$ with topological boundary condition $\langle L(B)|$.}
    \label{Fig:compactM4}
\end{figure}
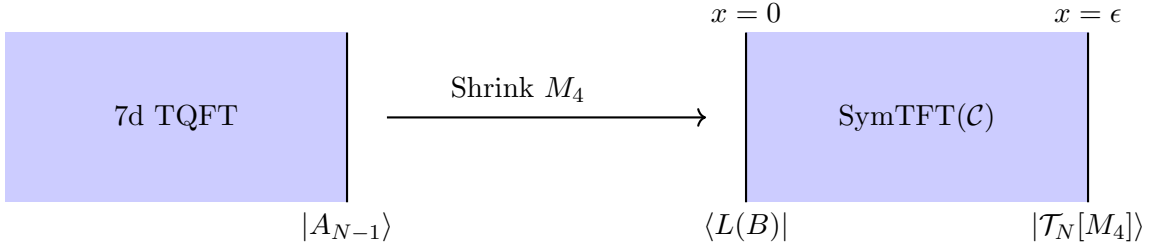

\subsection{SymTFT of $T_N[M_4]$}

Let $\{\zeta_i\}$ be a basis of $H_2(M_4,\bZ)$ with with $i=1,2,\ldots, r$.  
Compactification of the 7d action \eqref{eqn:symTFT-7d} leads to the following action in 3d
\begin{equation} \label{eq:A3d}
    S_{3d} = \frac{N}{4\pi} \sum_{i,j} Q^{ij} \int_{W_3}  a_i \wedge d a_j,
\end{equation}
where 
$Q$ is the intersection form of $M_4$ and 
$a_i$ are the 1-form gauge fields 
\begin{equation} \label{Eqn:afromc}
    a_i = \int_{\zeta_i} c, \qquad i=1,2, \ldots, r\,.
\end{equation}
For rational complex surfaces $M_4$, the classification of $Q$-matrices is well known. For each $r\neq 2$, the matrix $Q$ has to be the diagonal matrix with signature $(1,r-1)$:
\be
Q=\bp 1 & 0 & \dots & 0\\0 & -1 & \dots & 0\\ \vdots & \vdots & \ddots & \vdots\\ 0 & 0 & \dots & -1\ep\,.
\ee
For $r=2$, we have 
\be
Q=\bp 0 & 1\\1 & 0\ep 
\ee
if $M_4$ is a Hirzebruch surface $\mb{F}_l$ with even $l$, and 
\be
Q=\bp 0 & 1\\1 & -1\ep 
\ee
(or equivalently the diagonal matrix with signature $(1,1)$) if $M_4$ is a Hirzebruch surface with odd $l$.

We next define the Chern-Simons level matrix 
\begin{equation}
    K^{i,j} \equiv N Q^{i,j}.
\end{equation}
$K$ is an $r \times r$ symmetric matrix with integer entries where $r$ is the rank of $H^2(M_4,\mathbb{Z})$.
The 1-form defect group of the 3d theory \eqref{eq:A3d} can be obtained by finding the Smith normal form of K, that is finding matrices $P, R \in SL(r,\mathbb{Z})$ such that 
\begin{equation}
    P^t K R = D, 
\end{equation}
with $D$ a diagonal matrix of the form
\begin{equation}
    D = \left(\begin{array}{ccc}d_1 & ~ & ~\\ ~ & \ddots & ~\\ ~ & ~ & d_n\end{array}\right),
\end{equation}
with all elements $d_1, \ldots, d_n \in \mathbb{Z}$. Then the defect group is determined to be
\begin{equation}
    \mathscr{D} = \bigoplus_{i=1}^r \mathbb{Z}_{d_i}.
\end{equation}
The line operators can be obtained from 7d by 
\begin{equation} \label{Eqn:p1p1L}
    L_{\vec{\alpha}}=e^{\frac{i2\pi}{N} \int_{\gamma \times M_2} c} = e^{\frac{i2\pi \vec{\alpha}}{N} \cdot \int_{\gamma}\vec{a}}
\end{equation}
where $\vec{\alpha}=(\alpha_1,\alpha_2,\ldots,\alpha_{r})$ denotes the charge of the line defect. 
% %
S- and T-matrices of the corresponding TQFT are given by
\begin{equation}
    S(\vec{\alpha},\vec{\beta}) \equiv \frac{B(\vec{\alpha},\vec{\beta})}{\sqrt{|\mathscr{D}|}}, \quad T(\vec{\alpha},\vec{\beta}) \equiv \theta_{\vec{\alpha}} e^{-2\pi i c/24} \delta_{\vec{\alpha}\vec{\beta}}, 
\end{equation}
where $\vec{\alpha}, \vec{\beta} \in \mathscr{D}$ are Anyons, $B(\vec{\alpha},\vec{\beta})$ is the braiding matrix
\begin{equation}
    B(\vec{\alpha},\vec{\beta}) \equiv \exp\left[2\pi i \vec{\alpha}^t K^{-1} \vec{\beta}\right],
\end{equation}
and
\begin{equation}
    \theta(\vec{\alpha}) \equiv \exp[2\pi i \vec{\alpha}^t K^{-1} \vec{\alpha}], 
    % \quad c = \sign(K).
\end{equation}
are the topological spin of anyons $\vec{\alpha}$.

\begin{figure}
    \centering
    \begin{tikzpicture}[scale=0.75]
    \fill[blue!20] (4,0) rectangle (10,3);
        \draw[very thick, blue!80] (15,0) -- (15,3);
	\draw[thick] (4,0) -- (4,3);
        \draw[thick] (10,0) -- (10,3);
        \draw[thick,  ->] (10.7,1.5) -- (14.3,1.5);
        \node[above] at (12.5,1.6) {$\epsilon \to 0$};
	\node at (7,1.5) {SymTFT($\cC$)};
 \node[below] at (15,0) {$Z_{\cT_L}[B]$};
 \node[below] at (4,0) {$\langle L(B)|$};
 \node[below] at (10,0) {$|\cT_N[M_4]\rangle $}; 
 \node[above] at (15,3) {$x=0$};
 \node[above] at (4,3) {$x=0$};
 \node[above] at (10,3) {$x=\epsilon $}; 
	\end{tikzpicture}
    \caption{The 2d absolute theory is obtained by shrinking the interval.}
    \label{Fig:symTFT}
\end{figure}
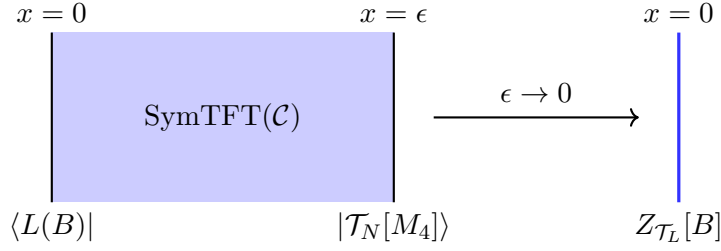

\paragraph{3d TQFT as SymTFT}

Choosing a polarization $L$ can be understood as putting the 3d TQFT on $\Sigma_2 \times I_{(0,\epsilon)}$. As shown in Figure \ref{Fig:compactM4}, the left boundary is the topological one specified by $L$ and the right one is dynamical encoding the dynamics of $T_N[M_4]$. 
In fact, this 3d TQFT defines the SymTFT of $T_N[M_4]$. Given a topological boundary condition, the absolute theory is obtained by shrinking the interval in Figure \ref{Fig:symTFT}.

The SymTFT defines the 0-form symmetries of $T_N[M_4]$ denoted by a group $G$ when it is invertible and by a fusion category $\cC$ in the general setting. 
As studied in \cite{Thorngren:2019iar, Lin:2022dhv, Kaidi:2022cpf}, the SymTFT is a Dijkgraaf-Witten (DW) theory \cite{Dijkgraaf:1989pz} when the symmetries are invertible. However, for non-invertible symmetries, the SymTFT is the Turaev-Viro theory on $\cC$ or equivalently the Reshetikhin-Turaev
theory on the Drinfeld center $Z(\cC)$.

\paragraph{Choice of Polarization and topological boundary conditions.}

As observed in \cite{Gukov:2020btk}, the choice of maximal isotropic sublattices corresponds to the different topological boundary conditions in the 3d TQFT. 
Maximal isotropic sublattices $L$ then correspond to those sublattices whose elements have trivial braiding with each other:
\begin{equation}
B(\vec{\alpha},\vec{\beta}) = 1 \quad \textrm{for} \quad \vec{\alpha}, \vec{\beta} \in L.
\end{equation}
As we have seen that different choices of $L$ lead to different absolute theories of $T_N[M_4]$.

When we make a choice of a maximal isotropic sublattice $L$ in 
$H_2(M_4,\mathbb{Z}_N)$, then we are free to choose any element of $H_1(\Sigma_2,\mathbb{Z}_N)$. This allows us to specify line defects in the boundary theory along any direction in $\Sigma_2$. This gives rise to an absolute theory on $\Sigma_2$ whose partition function with background fields $x \in H^1(\Sigma_2,L)$ we denote by $Z_{T_{N,L}[M_4]}[\Sigma_2;x]$. The corresponding wavefunctions in $\mathcal{H}_L$ are then labeled by 
\begin{equation}
    |\Psi_L\rangle = \sum_{x \in H^1(\Sigma_2,L)} Z_L[\Sigma_2;x]|x\rangle,
\end{equation}
where we have abbreviated the partition function of the 2d theory with the choice of polarization $L$ as $Z_L[\Sigma_2;x] \equiv Z_{T_L[M_4]}[\Sigma_2,x]$ with a background field $x \in H^1(\Sigma_2,L)$ turned on. Elements of $L$ can then be viewed as discrete versions of $x$-coordinates while elements in $L^{\perp}$ are discrete versions of $p$-coordinates. A \textit{Dirichlet}-boundary condition then amounts to pairing $|\Psi_L\rangle$ with a coordinate-state $\langle D_X|$ given by
\begin{equation}
    \langle D_X| = \sum_{x \in H^1(\Sigma_2,L)} \langle x| \delta(x - X),
\end{equation}
such that,
\begin{equation}
    \langle D_X | \Psi_L\rangle = Z_L[\Sigma_2;X].
\end{equation}
A \textit{Neumann} boundary condition then amounts to switching to momentum eigenstates $\langle N_P |$, where $P \in L^{\perp} \otimes H_1(\Sigma_2,\mathbb{Z})$, given by
\begin{equation}
    \langle N_P | = \frac{1}{|H^1(\Sigma_2,L)|} \sum_{x \in H^1(\Sigma_2,L)}e^{\frac{2\pi i}{N}\langle x, P\rangle} \langle x | .
\end{equation}

\paragraph{SPT phases.}

We can define Wilson surfaces in 6d which become line defects in the 2d theories we are after. We subdivide between two different types of Wilson surfaces, namely
\begin{eqnarray}
    \Phi_i(\gamma) & \equiv & \Phi(\gamma \times M_{2,i})=e^{\frac{2\pi i}{N}\oint_{\gamma \times M_{2,i}} c}, \nonumber \\
    \widehat{\Phi}_i(\gamma) & \equiv & \Phi(\gamma \times \widehat{M}_{2,i}) = e^{\frac{2\pi i}{N}\oint_{\gamma \times \widehat{M}_{2,i}}c},
\end{eqnarray}
where $M_{2,i} \in L$ and $\widehat{M}_{2,i} \in L^{\perp}$. The operators $\Phi_i$ do not change a given Dirichlet boundary condition, while the $\widehat{\Phi}_i$ act as \textit{discrete translation} operators and create line defects on the boundary. But note that the $\widehat{\Phi}_i$ crucially depend on the choice of representative of elements in $L^{\perp}$. The choice of a different representative amounts to shifting
\begin{equation}
    \widehat{M}_{2,i} \rightarrow \widehat{M}_{2,i} + \sum_j k_{ij} M_{2,j}, \quad k_{ij} \in \bZ.
\end{equation}
Using the quantum torus algebra \eqref{eq:qtorus}, under the above shift one has,
\begin{equation}
    \widehat{\Phi}_i \rightarrow {\widehat{\Phi}_i}' =  \Phi(\gamma_i \times \widehat{M}_{2,i} + \sum_j k_{ij} \gamma_i \times M_{2,j}) = e^{\frac{2\pi i}{N}\sum_j k_{ij} \langle \widehat{M}_{2,i},M_{2,j}\rangle \int_{\Sigma_2}\frac{A_i \cup A_i}{2}} \widehat{\Phi}_i(\gamma_i) \times \prod_j {\Phi_j}(\gamma_i)^{k_{ij}},
\end{equation}
where $A_i$ is the Poincare dual of $\gamma_i$. Similarly to \cite{Bashmakov:2022uek}, in a product $\prod_i {\widehat{\Phi}_i}'(\gamma_i)$ we can then first use the above splitting and then successively commute all $\Phi_j$ operators past the $\widehat{\Phi}_i$ operators and thus pick up an SPT phase,
\begin{equation}
    \prod_i \widehat{\Phi}'_i(\gamma_i) = \exp\left[\frac{2\pi i}{N} \sum_{i,j} k_{ij} \langle \widehat{M}_{2,i},M_{2,j}\rangle \int_{\Sigma_2} \left(\frac{A_i \cup A_i}{2} + A_i \cup A_j\right)\right] \prod_i \widehat{\Phi}_i(\gamma_i)~.
\end{equation}
where the factor $\frac{A_i \cup A_i}{2}$ can be understood as the possible quadratic refinement on $\Sigma_2$.

\paragraph{Topological manipulations}

Suppose the 2d absolute theories have a non-anomalous discrete symmetry $G$. There are three kinds of 2d topological manipulations that will transform between these theories. The first one is the gauging of subgroups $H\subset G$
\begin{equation}
    Z_{\cT/G}[A] \sim \sum  Z_{\cT}[a] e^{\frac{2\pi i}{N} \langle a,A \rangle}\,,
\end{equation}
where $a$ and $A$ are background fields for $H$ and $\hat{H}$ where $\hat{H}$ is the quantum symmetry after gauging, and $\langle a,A \rangle$ is the standard pairing on $\Sigma_2$.

The second topological manipulation is stacking the theory with an SPT phase $v_2 \in H^2(G, U(1))$. When $\cT$ is spin, one can also stack the fermionic SPT phase \cite{Gaiotto:2020iye}, for example, the Arf invariant in 1+1 dimension. With these SPTs, we can have orbifolding a subgroup $H \subset G$  
\begin{equation}
   Z_{\cT/_{v_2}G}[A] \sim \sum  Z_{\cT}[a] e^{\frac{2\pi i}{N} \langle a,A \rangle} \epsilon_{v_2}~,
\end{equation}
where $\epsilon_{v_2}$ is the action of the SPT phase.

Note that it is sufficient to determine if $T_N[M_4]$ is spin or not by studying the spin structure of 4-manifolds. 
In the dimensional reduction $M_6 = M_4  \times \Sigma_2$, the second Stiefel–Whitney class decomposes as $\omega_2(M_6) = \omega_2(M_4) + \omega_2(\Sigma_2)$. Obviously, 6d SCFTs are spin with $\omega_2(M_6)=0$. Thus, the existence of spin structure on $\Sigma_2$ requires that  $\omega_2(M_4)$ is also trivial \footnote{Assume that both $M_4$ and $\Sigma$ are orientable.}, i.e. $M_4$ is spin. A 4-manifold $M_4$ is spin if and only if all its self-intersection numbers are even \cite{spin4manifold}. For example, $\bP^1 \times \bP^1$ and the connected sum of them are spin, Hirzebruch surface $\mb{F}_l$ is spin when $l$ is even, and Del Pezzo surfaces are not spin.

The third topological manipulation is the permutation of the symmetry lines in $\cT$.   Notice that this manipulation only changes the way how the symmetry is coupled to the background fields on $\Sigma_2$ and will not lead to new global variants of $\cT$. We confirm this point by analyzing the topological defect lines in a theory with $\bZ_N$ symmetry in the next section.

The operations of gauging and stacking SPT phases and their composition are expected to generate all the global variants of $\cT$. These global variants are closed under these topological manipulations. If one only considers the gauging operations, then these theories and the associated operations form the {\it orbifold groupoids} \cite{Gaiotto:2020iye}. As we will see, these different global variants are different boundary conditions of the 3d SymTFT and the topological manipulations are determined by the automorphism group or (0-form symmetry) of the SymTFT.

\paragraph{Dualities from 4-manifold}

In class S theory, the mapping class group of torus leads to the $SL(2,\bZ)$ Montonen–Olive duality \cite{Montonen:1977sn, Witten:1995zh, Vafa:1997mh}. 
The mapping class group of 4-manifold $\MCG(M_4)$ is given by  
%by the automorphism group of $H_2(M_4,\bZ)$, i.e. 
%
\begin{equation} \label{Eqn:autZ}
    P^t Q P = Q,\; \qquad P \in GL(r,\bZ),
\end{equation}
where $r$ is the rank of the intersection form $Q$. 
Similarly, we expect that $\MCG(M_4)$ will give rise to Montonen–Olive-like dualities for 2d theories.

\paragraph{Global variants.}

The automorphism group or discrete 0-form symmetries of the 3d SymTFT denoted by $\Aut_{\bZ_N}(Q)$ can be determined by  
\begin{equation} \label{Eqn:autZn}
    T^t Q T = Q,\; \qquad T \in GL(r,\bZ_N),
\end{equation}
where $r$ is the rank of the intersection form $Q$. These elements in $\Aut_{\bZ_N}(Q)$ transforming different absolute theories or global variants of $T_N[M_4]$ correspond to the 2d topological manipulations \cite{Gaiotto:2020iye}. 
In particular, this group can be decomposed as 
\begin{equation} \label{Eqn:AutQN-decomp}
    \Aut_{\bZ_N}(Q) = \Aut(G) \times \cO_N(Q)
\end{equation}
where $\Aut(G)$ is the automorphism group of the symmetry of $T_4[M_4]$ that corresponds to permutations of the 2d symmetry lines while the group $\cO_N(Q)$ corresponds to the different ways of gauging and stacking possible SPT phases, which give rise to different global variants of $T_N[M_4]$. 
Thus, the number of global variants is simply given by 
\begin{equation}
    d(N) = \frac{|\Aut_{\bZ_N}(Q)|}{|\Aut(G)|}= |\cO_N(Q)|
\end{equation}
There is a similar result for the class S theory \cite{Bashmakov:2022uek}.

One can associate each global variant with a matrix $M \in \cO_N(Q)$. In fact, these matrix representations can also be obtained from the data of $L$ and $L^{\perp}$. These matrices are closed under dualities and topological manipulations discussed above. The action of duality is from the left,
\begin{equation} \label{Eqn:actF}
    M \to F^t M, \hspace{2cm} F \in \MCG(M_4),
\end{equation}
while a topological manipulation acts from the right,
\begin{equation} \label{Eqn:actG}
    M \to  M G, \hspace{2cm} G \in \Aut_{\bZ_N}(Q)~.
\end{equation}
The actions of dualities and topological manipulations on global variants of $T_N[M_4]$ will play an important role in realizing topological defects later.

\section{6d $\cN=(2,0)$ SCFTs on $\bP^1 \times \bP^1$} \label{sec:3}

In this section, we will study the theory from the compactification of the 6d $\cN=(2,0)$ theories of type $A_{N-1}$ on $M_4 = \bP^1 \times \bP^1$. 
By choosing the maximal isotropic sublattice $L$, different absolute theories of $T_N[M_4]$ are obtained on the boundary of the $\bZ_N$ gauge theories. 
Using this SymTFT, we will study their global variants (when SPT phases are 
considered), and analyze the symmetries and possible anomalies.

\subsection{$\bZ_N$ gauge theory}

First, we will introduce the $\bZ_N$ gauge theory which is the 3d SymTFT of $T_N[\bP^1 \times \bP^1]$. 
The homology of $\bP^1 \times \bP^1$ is 
\begin{equation}
    H_{*}(\bP^1 \times \bP^1,\bZ) = \{ \bZ,0,\bZ^2,0,\bZ \},
\end{equation}
with intersection form 
\begin{equation} \label{Eqn:interP1P1}
Q=    
\begin{pmatrix}
0 & 1  \\
1 & 0
\end{pmatrix}.
\end{equation}
Let $b$ and $f$ be a basis of $H_2(\bP^1\times \bP^1,\bZ)$, with intersection numbers
\be
b^2=0\ ,\ f^2=0\ ,\ b\cdot f=1\,.
\ee By the equation (\ref{Eqn:afromc}), one can define the following two 1-form gauge fields,
\begin{equation*}
    a = \int_{b} c, \hspace{0.5 in} \hat a = \int_{f} c~.
\end{equation*}
Integrating over $\bP^1\times \bP^1$, the 3d action becomes 
\begin{eqnarray}
    S_{3d} &=&  \frac{N}{4\pi} \int_{W_3}  a \wedge d\widehat{a} + \widehat{a} \wedge d a\\
    &=& \frac{2 \pi}{N} \int_{W_3} a  \cup \delta  \widehat{a}\;, \nonumber
\end{eqnarray}
where $W_3= \Sigma_2\times I_{[0,\varepsilon]}$ is a slab. Let $x$ be the coordinate of the interval $I_{[0,\varepsilon]}$, then the two boundaries are located at $\Sigma_2|_{x=\varepsilon}$ and $\Sigma_2|_{x=0}$ corresponding to the topological and dynamical boundary, respectively. 
Notice that this Chern-Simons action has the form of a $\bZ_N$ discrete gauge theory. The gauge fields can also be written in terms of the $\bZ_N$-valued 1-cochains as $a \to \frac{2 \pi i}{N} a$.

The 3d $\bZ_N$ discrete gauge theory has line operators, which can be obtained from discrete Wilson surfaces as follows, 
\begin{eqnarray}
\label{eq:beforegaugingLs}
L_{(e,m)}(\gamma) =  \exp\left(\frac{2\pi i}{N} \oint_{\gamma\times M_2} c \right) = \exp\left(\frac{2\pi i}{N} \oint_\gamma e a\right) \exp\left( \frac{2\pi i}{N} \oint_\gamma m \widehat{a}\right)~,
\end{eqnarray}
where $M_2 = e b + m f \in H_2(\bP^1 \times \bP^1,\bZ_N)$ and $ (e, m) \in \bZ_N \times \bZ_N$ are the electric/magenatic charges. 
The topological spin of the line operator is
\begin{equation}
    \theta \left( L_{(e,m)} \right) = \exp \left( \frac{4\pi i}{N}em\right).
\end{equation}
Notice that $L_{(1,0)}$ and $L_{(0,1)}$ together generate a $\bZ_N^{(1)}\times \bZ_N^{(1)}$ 1-form symmetry for the 3d SymTFT.

The fusion rule between two distinct line defects is given by  
\begin{equation}
    L_{(e,m)}(\gamma) \times L_{(e',m')}(\gamma) = L_{(e+e',m+m')}(\gamma)~.
\end{equation}
The braiding between them is 
\begin{equation}
    L_{(e,m)}(\gamma) L_{(e',m')}(\gamma') = \exp\left(-{2 \pi i \over N} (e m' +  m e') \langle \gamma, \gamma'\rangle \right) L_{(e',m')}(\gamma')  L_{(e,m)}(\gamma) ~,
\end{equation}
where $\langle \gamma, \gamma'\rangle$ represents the intersection number between $\gamma$ and $\gamma'$ on $\Sigma_2$.

As shown in Figure \ref{Fig:symTFT}, to obtain an absolute 2d theory, one needs to specify a topological boundary condition in $\bZ_N$ gauge theory and then shrink the slab. One can take the Dirichlet-boundary condition and half of the line operators will survive on the boundary generating a $\bZ_N$ 0-form symmetry for the 2d theory. One can also take other topological boundary conditions giving rise to different global variants of $T_N[\bP^1 \times \bP^1]$. As we will see later, these theories all have $\bZ_N$ 0-form symmetry and can be related to each other by topological manipulations.

In general, a $\bZ_N$ discrete gauge theory is the SymTFT for theories with invertible $\bZ_N$ symmetry with fusion category denoted by $\text{Vec}_{\bZ_N}$. One can see that, as the $\bZ_N$ discrete gauge theory has $N^2$ lines, while the fusion category $\text{Vec}_{\bZ_N}$ only admits $N$ lines, the 3d SymTFT is the quantum double of the categorical symmetry in 2d.

\subsection{Orbifold groupoid and global variants}

In this subsection, we will study how many different absolute 2d theories can be obtained by choosing a suitable polarization on $M_4$. The discrete isometry of $M_4$ usually leads to interesting operations on these absolute theories. These operations act on the polarizations and can transform these 2d absolute theories between each other. Our analysis using the polarization matches with the analysis from the 2d field theories. These different absolute theories can be related by gauging and stacking the SPT phase. We take $N=2$, $N=p$ prime numbers, $N=4$ and $N=6$ to illustrate our results.

\paragraph{Topological manupulations.}

Consider a 2d theory $\cX$ with an anomaly free $\bZ_N$ zero-form global symmetry on a closed two-dimensional spacetime $\Sigma_2$. 
One can introduce two topological operations 
\begin{itemize}
    \item gauging 0-form symmetry $\bZ_N$ denoted by $\sigma$:
    
    % We denote the $\bZ_N$  background gauge field as $A$, and the partition function as $Z_{T[M_4]}[T^2,A]$. Gauging $\bZ_N$ gives a new theory $T[M_4]/\bZ_N$, 
\begin{eqnarray}\label{eq:2dgauging}
 Z_{T_N[M_4]/\bZ_N}[\Sigma_2,A] \sim 
 %\frac{1}{|H^0(T^2, \bZ_N)|} 
 \sum_{a\in H^1(\Sigma_2, \bZ_N)} Z_{T_N[M_4]}[\Sigma_2,a]\, e^{ \frac{2\pi i}{N} \int_{\Sigma_2} a A}~,
\end{eqnarray}
where now $A\in H^1(\Sigma_2, \widehat \bZ_N)$ is the background field of the quantum symmetry $\widehat\bZ_N$ after gauging.

\item In our case, there is no bosonic SPT phase since $H^2(\bZ_N, U(1)) = 0$. But, when the theory has the spin structure, we can stack a fermionic SPT phase, i.e. Arf invariant \cite{Karch:2019lnn, Ji:2019ugf}. 

\end{itemize}

\paragraph{Duality}

The automorphism group of $\bP^1 \times \bP^1$  is the matrix that preserves the quadratic form defined by $Q$ with action on $H_2(\bP^1\times \bP^1,\bZ)$. 
It turns out that these matrices are elements of the $\MCG(\bP^1 \times \bP^1)$\footnote{Precisely, the automorphism group $\MCG(\bP^1 \times \bP^1)$ is $O(II_{1,1},\bZ)$ where $II_{1,1}$ is the rank-2 even unimodular matrix. 
% intersection form is congruent to diagonal matrix $(2,-2)$, which is not $O(1,1,\bZ)$.
}, which is isomorphic to $\bZ^2_2$ given by 
\begin{equation} \label{Eqn:p1p1duality}
I=\left(
\begin{array}{cc}
 1 & 0 \\
 0 & 1 \\
\end{array}
\right), \quad 
-I=\left(
\begin{array}{cc}
 -1 & 0 \\
 0 & -1 \\
\end{array}
\right), \quad 
s=
\left(
\begin{array}{cc}
 0 & 1 \\
 1 & 0 \\
\end{array}
\right),\quad 
-s=
\left(
\begin{array}{cc}
 0 & -1 \\
 -1 & 0 \\
\end{array}
\right)
\end{equation}
These matrices correspond to the switch of two $\bP^1$'s and the flip of their orientation. The duality is generated by the element $s$.

\paragraph{For $N=2$:}

As discussed in the previous section, to obtain absolute theories, one needs to specify a maximal isotropic sublattice $L \subset H_2(\bP^1 \times \bP^1,\bZ_2)= (\bZ_2)^2$, i.e. a $2\times 2$ integral lattice. 
Besides $L$, the other piece of information is the choice of the elements in $L^{\perp}$, which determines the couplings of the background fields and possible stacking of the SPT phase.

With the inner product in equation \eqref{Eqn:isotropyM4}, we find the following three maximal isotropic sub-lattices,
\begin{equation} \label{Eqn:LN2}
\begin{array}{cccc}
L_1= \{(0,0), & (0,1)\} &\to &\bZ_2\\
L_2=\{ (0,0), & (1,0)\} &\to &\widehat{\bZ}_2 \\
L_3 = \{ (0,0), & (1,1)\}&\to &\bZ_2^f
\end{array}
\end{equation}
Thus, we have three 2d absolute theories. 
As a theory on the topological boundary of $\bZ_2$ discrete gauge theory, all these three theories have $\bZ_2$ symmetry. We will label them by $\bZ_2$, $\widehat{\bZ}_2$, and $\bZ_2^f$.
The physical meaning of these notations will become clear later.

From the equation \eqref{Eqn:autZn}, the automorphism group is $\Aut_{\bZ_2}(Q)=S_3$, with generators
\begin{equation*}
\sigma=\left(
\begin{array}{cc}
 0 & 1 \\
 1 & 0 \\
\end{array}
\right),
\qquad 
\tau=\left(
\begin{array}{cc}
 1 & 1 \\
 0 & 1 \\
\end{array}
\right).
\end{equation*}
This automorphism group determines the 0-form symmetries of the SymTFT, which transform between different topological boundary conditions, or in other words, between different absolute theories by topological manipulations \cite{Gukov:2020btk, Gaiotto:2020iye}. 

\begin{figure}
\centering
\begin{tikzpicture}[scale=1.5]
\draw node at (0,0) {$\bZ_2$};
\draw node at (2,0) {$\widehat{\bZ}_2$};
\draw node at (4,0) {$\bZ_2^f$};
\draw [<-,blue] (.3,0) -- (1.7,0);
\draw [<-,blue] (2.3,0) -- (3.7,0);
\draw [->,blue] (0,.2) arc (120:60:4);
\draw node at (2,0.9) {${\color{blue} g}$};
\draw node at (1, 0.2) {$\color{blue} g$};
\draw node at (3, 0.2) {$\color{blue} g$};
\end{tikzpicture}
\caption{Orbifold groupoids for $T_2[\bP^1 \times \bP^1]$ with $\bZ_2$ symmetry. The map $g$ represents the topological manipulation gauging $\bZ_2$ up to an SPT phase. }
\label{Fig:N2f1}
\end{figure}
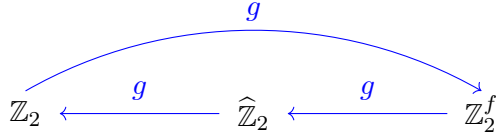

We find $g \in S_3$ acts transitively on three maximal isotropic sublattices and transforms different absolute theories as in Figure \ref{Fig:N2f1} with 
\begin{equation*}
    g  = \sigma \tau =\left(
\begin{array}{cc}
 1 & 1 \\
 1 & 0 \\
\end{array}
\right),
\end{equation*}
This gives the orbifold groupoid for a fermionic theory with $\bZ_2$ symmetries \cite{Gaiotto:2020iye}. It implies that the theory $T_2[\bP^1 \times \bP^1]$ is fermionic, which makes sense because the 4-manifold $\bP^1 \times \bP^1$ has spin structure.

For 2d fermionic theories with non-anomalous $\bZ_2$ symmetry, up to SPT phases, there are indeed three absolute theories \cite{Karch:2019lnn}. In particular, the theories $\bZ_2$ and $\widehat{\bZ_2}$ are related by gauging with $\widehat{\bZ_2}$ being the quantum symmetry. The theory $\bZ_2$ and $\bZ_2^f$ are related by fermionization/bosonization. Thus, with the help of SymTFT, we are able to determine the orbifold groupoid of $T_2[\bP^1 \times \bP^1]$ and the predictions are consistent with the field-theoretic analysis made in \cite{Gaiotto:2020iye}.

The maximal isotropic sublattice defines three  absolute theories. 
To obtain the global variants of $T_2[\bP^1 \times \bP^1]$, one needs also to specify the representatives in $L^{\perp}$. For example, consider the theory $\bZ_2$ defined by $L_1$. The complement of it is given by $L_1^{\perp}=H_2(M_4,\bZ_2)/L_1$, which contains two equivalent classes 
\begin{equation*}
    [(0,0)]=\{(0,0),(0,1)\}, \qquad [(1,0)]=\{(1,0),(1,1)\}.
\end{equation*}
The choice of representative in $[(1,0)]$ determines the possible stacking of the SPT phase.  
As studied in \cite{Bashmakov:2022uek}, the choice of $(1,0)$ implies that the theory does not stack an SPT phase denoted by $(\bZ_2)_0$ while the choice of $(1,1)$ means that the 2d theory is stacked with an SPT phase denoted by $(\bZ_2)_1$. 
The choice of representative in $L^{\perp}$ is denoted by $B^L$. Similarly, depending on whether an SPT phase is present, there are two global variants for each maximal isotropic sublattice.
In the following, we label them using the subscript 0/1 to denote if there are SPT phases stacked.

Thus, we find a total of $6$ global variants for $T_2[\bP^1 \times \bP^1]$ specified by $(L,B)$. As observed in \cite{Bashmakov:2022uek}, there is a prescription to assign the global variants to the matrix in the automorphism group as 
\begin{equation} \label{eqn:presM}
    M_{L,B}=\left(K_L,K_B \right),
\end{equation}
where $K_L$ is a vector containing the lattice point generating the polarization $L$ and $K_B $ denotes the representative of the lattice point in $L^{\perp}$. In this way, one can associate each global form with a $2\times 2$ matrix.

The automorphism groups $\Aut_{\bZ_2}(Q)$ and $\MCG(\bP^1 \times \bP^1)$ transform among these global variants, which correspond to perform the topological manipulations $(\sigma, \tau)$ and dualities $s$ to these 2d theories. By the action from the equation \eqref{Eqn:actF} and \eqref{Eqn:actG}, we find these global forms transform according to Figure (\ref{Fig:N2f2}). As one can check this matches the field theoretical analysis in \cite{Karch:2019lnn}. Thus, we have found all global variants of $T_2[\bP^1 \times \bP^1]$ and identified how they transform under topological manipulations and duality.

\begin{figure}
    \centering
   \begin{tikzpicture}[scale=1.5]
\draw node at (0,0) {$(\bZ_2)_0$};
\draw node at (3,0) {$(\widehat{\bZ}_2)_0$};
\draw node at (6,0) {$(\bZ_2^f)_0$};
\draw node at (0,2) {$(\bZ_2)_1$};
\draw node at (3,2) {$(\widehat{\bZ}_2)_1$};
\draw node at (6,2) {$(\bZ_2^f)_1$};
\draw [<->,orange] (.5,0) -- (2.5,0);
\draw [<->,orange] (.5,2) -- (2.5,2);
% \draw [<->,blue] (3.5,2) -- (5.5,2);
%%
\draw [<->,blue] (0.05,0.2) -- (0.05,1.8);
\draw [<->,blue] (3,0.2) -- (3,1.8);
\draw [<->,blue] (5.95,0.2) -- (5.95,1.8);
\draw [<->,orange] (6.05,0.2) -- (6.05,1.8);
\draw [<->,blue] (0.2,1.8) -- (5.8,0.2);
\draw [<->,blue] (0,-.2) arc (-120:-60:3);
\draw [<->,blue] (3,2.2) arc (120:60:3);
\draw node at (1.5,.2) {${\color{orange} s}$};
\draw node at (1.5,2.2) {${\color{orange} s}$};
\draw node at (1.5,-.45) {${\color{blue} \sigma}$};
\draw node at (4.5,2.75) {${\color{blue} \sigma}$};
\draw node at (0.2,1) {${\color{blue} \tau}$};
\draw node at (5.8,1) {${\color{blue} \tau}$};
\draw node at (6.2,1) {${\color{orange} s}$};
\draw node at (2.8, 0.6) {$\color{blue} \tau$};
\draw node at (4, 0.9) {$\color{blue} \sigma$};
\node[below] at (0,-0.5) {$\left(\begin{matrix} 0 & 1 \\ 1 & 0 \end{matrix}\right)$};
\node[above] at (0,2.5) {$\left(\begin{matrix} 0 & 1 \\ 1 & 1 \end{matrix}\right)$};
  \node[below] at (3,-0.5) {$\left(\begin{matrix} 1 & 0 \\ 0 & 1 \end{matrix}\right)$};
 \node[above] at (3,2.5) {$\left(\begin{matrix} 1 & 1 \\ 0 & 1 \end{matrix}\right)$};
 \node[below]  at (6,-0.5) {$\left(\begin{matrix} 1 & 0 \\ 1 & 1 \end{matrix}\right)$};
 \node[above] at (6,2.5) {$\left(\begin{matrix} 1 & 1 \\ 1 & 0 \end{matrix}\right)$};
\end{tikzpicture}
    \caption{Web of transformations for $T_2[\bP^1 \times \bP^1]$. The transformations
in orange are the duality transformations.
The transformations in blue are topological manipulations.}
    \label{Fig:N2f2}
\end{figure}

\paragraph{For $N=p>2$:}

For prime number $N=p>2$, there are two maximal isotropic sublattices 
\begin{equation} \label{Eqn:LN3}
\begin{array}{cccc}
L_1 = \{(0,0),(1,0),\ldots, (p-1,0)\} &\to &\bZ_p\\
L_2 = \{(0,0),(0,1),\ldots, (0,p-1)\} &\to &\widehat{\bZ}_p 
\end{array}
\end{equation}
which defines two absolute theories with $\bZ_p$ symmetry. We will denote them by $\bZ_p$ and $\widehat{\bZ}_p$ because as we will see that they are related by $\bZ_p$ gauging.

One can show that $L_1$ and $L_2$ are the only two maximal isotropic sublattices. Let's consider the sublattice generated by a lattice point $(e,m)$ other than $(0,0)$ in $\bZ^2_p$. Thus, the sublattice contains points $(e',m')$ satisfying $(e',m') = k (e,m)$ with $k\in \bZ_p^{\times}$. The inner product between these two points is 
\begin{equation} \label{Eqn:p1p1prime}
    2kem = 0, \hspace{0.3cm} \text{mod}\; p .
\end{equation}
For prime $p$, the only solution is either $e=0$ or $m=0$, which gives the two maximal isotropic sublattices $L_1$ and $L_2$. One can also consider the sublattices generated by two or more linear independent points in $\bZ^2_p$. However, in this case, one always gets the full lattice, which is obviously not isotropic.

We find the automorphism group $\Aut_{\bZ_p}(Q)$ is the Dihedral group $D_{2(p-1)}$ defined by 
\begin{equation}
    D_{2(p-1)} = \langle r,\sigma | r^{p-1}=\sigma^2=(\sigma r)^2=1\rangle
\end{equation}
The order is $2(p-1)$ and the two generators are 
\begin{equation*}
r=\left(
\begin{array}{cc}
 r_1 & 0 \\
 0 & r_2 \\
\end{array}
\right),
\qquad 
\sigma=\left(
\begin{array}{cc}
 0 & 1 \\
 1 & 0 \\
\end{array}
\right)~,
\end{equation*}
where $r_1$, $r_2$ are integers coprime to $p$ and satisfy $r_1r_2=1 $, $\text{mod}\;p$. According to equation \eqref{Eqn:AutQN-decomp}, this group can be decomposed as 
\begin{equation}
    \Aut(\bZ_p) = \bZ^{\times}_p  = \langle r \rangle, \qquad  \cO_p(Q)= \bZ_2 = \langle \sigma \rangle.
\end{equation}
Here $\bZ_2$ transforms between Dirichlet and Neumann boundary conditions in the bulk theory corresponding to performing $\bZ_p$ gauging to 2d theories. However, $\bZ^{\times}_p$ will not give new global variants. As we will see below, they correspond to different ways to turn on the background fields of the same global variant on $\Sigma_2$.

Thus, there are two absolute theories and they transform into each other by gauging $\sigma$. 
Note that in this case, it is not possible to stack the Arf invariant as the generator $\tau$ which was present in the $N=2$ case is missing for $N=p > 2$ and $p$ prime, so we have only two global variants and we can assign two $2\times 2$ matrices in $\langle \sigma \rangle$. Taking into account the duality $s$, 
we plot the orbifold groupoid and global variants in Figure \ref{Fig:N3f1}. This is consistent with the result in \cite{Gaiotto:2020iye}. 

\begin{figure}
\centering
\begin{tikzpicture}[scale=1.5]
\draw node at (0,0) {$\bZ_p$};
\draw node at (3,0) {$\widehat{\bZ}_p$};
% \draw [<->,orange] (.3,0) -- (2.7,0);
\draw [<->,blue] (0,.3) arc (120:60:3);
\draw [<->,orange] (0,-.3) arc (-120:-60:3);
 \draw node at (1.5,-0.5) {${\color{orange} s}$};
 \draw node at (1.5,.5) {${\color{blue} \sigma}$};
\node[left] at (-0.3,0) {$\left(\begin{matrix} 0 & 1 \\ 1 & 0 \end{matrix}\right)$};
\node[right] at (3.3,0) {$\left(\begin{matrix} 1 & 0 \\ 0 & 1 \end{matrix}\right)$};
\end{tikzpicture}
    \caption{Web of transformations for $T_p[\bP^1 \times \bP^1]$. The transformations
in orange are the duality transformations.
The transformations in blue are topological manipulations.}
    \label{Fig:N3f1}
\end{figure}

\paragraph{Topological defect lines:}

The automorphism group implies that there are actually $2(p-1)$ orbifolding theories from the 6d perspective. Indeed, from the 2d viewpoint, we have exactly $2(p-1)$ ways to orbifold a 2d theory with $\mathbb Z_p$ symmetries, once we turn on the background gauge field for the $\mathbb Z_p$ symmetries, and the $D_{2(p-1)}$ automorphism group will be faithfully manifest. This point can be verified at the level of partition functions on the  torus. Recall that, for a 2d CFT denoted as $Z_p$, the $\mathbb Z_p$ symmetries are identical to $p$ different topological defect lines (TDLs). We can put these TDLs along either temporal or spatial directions, and thus overall there are $p^2$ numbers of defect partition functions denoted by $Z_p(a_1,a_2)$, where $a_{i}\in\mathbb Z_p$ are holonomies with respect to $\mathbb Z_p$ along two cycles of the torus, and label the different types of TDLs. For a given theory $Z_p$, we have a collection of partition functions dressed with these TDLs as $\left\{Z_p(a_1,a_2)\right\}_{a_i\in\mathbb Z_p}$. Now having $Z_p$ at hand, we spell out the orbifolding theories as
\begin{align}
    \hat Z^k_p(b_1,b_2)=\frac{1}{p}\sum_{a_i\in\mathbb Z_p}Z_p(a_1,a_2)\,\omega_{p,k}^{a_1 b_2-a_2 b_1}\,,
\quad {\rm or} \quad
Z_p(a_1,a_2)=\frac{1}{p}\sum_{b_i\in\mathbb Z_p}\hat{Z}^k_p(b_1,b_2)\,\omega_{p,k}^{b_1 a_2-b_2 a_1}\,,
\label{eq:orbifold}
\end{align}
where the $b_i$ label the types of TDLs with respect to the quantum symmetries $\hat{\mathbb Z}_p$ in $\hat Z_p$, and 
\begin{align}
\omega_{p,k}\equiv e^{\frac{2\pi i k}{p}}\,,\qquad {\rm with}\quad 1\leq k < p\,.
\end{align} 
Therefore, starting from $Z_p$, one has $(p-1)$-ways to orbifold it, denoted by $\hat{Z}^k_p$ with a collection of $\left\{\hat Z^k_p(a_1,a_2)\right\}_{a_i\in \hat{\mathbb Z}_p}$ defect partition functions. One can continue this operation from one of the resulted $p-1$ orbifolding theories, say for example $\hat{Z}^1_p$. But notice that now there are only $(p-2)$-ways to obtain new orbifolding theories from it, as one way will transform $\hat{Z}^1_p$ back to $Z_p$ from \eqref{eq:orbifold}. Overall, there are
\begin{align}
    N_p=1+(p-1)+(p-2)=2(p-1)
\end{align}
orbifolding theories corresponding to the group elements in $D_{2(p-1)}$. It is not hard to show that there are no more new orbifolding theories apart from the $N_p$ ones obtained this way.

For example, in the case of $p=3$, we have $4$ orbifolding theories given below
\begin{align}
\begin{gathered}
\begin{tikzpicture}[scale=1.5]
\draw node at (0,0) {$Z_3$};
\draw node at (2,0) {$\hat Z^1_3$};
\draw node at (0,2) {$\hat Z^2_3$};
\draw node at (2.8,2) {$\left(\widehat{\hat Z^2_3}\right)^1 =\left(\widehat{\hat Z^1_3}\right)^2$};
\draw [<->,blue] (.2,0) -- (1.8,0);
\draw [<->,blue] (.2,2) -- (1.8,2);
\draw [<->,red] (-0.05,0.2) -- (-0.05,1.8);
\draw [<->,red] (2.05,0.2) -- (2.05,1.65);
%
%\draw node at (2,-.2) {${\color{blue} \sigma_4}$};
%\draw node at (2,2.2) {${\color{blue} \sigma_4}$};
%\draw node at (-0.2,1) {${\color{blue} f}$};
%\draw node at (4.2,1) {${\color{blue} f}$};
\draw [<->,orange] (.2,0.2) -- (1.8,1.8);
\draw [<->,orange] (0.2,1.8) -- (1.8,0.2);
\end{tikzpicture}
\end{gathered}
\,,
\end{align}
where the red, blue, and orange lines denote the orbifolding with respect to $\omega_{3,1}$, $\omega_{3,2}$ and the charge conjugation operation $\mathcal C$,
\begin{align}
    \mathcal C:\quad Z_p(a_1, a_2)\longrightarrow Z_p(-a_1,-a_2)\,.
\label{eq:Cconj}
\end{align}
One can honestly check that
\begin{align}
    Z_3(a_1,a_2)=\left(\widehat{\hat Z^2_3}\right)^1(-a_1,-a_2)\,,\quad {\rm and}\quad
    \hat Z^2_3(a_1,a_2)=\hat Z^1_3(-a_1,-a_2)\,.
\end{align}
Therefore, together with the identity operation, $\left\{1,\,\omega_{3,1},\,\omega_{3,2},\, \mathcal C\right\}$ are precisely identified with the automorphism group $\mathbb Z_2\times\mathbb Z_2$ for $p=3$.

Below, we also draw the diagram of the orbifolding theories for $p=5$, as elements of the first non-abelian group $D_8$,
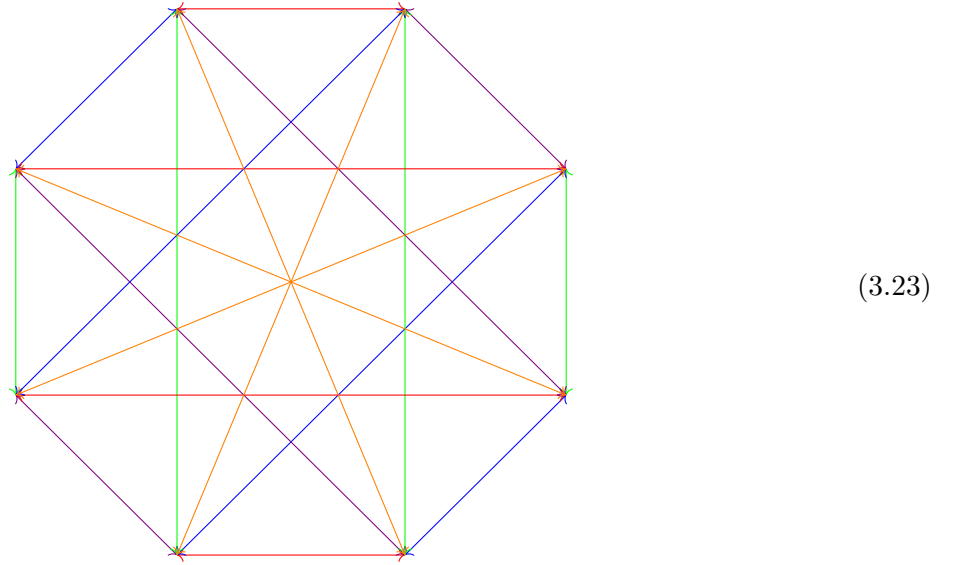
\begin{align}
\begin{gathered}
\begin{tikzpicture}[scale=1.5]
\draw [<->,red] (0,0) -- (2,0);
\draw [<->,blue] (2,0) -- (3.414,1.414);
\draw [<->, green] (3.414,1.414) -- (3.414,3.414);
\draw [<->, violet] (3.414,3.414) -- (2, 4.828);
\draw [<->, red] (2, 4.828) -- (0,4.828);
\draw [<->, blue] (0,4.828)--(-1.414,3.414);
\draw [<->, green] (-1.414,3.414)--(-1.414,1.414);
%-----
\draw [<->, violet] (0,0)--(-1.414,1.414);
\draw [<->, blue] (0,0)--(3.414, 3.414);
\draw [<->, green] (0,0)--(0, 4.828);
%-----
\draw [<->, violet] (2,0) -- (-1.414,3.414);
\draw [<->, green] (2,0) -- (2,4.828);
%------
\draw [<->, red] (3.414,1.414) -- (-1.414,1.414);
\draw [<->, violet] (3.414,1.414) -- (0,4.828);
%-----
\draw [<->, red] (3.414,3.414) -- (-1.414,3.414);
%-----
\draw [<->, blue] (2,4.828) -- (-1.414,1.414);
%-----
\draw [<->, orange] (0,0) -- (2, 4.828);
\draw [<->, orange] (2,0) -- (0, 4.828);
\draw [<->, orange] (3.414,1.414) -- (-1.414, 3.414);
\draw [<->, orange] (3.414,3.414) -- (-1.414, 1.414);
\end{tikzpicture}
\end{gathered}
\label{fig:p=5}
\end{align}
where the 8 vertices denote the 8 orbifolding theories, and the red, blue, purple, green and orange lines represent the orbifolding action with respect to $\omega_{5,1},\,\omega_{5,2},\,\omega_{5,3},\,\omega_{5,4}$ and charge conjugation $\mathcal C$ defined in \eqref{eq:Cconj}. Following the orbifolding trajectories in \eqref{fig:p=5}, one can convince oneself that $\omega_{5,1}$ and $\omega_{5,2}$ generate the whole diagram and satisfy the following relations,
\begin{align}
\omega_{5,1}^2=\omega_{5,2}^2=\left(\omega_{5,1}\cdot\omega_{5,2}\right)^4=1\,.
\end{align}
Therefore, one can identify the orbifolding groupoid as
\begin{align}
D_8=\left\langle r=\omega_{5,1}\cdot\omega_{5,2},\,s=\omega_{5,1}\vert\,r^4=s^2=\left(rs\right)^2=1\right\rangle~.
\end{align}

The above groupoid structure can be easily generalized to abitrary odd prime number $p$, which is generated by
\begin{align}
    D_{2(p-1)}=\left\langle r=\omega_{5,1}\cdot\omega_{5,2},\,s=\omega_{5,1}\vert\,r^{p-1}=s^2=\left(rs\right)^2=1\right\rangle~.
\end{align}
Once we turn off the background gauge field, the above diagram collapses back to \eqref{Fig:N3f1} corresponding the $\mathbb Z_2$ automorphism subgroup as discussed before. It simply implies that the $p-1$ different way of orbifolding just give the same orbifolded theory up to the automorphism group ${\rm Aut}(\mathbb Z_p)=\mathbb Z_p^\times$.

\paragraph{For $N=4$:}

We find the following 5 maximal isotropic sublattices 
\begin{equation} \label{Eqn:LN4}
\begin{array}{cccccc}
L_1= \{(0,0), & (0,1), & (0,2), & (0,3)\}  &\to &\bZ_4\\
L_2=\{ (0,0), & (1,0), & (2,0), & (3,0) \} &\to &\widehat{\bZ}_4 \\
L_3 = \{ (0,0), & (0,2), & (2,1), & (2,3) \}&\to &\bZ_4^f\\
L_4 =\{(0,0), & (2,0), & (1,2), & (3,2) \}&\to &\widehat{\bZ}_4^f\\
L_5 = \{ (0,0), & (0,2), & (2,0), & (2,2) \} &\to &(\bZ_2\times\widehat{\bZ}_2 )_{\mu_3}\\
\end{array}
\end{equation}
It implies that we have 5 different absolute theories. We will label them with their symmetries and possible anomalies. The first four theories have $\bZ_4$ symmetry while the last one has anomalous $\bZ_2\times\widehat{\bZ}_2$and will be discussed later.

The $\bZ_4$ symmetry is defined by the non-trivial extension of $\bZ_2$ by $\bZ_2$
\begin{equation} \label{Eqn:z4}
    1 \to \bZ_2 \to \bZ_4 \to \bZ_2 \to 1~.
\end{equation}
This central extension is determined by the cohomology classes $\kappa \in H^2(\bZ_2,\bZ_2)=\bZ_2$. When $\kappa$ is non-trivial, the extension gives $\bZ_4$. Gauging $\bZ_4$, one obtains another absolute theory with quantum symmetry given by 
\begin{equation}\label{Eqn:z4hat}
    1 \to \widehat{\bZ}_2 \to \widehat{\bZ}_4 \to \widehat{\bZ}_2 \to 1
\end{equation}
Besides that, since $\bZ_4$ has a subgroup $\bZ_2$, one can gauge $\bZ_4$ with an Arf invariant stacked. In this way, one obtains a fermionic theory with symmetry 
\begin{equation}\label{Eqn:z4f}
    1 \to \bZ_2^f  \to \bZ_4^f \to  \bZ_2 \to 1
\end{equation}
where $\bZ_2 = (-1)^F$. 
Similarly, one can gauge $\widehat{\bZ}_4$ with an Arf invariant and the theory so obtained is also fermionic with symmetry 
\begin{equation}\label{Eqn:z4fhat}
    1 \to \bZ_2^f \to \widehat{\bZ}_4^f \to  \widehat{\bZ}_2 \to 1
\end{equation}
We label the first four absolute theories using the symmetries defined above.

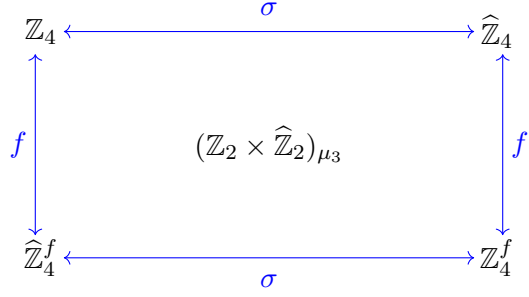
\begin{figure}
    \centering
    \begin{tikzpicture}[scale=1.5]
\draw node at (0,0) {$\widehat{\bZ}_4^f$};
\draw node at (4,0) {$\bZ_4^f$};
\draw node at (2,1) {$(\bZ_2 \times \widehat{\bZ}_2)_{\mu_3}$};
\draw node at (0,2) {$\bZ_4$};
\draw node at (4,2) {$\widehat{\bZ}_4$};
\draw [<->,blue] (.2,0) -- (3.8,0);
\draw [<->,blue] (.2,2) -- (3.8,2);
\draw [<->,blue] (-0.05,0.2) -- (-0.05,1.8);
\draw [<->,blue] (4.05,0.2) -- (4.05,1.8);
\draw node at (2,-.2) {${\color{blue} \sigma}$};
\draw node at (2,2.2) {${\color{blue} \sigma}$};
\draw node at (-0.2,1) {${\color{blue} f}$};
\draw node at (4.2,1) {${\color{blue} f}$};
\end{tikzpicture}
    \caption{Orbifold groupoids for $T_4[\bP^1 \times \bP^1]$ with $\bZ_2$ symmetry. The map $\sigma_4$ represents the topological manipulation gauging $\bZ_4$ and $f$ is the gauging $\bZ_2$ up to an SPT phase.}
    \label{Fig:N4f1}
\end{figure}

The automorphism group is $\Aut_{\bZ_4}(Q) =\bZ_2 \times D_8$ with $\Aut(\bZ_4)=\bZ_2$ and $\cO_{4}(Q)=D_8$. 
Here, $\bZ_2$ represents different ways to turn on the background fields on $\Sigma_2$ while $D_8$ is generated by 
\begin{equation*}
\sigma=\left(
\begin{array}{cc}
 0 & 1 \\
 1 & 0 \\
\end{array}
\right), 
\qquad 
\tau=\left(
\begin{array}{cc}
 1 & 2 \\
 0 & 1 \\
\end{array}
\right)~,
\end{equation*}
and transforms different maximal isotropic sublattices into each other which represents 2d topological manipulations.

In particular, with $\sigma$ and $f = \tau \sigma \tau$, we find these theories transform according to Figure \ref{Fig:N4f1}. 
As we can observe, $L_5$ is a singlet with respect to $D_8$, while the square, which consists of the remaining set $\left\{L_1,\,L_2,\,L_3,\,L_4\right\}$, furnishes a two-dimensional irreducible rep. of $D_8$. Note that $\Aut_{\bZ_4}(Q)$ only contains the operation of gauging $\bZ_4$, not $\bZ_2$, so the theory specified by $L_5$ is isolated and cannot be related with other theories by topological manipulations defined by $\Aut_{\bZ_4}(Q)$. Besides that, our result reproduces the orbifold groupoid for the theory with non-anomalous $\bZ_4$ symmetry studied in \cite{Gaiotto:2020iye}.

Now, let's take into account the SPT phase for each of these theories. Thus, for theories with $\bZ_4$ symmetry, there are $8$ global variants, which can be transformed to each other by two basic topological manipulations, $\bZ_4$ 
 gauging $\sigma$ and stacking Arf invariant $\tau$. Following the same logic, one can assign each global variant with a $2 \times 2$ $D_8$ matrix $M$. 
 Then, one can study how they transform under the topological operations $\sigma$ and $\tau$, and duality $s$. The result is plotted in Figure \ref{Fig:N4f2}.

\begin{figure}
    \centering
    \begin{tikzpicture}[scale=1.5]
\draw node at (0,0) {$(\bZ_4)_0$};
\draw node at (0,2) {$(\widehat{\bZ}_4)_0$};
\draw node at (3,0) {$(\bZ_4)_1$};
\draw node at (3,2) {$(\widehat{\bZ}_4)_1$};
\draw node at (6,0) {$(\widehat{\bZ}_4^f)_0$};
\draw node at (6,2) {$(\bZ_4^f)_1$};
\draw node at (9,0) {$(\widehat{\bZ}_4^f)_1$};
\draw node at (9,2) {$(\bZ_4^f)_0$};
\node[below] at (0,-0.3) 
{$\left(\begin{matrix} 0 & 1 \\ 1 & 0 \end{matrix}\right)$};
\node[below] at (3,-0.3) 
{$\left(\begin{matrix} 0 & 1 \\ 1 & 2 \end{matrix}\right)$};
\node[below] at (6,-0.3) 
{$\left(\begin{matrix} 1 & 0 \\ 2 & 1 \end{matrix}\right)$};
\node[below] at (9,-0.3) 
{$\left(\begin{matrix} 1 & 2 \\ 2 & 1 \end{matrix}\right)$};
\node[above] at (0,2.3) 
{$\left(\begin{matrix} 1 & 0 \\ 0 & 1 \end{matrix}\right)$};
\node[above] at (3,2.3) 
{$\left(\begin{matrix} 1 & 2 \\ 0 & 1 \end{matrix}\right)$};
\node[above] at (6,2.3) 
{$\left(\begin{matrix} 2 & 1 \\ 1 & 0 \end{matrix}\right)$};
\node[above] at (9,2.3) 
{$\left(\begin{matrix} 2 & 1 \\ 1 & 2 \end{matrix}\right)$};
\draw [<->,orange] (0.05,0.3) -- (0.05,1.7);
\draw [<->,orange] (3,0.3) -- (3,1.7);
\draw [<->,orange] (6,0.3) -- (6,1.7);
\draw [<->,orange] (8.95,0.3) -- (8.95,1.7);
\draw [<->,blue] (-0.05,0.3) -- (-0.05,1.7);
\draw [<->,blue] (9.05,0.3) -- (9.05,1.7);
\draw [<->,blue] (0.4,0.0) -- (2.6,0.0);
\draw [<->,blue] (0.4,2) -- (2.6,2);
\draw [<->,blue] (3.4,0.0) -- (5.6,0.0);
\draw [<->,blue] (3.4,2) -- (5.6,2);
\draw [<->,blue] (6.4,0.0) -- (8.6,0.0);
\draw [<->,blue] (6.4,2) -- (8.6,2);
\draw node at (0.2,1) {${\color{orange} s}$};
\draw node at (3.15,1) {${\color{orange} s}$};
\draw node at (5.85,1) {${\color{orange} s}$};
\draw node at (8.8,1) {${\color{orange} s}$};
\draw node at (-0.2,1) {${\color{blue} \sigma}$};
\draw node at (9.2,1) {${\color{blue} \sigma}$};
\draw node at (1.5,0.15) {${\color{blue} \tau}$};
\draw node at (1.5,2.15) {${\color{blue} \tau}$};
\draw node at (4.5,0.15) {${\color{blue} \sigma}$};
\draw node at (4.5,2.15) {${\color{blue} \sigma}$};
\draw node at (7.5,0.15) {${\color{blue} \tau}$};
\draw node at (7.5,2.15) {${\color{blue} \tau}$};
\end{tikzpicture}
\caption{Web of transformations for $T_4[\bP^1 \times \bP^1]$. The transformations
in orange are the duality transformations.
The transformations in blue are topological manipulations.}
    \label{Fig:N4f2}
\end{figure}

\subsubsection*{Mixed anomaly}

There is one more absolute theory described by $L_5$. It can be obtained from the $\bZ_4$ ($\widehat{\bZ}_4$) theory by $\bZ_2$ ($\widehat{\bZ}_2$) gauging. Since the extension class is non-trivial, it has anomalous $\bZ_2 \times \bZ_2$ symmetry with mixed anomaly given by \cite{Tachikawa:2017gyf} 
\begin{equation} \label{Eqn:anomalyz4}
    \mu_3 = \int_{W_3} \widehat{A} \cup \kappa(A)~,
\end{equation}
where $\widehat{A} \in H^1(\Sigma_2,\widehat{\bZ}_2)$ and $A \in H^1(\Sigma_2,\bZ_2)$ are the background connections of $\widehat{\bZ}_2$ and $\bZ_2$ respectively, and $\kappa$ is the extension class.

This anomaly can be detected from the 3d SymTFT. 
For the topological boundary condition specified by $L_5$, the terminal lines are $\{L_{(0,0)},L_{(0,2)},L_{(2,0)},L_{(2,2)}\}$. 
The non-terminal lines are given by the equivalence classes in $L_5^{\perp}$,
\begin{equation} \label{Eqn:anomalyz4line}
\begin{array}{cccc}
B_1 = \{L_{(0,0)},L_{(0,2)},L_{(2,0)},L_{(2,2)}\}, \\
B_2 = \{L_{(0,1)},L_{(0,3)},L_{(2,1)},L_{(2,3)}\}, \\
B_3 = \{L_{(1,0)},L_{(3,0)},L_{(1,2)},L_{(3,2)}\}, \\
B_4 = \{L_{(1,1)},L_{(1,3)},L_{(3,1)},L_{(3,3)}\}. 
\end{array}
\end{equation}
After computing the braiding of lines from $B_2$ and $B_3$, we find all choices of lines have non-trivial braiding. This implies a mixed anomaly between $\widehat{\bZ}_2 \times \bZ_2$ \cite{Kaidi:2023maf}.

\paragraph{For $N=6$:}

There are 6 maximal isotropic sublattices 
\begin{equation} \label{Eqn:LN6}
\begin{array}{cccccccc}
L_1=\{ (0,0), & (0,1), & (0,2), & (0,3), & (0,4) ,& (0,5)\} &\to 
&(\bZ_6)_{++} = \bZ_3 \times \bZ_2\\
L_2 = \{ (0,0), & (2,0), & (4,0), & (0,3), & (2,3), & (4,3) \}&\to 
&(\bZ_6)_{-+} = \widehat{\bZ}_3 \times \bZ_2\\ 
L_3 = \{ (0,0), & (0,2), & (0,4), & (3,0), & (3,2), & (3,4) \}&\to 
&(\bZ_6)_{+-} = \bZ_3 \times \widehat{\bZ}_2 \\
L_4 = \{ (0,0), & (1,0), & (2,0), & (3,0), & (4,0) ,& (5,0)\} &\to 
&(\bZ_6)_{--} = \widehat{\bZ}_3 \times \widehat{\bZ}_2\\
L_5 = \{ (0,0), & (0,2), & (0,4), & (3,1), & (3,3), & (3,5)\} &\to 
&(\bZ_6)_{+f} = \bZ_3 \times \bZ_2^f\\
L_6 = \{ (0,0), & (2,0), & (4,0), & (1,3), & (3,3), & (5,3) \}&\to 
&(\bZ_6)_{-f} = \widehat{\bZ}_3 \times \bZ_2^f
\end{array}
\end{equation}
These maximal isotropic sublattices define six absolute theories with $\bZ_6$ symmetry. Since $\bZ_6 = \bZ_2 \times \bZ_3$, we can also label them using their subgroups. These absolute theories look like the tensor product of absolute theories for $N=2$ and $N=3$. Depending on the different behavior of the $\bZ_2$ and $\bZ_3$ factors, one has 6 absolute theories denoted by $(\bZ_6)_{\pm,\pm/f}$, where $+/-$ represents whether a subgroup is gauged or not and $f$ means whether one has performed the fermionization operation on the $\bZ_2$ factor.

The automorphism group is $\Aut_{\bZ_6}(Q) = \bZ^2_2 \times S_3$, where $\Aut(\bZ_6)=\bZ_2$ accounts for different ways to couple the background fields and $\cO_{6}(Q)=S_3 \times \bZ_2$ transforms between different theories and encodes the possible 2d topological manipulations. The topological manipulations for a theory with $\bZ_6$ symmetry can be understood through the manipulations on its subgroups $\bZ_2$ and $\bZ_3$. The generators of $\Aut_{\bZ_6}(Q)$ are
\begin{equation} \label{Eqn:topManN6}
\sigma_3=\left(
\begin{array}{cc}
 0 & 1 \\
 1 & 0 \\
\end{array}
\right),
\quad 
\sigma_2=\left(
\begin{array}{cc}
 4 & 3 \\
 3 & 4 \\
\end{array}
\right),
\quad 
\tau=\left(
\begin{array}{cc}
 1 & 3 \\
 0 & 1 \\
\end{array}
\right),
\end{equation}
where $\sigma_3$/$\sigma_2$ denote gauging of $\bZ_3$/$\bZ_2$ and $\tau$ is denotes stacking the Arf invariant. With $\sigma_3$ and $\tilde \sigma_2 = \tau \sigma_2$, one can obtain the orbifold groupoid in Figure \ref{Fig:N6f1}, which can be identified by the tensor product of the orbifold groupoid of $\bZ_2$ and $\bZ_3$ in Figure \ref{Fig:N2f1} and Figure \ref{Fig:N3f1}.

Consider the possible stacking of the Arf invariant. We have 12 global variants. By the same procedure, one can associate each of them with a $\cO_{6}(Q)$ matrix. Using the representation of topological manipulations in \eqref{Eqn:topManN6} and duality operation in \eqref{Eqn:p1p1duality}, one can study the transformation among these global variants. The result is plotted in Figure \ref{Fig:N6f2}. We find that the topological manipulations drawn in blue are simply the direct product of the diagram for $\bZ_2$ in Figure \ref{Fig:N2f2} and $\bZ_3$ in Figure \ref{Fig:N3f1}. 
This diagram can be simplified by combining the operation of $\sigma_3$ and $\sigma_2$ which defines a $\bZ_6$ gauging below  
\begin{equation}
    \sigma = \sigma_3 \sigma_2 = \left(
\begin{array}{cc}
 0 & 1 \\
 1 & 0 \\
\end{array}
\right)
\end{equation}
The diagram of the global variants plotted using $\sigma$ and $\tau$ is in Figure \ref{Fig:N6f3}.

\begin{figure}
    \centering
    \begin{tikzpicture}[scale=1.5]
\draw node at (0,0) {$\bZ_3 \times \bZ_2$};
\draw node at (3,0) {$\bZ_3 \times \widehat{\bZ}_2$};
\draw node at (6,0) {$\bZ_3 \times \bZ_2^f$};
\draw node at (0,2) {$\widehat{\bZ}_3 \times \bZ_2$};
\draw node at (3,2) {$\widehat{\bZ}_3 \times \widehat{\bZ}_2$};
\draw node at (6,2) {$\widehat{\bZ}_3 \times \bZ_2^f$};
\draw [->,blue] (.6,0) -- (2.4,0);
\draw [->,blue] (3.6,0) -- (5.4,0);
\draw [->,blue] (.6,2) -- (2.4,2);
\draw [->,blue] (3.6,2) -- (5.4,2);
\draw [<-,blue] (0,-.2) arc (-120:-60:6);
\draw [<-,blue] (0,2.2) arc (120:60:6);
\draw [<->,blue] (-0.05,0.2) -- (-0.05,1.8);
\draw [<->,blue] (3,0.2) -- (3,1.8);
\draw [<->,blue] (6.05,0.2) -- (6.05,1.8);
\draw node at (1.5,.2) {${\color{blue} \tilde\sigma_2}$};
\draw node at (4.5,.2) {${\color{blue} \tilde\sigma_2}$};
\draw node at (1.5,2.2) {${\color{blue} \tilde\sigma_2}$};
\draw node at (4.5,2.2) {${\color{blue} \tilde\sigma_2}$};
% \draw node at (1.5,-.45) {${\color{blue} \sigma}$};
% \draw node at (4.5,2.75) {${\color{blue} \sigma}$};
\draw node at (-0.2,1) {${\color{blue} \sigma_3}$};
% \draw node at (0.2,1) {${\color{blue} \tau}$};
% \draw node at (5.8,1) {${\color{blue} \tau}$};
\draw node at (3.2,1) {${\color{blue} \sigma_3}$};
\draw node at (6.2,1) {${\color{blue} \sigma_3}$};
\draw node at (3,2.8) {${\color{blue} \tilde\sigma_2}$};
\draw node at (3,-0.8) {${\color{blue} \tilde\sigma_2}$};
\end{tikzpicture}
    \caption{Orbifold groupoids for $T_6[\bP^1 \times \bP^1]$ with $\bZ_2$ symmetry. The map $\sigma_3$ represents the topological manipulation gauging $\bZ_3$ and $\tilde \sigma_2$ is the gauging $\bZ_2$ up to an SPT phase.}
    \label{Fig:N6f1}
\end{figure}

\begin{figure}
    \centering
    \begin{tikzpicture}[scale=1.5]
\draw node at (0,0) {$(\bZ_3 \times \bZ_2)_1$};
\draw node at (2,0) {$(\bZ_3 \times \bZ^f_2)_0$};
\draw node at (4,0) {$(\bZ_3 \times \bZ^f_2)_1$};
\draw node at (6,0) {$(\bZ_3 \times \widehat{\bZ}_2)_1$};
\draw node at (0,4.5) {$(\widehat{\bZ}_3 \times \bZ_2)_1$};
\draw node at (2,4.5) {$(\widehat{\bZ}_3 \times \bZ^f_2)_0$};
\draw node at (4,4.5) {$(\widehat{\bZ}_3 \times \bZ^f_2)_1$};
\draw node at (6,4.5) {$(\widehat{\bZ}_3 \times \widehat{\bZ}_2)_1$};
\draw node at (0,1.5) {$(\bZ_3 \times \bZ_2)_0$};
\draw node at (0,3) {$(\widehat{\bZ}_3 \times \bZ_2)_0$};
\draw node at (6,1.5) {$(\bZ_3 \times \widehat{\bZ}_2)_0$};
\draw node at (6,3) {$(\widehat{\bZ}_3 \times \widehat{\bZ}_2)_0$};
\node[below] at (0,-0.3) 
{$\left(\begin{matrix} 0 & 1 \\ 1 & 3 \end{matrix}\right)$};
\node[below] at (2,-0.3) 
{$\left(\begin{matrix} 3 & 4 \\ 1 & 3 \end{matrix}\right)$};
\node[below] at (4,-0.3) 
{$\left(\begin{matrix} 3 & 1 \\ 1 & 0 \end{matrix}\right)$};
\node[below] at (6,-0.3) 
{$\left(\begin{matrix} 3 & 1 \\ 4 & 3 \end{matrix}\right)$};
\node[above] at (0,4.8) 
{$\left(\begin{matrix} 4 & 3 \\ 3 & 1
\end{matrix}\right)$};
\node[above] at (2,4.8) 
{$\left(\begin{matrix} 1 & 0 \\ 3 & 1 \end{matrix}\right)$};
\node[above] at (4,4.8) 
{$\left(\begin{matrix} 1 & 3 \\ 3 & 4 \end{matrix}\right)$};
\node[above] at (6,4.8) 
{$\left(\begin{matrix} 1 & 3 \\ 0 & 1 \end{matrix}\right)$};
\node[left] at (-1,1.5) 
{$\left(\begin{matrix} 0 & 1 \\ 1 & 0 \end{matrix}\right)$};
\node[left] at (-1,3) 
{$\left(\begin{matrix} 4 & 3 \\ 3 & 4 \end{matrix}\right)$};
\node[right] at (7,1.5) 
{$\left(\begin{matrix} 3 & 4 \\ 4 & 3 \end{matrix}\right)$};
\node[right] at (7,3) 
{$\left(\begin{matrix} 1 & 0 \\ 0 & 1 \end{matrix}\right)$};
\draw [<->,blue] (0,0.3) -- (0,1.2);
\draw [<->,blue] (0,1.8) -- (0,2.7);
\draw [<->,blue] (0,3.3) -- (0,4.2);
\draw [<->,blue] (6,0.3) -- (6,1.2);
\draw [<->,blue] (6,1.8) -- (6,2.7);
\draw [<->,blue] (6,3.3) -- (6,4.2);
\draw [<->,blue] (0.7,0.0) -- (1.3,0.0);
\draw [<->,blue] (2.7,0.0) -- (3.3,0.0);
\draw [<->,blue] (4.7,0.0) -- (5.3,0.0);
\draw [<->,blue] (0.7,4.5) -- (1.3,4.5);
\draw [<->,blue] (2.7,4.5) -- (3.3,4.5);
\draw [<->,blue] (4.7,4.5) -- (5.3,4.5);
\draw [<->,blue] (0.7,1.5) -- (5.3,1.5);
\draw [<->,blue] (0.7,3) -- (5.3,3);
\draw [<->,blue] (0.7,1.5) -- (5.3,1.5);
\draw [<->,blue] (0.7,3) -- (5.3,3);
\draw [<->,blue] (2,0.3) -- (2,4.2);
\draw [<->,blue] (4,0.3) -- (4,4.2);
\draw [<->,blue] (-0.3,0.3) arc  (210:150:4); 
\draw [<->,blue] (6.3,0.3) arc (-30:30:4);
\draw [<->,orange] (0.3,0.3) -- (5.7,4.2);
\draw [<->,orange] (0.3,4.2) -- (5.7,0.3);
\draw [<->,orange] (0.3,1.7) -- (5.7,2.7);
\draw [<->,orange] (0.3,2.7) -- (5.7,1.7);
\draw [<->,orange] (2.3,0.3) -- (3.7,4.2);
\draw [<->,orange] (3.7,0.3) -- (2.3,4.2);
% %
\draw node at (2.2,3.9) {${\color{orange} s}$};
\draw node at (3.8,3.9) {${\color{orange} s}$};
\draw node at (0.5,3.8) {${\color{orange} s}$};
\draw node at (5.5,3.8) {${\color{orange} s}$};
\draw node at (0.5,2.5) {${\color{orange} s}$};
\draw node at (5.5,2.5) {${\color{orange} s}$};
\draw node at (1,-0.2) {${\color{blue} \sigma_2}$};
\draw node at (3,-0.2) {${\color{blue} \tau}$};
\draw node at (5,-0.2) {${\color{blue} \sigma_2}$};
\draw node at (1,4.7) {${\color{blue} \sigma_2}$};
\draw node at (3,4.7) {${\color{blue} \tau}$};
\draw node at (5,4.7) {${\color{blue} \sigma_2}$};
\draw node at (-0.2,0.75) {${\color{blue} \tau}$};
\draw node at (-0.2,2.25) {${\color{blue} \sigma_3}$};
\draw node at (-0.2,3.75) {${\color{blue} \tau}$};
\draw node at (6.2,0.75) {${\color{blue} \tau}$};
\draw node at (6.2,2.25) {${\color{blue} \sigma_3}$};
\draw node at (6.2,3.75) {${\color{blue} \tau}$};
\draw node at (7.1,2.25) {${\color{blue} \sigma_3}$};
\draw node at (-1.1,2.25) {${\color{blue} \sigma_3}$};
\draw node at (1.7,0.6) {${\color{blue} \sigma_3}$};
\draw node at (4.3,0.6) {${\color{blue} \sigma_3}$};
\draw node at (1,3.2) {${\color{blue} \sigma_2}$};
\draw node at (1,1.3) {${\color{blue} \sigma_2}$};
\end{tikzpicture}
\caption{Web of transformations for $T_6[\bP^1 \times \bP^1]$. The transformations
in orange are the duality transformations.
The transformations in blue are topological manipulations.}
\label{Fig:N6f2}
\end{figure}

\begin{figure}
    \centering
    \begin{tikzpicture}[scale=1.5]
\draw node at (0,0) {$(\bZ_6)_{++1}$};
\draw node at (2,0) {$(\bZ_6)_{-f0}$};
\draw node at (4,0) {$(\bZ_6)_{-f1}$};
\draw node at (6,0) {$(\bZ_6)_{+-1}$};
\draw node at (0,4.5) {$(\bZ_6)_{--1}$};
\draw node at (2,4.5) {$(\bZ_6)_{+f1}$};
\draw node at (4,4.5) {$(\bZ_6)_{+f0}$};
\draw node at (6,4.5) {$(\bZ_6)_{-+1}$};
\draw node at (0,1.5) {$(\bZ_6)_{++0}$};
\draw node at (0,3) {$(\bZ_6)_{--0}$};
\draw node at (6,1.5) {$(\bZ_6)_{+-0}$};
\draw node at (6,3) {$(\bZ_6)_{-+0}$};
\node[left] at (-0.5,3) 
{$\left(\begin{matrix} 1 & 0 \\ 0 & 1 \end{matrix}\right)$};
\node[left] at (-0.5,1.5) 
{$\left(\begin{matrix} 0 & 1 \\ 1 & 0 \end{matrix}\right)$};
\node[below] at (0,-0.3) 
{$\left(\begin{matrix} 0 & 1 \\ 1 & 3 \end{matrix}\right)$};
\node[below] at (2,-0.3) 
{$\left(\begin{matrix} 1 & 0 \\ 3 & 1 \end{matrix}\right)$};
\node[below] at (4,-0.3) 
{$\left(\begin{matrix} 1 & 3 \\ 3 & 4 \end{matrix}\right)$};
\node[below] at (6,-0.3) 
{$\left(\begin{matrix} 3 & 1 \\ 4 & 3 \end{matrix}\right)$};
\node[right] at (6.5,1.5) 
{$\left(\begin{matrix} 3 & 4 \\ 4 & 3 \end{matrix}\right)$};
\node[right] at (6.5,3) 
{$\left(\begin{matrix} 4 & 3 \\ 3 & 4 \end{matrix}\right)$};
\node[above] at (6,4.8) 
{$\left(\begin{matrix} 4 & 3 \\ 3 & 1 \end{matrix}\right)$};
\node[above] at (4,4.8) 
{$\left(\begin{matrix} 3 & 4 \\ 1 & 3 \end{matrix}\right)$};
\node[above] at (2,4.8) 
{$\left(\begin{matrix} 3 & 1 \\ 1 & 0 \end{matrix}\right)$};
\node[above] at (0,4.8) 
{$\left(\begin{matrix} 1 & 3 \\ 0 & 1 \end{matrix}\right)$};
\draw [<->,blue] (0,0.3) -- (0,1.2);
\draw [<->,blue] (0,1.8) -- (0,2.7);
\draw [<->,blue] (0,3.3) -- (0,4.2);
\draw [<->,blue] (6,0.3) -- (6,1.2);
\draw [<->,blue] (6,1.8) -- (6,2.7);
\draw [<->,blue] (6,3.3) -- (6,4.2);
\draw [<->,blue] (0.5,0.0) -- (1.5,0.0);
\draw [<->,blue] (2.5,0.0) -- (3.5,0.0);
\draw [<->,blue] (4.5,0.0) -- (5.5,0.0);
\draw [<->,blue] (0.5,4.5) -- (1.5,4.5);
\draw [<->,blue] (2.5,4.5) -- (3.5,4.5);
\draw [<->,blue] (4.5,4.5) -- (5.5,4.5);
\draw [<->,orange] (0.1,1.8) -- (0.1,2.7);
\draw [<->,orange] (5.9,1.8) -- (5.9,2.7);
\draw [<->,orange] (2,0.3) -- (2,4.2);
\draw [<->,orange] (4,0.3) -- (4,4.2);
\draw [<->,orange] (0.3,.3) arc (-60:60:2.2);
\draw [<->,orange] (5.7,.3) arc (240:120:2.2);
\draw node at (0.3,2.25) {${\color{orange} s}$};
\draw node at (5.7,2.25) {${\color{orange} s}$};
\draw node at (1.8,0.5) {${\color{orange} s}$};
\draw node at (4.2,0.5) {${\color{orange} s}$};
\draw node at (0.8,4) {${\color{orange} s}$};
\draw node at (5.2,4) {${\color{orange} s}$};
\draw node at (1,-0.2) {${\color{blue} \sigma}$};
\draw node at (3,-0.2) {${\color{blue} \tau}$};
\draw node at (5,-0.2) {${\color{blue} \sigma}$};
\draw node at (1,4.7) {${\color{blue} \sigma}$};
\draw node at (3,4.7) {${\color{blue} \tau}$};
\draw node at (5,4.7) {${\color{blue} \sigma}$};
\draw node at (-0.2,0.75) {${\color{blue} \tau}$};
\draw node at (-0.2,2.25) {${\color{blue} \sigma}$};
\draw node at (-0.2,3.75) {${\color{blue} \tau}$};
\draw node at (6.2,0.75) {${\color{blue} \tau}$};
\draw node at (6.2,2.25) {${\color{blue} \sigma}$};
\draw node at (6.2,3.75) {${\color{blue} \tau}$};
\end{tikzpicture}
\caption{Web of transformations for $T_6[\bP^1 \times \bP^1]$. The transformations
in orange are the duality transformations.
The transformations in blue are topological manipulations.}
\label{Fig:N6f3}
\end{figure}

\paragraph{For general $N$:}

Using the SymTFT, we are able to study the global variants of $T_N[\bP^1 \times \bP^1]$ for general $N$. 
The possible topological manipulations are determined by the automorphism group $\Aut_{\bZ_N}(Q)/\Aut(\bZ_N)$, where $\Aut(\bZ_N)=\bZ_{\phi(N)}$ and $\phi(N)$ is the Euler's totient function.  
These topological manipulations act transitively on the global variants. 
The number of global variants is 
\begin{equation}
    d(N) = |\cO_N(Q)|~.
\end{equation}
Assigning each global variant to a $\cO_N(Q)$ matrix, we can determine how they transform under the topological manipulations.

The automorphism group $\Aut_{\bZ_N}(Q)$ plays an important role in determining the global variants, and topological manipulations. Besides that, it also gives the 0-form symmetry of the SymTFT, in our case, a $\bZ_N$ gauge theory. We compute $\Aut_{\bZ_N}(Q)$ for $N=2,3\ldots 20$, and identify them with finite groups in Table \ref{Tab:p1p1Aut}. Note that, for odd $N$, our results match with the 0-form symmetry for the $\bZ_N$ gauge theory studied in \cite{Delmastro:2019vnj}. However, for even $N$, the 0-form symmetries from \cite{Delmastro:2019vnj} are subgroups of our result because the SymTFT from the compactification on $\bP^1 \times \bP^1$ is a spin DW theory admitting also fermionic topological boundary conditions \cite{Gaiotto:2020iye}. 
It is noted that, when $N=pq$ with ${\rm gcd}(p,q)=1$, ${\rm Aut}_{\bZ_{N}}(Q)$ can be factorized as direct product of ${\rm Aut}_{\bZ_p}(Q)$ and ${\rm Aut}_{\bZ_q}(Q)$.

\begin{table}[!htp] 
\renewcommand{\arraystretch}{1.3}
\begin{flushright}
  \begin{tabular}{c|cccccccccccccc}
    $N$ &
   2& 3& 4& 5& 6& 7& 8& 9& 10& 11 &12
    \\ \hline \strut
    $\Aut_{\bZ_N}(Q)$ &
    $S_3$& $\bZ_2^2$ & $D_8\times \bZ_2$& $D_8$& $S_3\times \bZ_2^2$& $D_{12}$& $D_8\times \mathbb Z_2^2$& $D_{12}$& $S_3\times D_8 $& $D_{20}$ & $D_8\times \mathbb Z_2^3$
      \end{tabular}
        \end{flushright}
\begin{flushright}
 \begin{tabular}{|cccccccccccccc}
    \hspace{0.01cm} 13& 14& 15& 16& 17& 18& 19& 20
    \\ \hline \strut
    
 \hspace{0.01cm} $D_{24}$ & $S_3\times D_{12}$& $D_{8}\times\mathbb Z_2^2$& $\left(\mathbb Z_4\times \mathbb Z_2\right)\rtimes \mathbb Z_2^2$ & $D_{34}$ &  $S_3\times D_{12}$ & $D_{36}$ & $D_{8}\times D_{8}\times\mathbb Z_2$
  \end{tabular}
  \end{flushright}
  \caption{The automorphism group $\Aut_{\bZ_N}(Q)$ of $\bP^1 \times \bP^1$ up to $N=20$.}
  \label{Tab:p1p1Aut}
\end{table}

\subsubsection*{Geometric perspective}

From the point of view of the 7d symTFT, the different choices of the maximal isotropic sublattice $L$ and $L^{\perp}$ are equivalent to the choice of handlebody of $M_5$ with $\partial M_5 = M_4$ \cite{Gukov:2020btk, Bashmakov:2022uek}. Similar to the solid tori, handlebodies in five dimensions are characterized by a "meridian" and a "longitude," which are a set of assignments of cycles in $H_2(M_4,\mathbb{Z}_N)$. The meridian extends to contractible cycles in $H_2(M_5, \mathbb{Z}_N)$, while the longitude extends to non-contractible cycles.

In our case, we will find the handlebodies for $\bP^1 \times \bP^1$. The number of handlebodies depends on the parameter $N$. For example, for $N=2$, we find the following three geometries with $\bP^1 \times \bP^1$ on the boundary
\begin{equation}
    M^{(1)}_5 = D^3 \times S^2, \quad  M^{(2)}_5 =  S^2 \times D^3, \quad  M^{(3)}_5 = S^2\times S^2 \times \bR ~. 
\end{equation}
which is equivalent to the three choices of maximal isotropic sublattice. For each of the above manifolds, one has two ways to choose longitude. In total, one can define $6$ handlebodies. However, for prime $N$ larger than 2, by the same argument discussed in \eqref{Eqn:p1p1prime} one cannot extend the $(1,1)$ cycle into a contractible cycle in $M_5$. Thus, only the first two manifolds are valid. Thus, one has two handlebodies in this case. 
% For example, in $M^{(1)}_5$, $\{e_1, e_1 + e_2 \}$ is the ``meridian" while $\{1,e_2\}$ are the ``longitude". It seems that the choice of the handle body by the choice of the meridian also works for $M_4$ considered here. 
% As it turns out the choice of the handle body corresponds to exactly the choice of $L$ in the following way. 

In this subsection, we studied the orbifold groupoid and global variants of the 2d theories from the warping of $N$ M5-brane on $\bP^1 \times \bP^1$. 
For the cases of prime $N$, we identify all global variants and possible topological manipulations. 
We also discuss how to generalize the result to the case when $N=pq$ is not prime, but a product of two primes $p$ and $q$ using two examples $N=4$ for $p,q$ are not coprime and $N=6$ for  $p,q$ are coprime. 
In general, given the prime factorization of $N$, we can apply the method discussed here recursively and find all global variants of $T_N[\bP^1 \times \bP^1]$. Finally, we identify the 0-form symmetries of spin $\bZ_N$ gauge theories up to $N=20$ from the perspective of geometry.

\subsection{Duality defects}

In this subsection, we will study the non-invertible symmetries of the theory $T_N[\bP^1\times \bP^1]$. Similar to the class S theories in 4d \cite{Bashmakov:2022uek}, the non-invertible symmetries can be realized by the combinations of topological manipulations and dualities at the special point in the conformal manifold. With the understanding of global variants of $T_N[\bP^1\times \bP^1]$ and their transformation properties studied in the previous subsection, we construct duality defects for the theories with prime $N$ and $N=4,6$. Besides that for $N=2$, there exist global variants with mixed anomaly between invertible symmetries. After gauging one of them, we find the same duality defect using the half-space gauging construction \cite{Kaidi:2021xfk}. 
Finally, we discussed how these defects are realized in the compactification of the 3d SymTFT. 

% Following \cite{Bashmakov:2022jtl}, we will show that there exist mixed anomalies between invertible 0-form symmetries $\bZ_N$, i.e. defect group of 6d SCFTs, and a subgroup $\bZ_2 \subset \Aut_{\bZ}(Q)$. After gauging $\bZ_N$, the subgroup $\bZ_2$ will become a duality defect \cite{Kaidi:2021gbs}. After that, we study the non-invertible symmetries using the web of global variants of $T_N[\bP^1\times \bP^1]$. As studied in \cite{Bashmakov:2022uek}, combinations of topological manipulations and dualities often lead to topological interfaces among the same variants. These interfaces become topological defects generating symmetries at the special point in the conformal manifold. 
% We perform a complete analysis of the possible non-invertible defects for prime $N$ and $N=4,6$. 

\paragraph{Couplings and fixed points}
In the compactification of M5-branes on $\bP^1 \times \bP^1$, the resulting 2d theory is a supersymmetric sigma model with target space the moduli space of magnetic monopoles \cite{Haghighat:2011xx, Haghighat:2012bm, Dedushenko:2017tdw}. 
% \footnote{The 2d theory also contains free non-compact bosons, chiral bosons, and fermions. See \cite{Dedushenko:2017tdw} for details.}
The target space contains a $U(1)$ isometry which for one M5-brane can be identified with a compact boson with radius $R$, where
$R$ depends on the conformal structure of the 4-manifold and describes the ratio of the sizes of the two $\bP^1$'s \cite{Dedushenko:2017tdw}. 
% When performing the operation of switching two $\bP^1$'s, it corresponds to T-duality for compact bosons. By gauging the $\bZ_2$ subgroup of $U(1)$, one obtains the Kramers-Wannier-like duality defect after the action of T-duality \cite{Ji:2019ugf}. 
For the compactification of $N$ M5-branes on $\bP^1 \times \bP^1$, 
%we know that the 2d theory is a CFT. 
% Since the coupling depends on the geometry of the 4-manifold, 
we expect the same coupling in the 2d theory $T_N[M_4]$.
From the equation \eqref{Eqn:p1p1duality}, the duality map changes the coupling constant into 
\begin{equation*}
    R \xrightarrow[]{-I} R, \qquad R \xrightarrow[]{s} \frac{1}{R}~.
\end{equation*}
%
% Thus, the first operation has no effect on the coupling constant while the second one will change eh
As we will see later, the subgroup $\bZ_2$ generated by $s$ will also be a subgroup of $\Aut_{\bZ_N}(Q)$, which will lead to duality defects for theories $T_N[\bP^1\times \bP^1]$ at $R=1$.

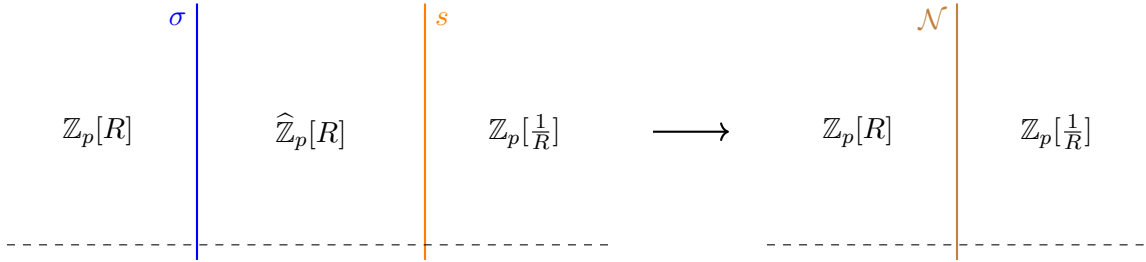
\begin{figure}[!tbp]
\centering
\begin{tikzpicture}[baseline=19,scale=1]
% \fill[black!30!green!15] (0,-0.2) rectangle (2.98,3.2);
\draw[thick,blue] (0,-0.2)--(0,3.2);
\draw[thick,orange] (3,-0.2)--(3,3.2);
\draw[thick,  ->] (6,1.5) -- (7,1.5);
\draw[brown, thick] (10,-0.2)--(10,3.2);
\draw[ dashed] (-2.5,0) -- (5.5,0);
\draw[ dashed] (7.5,0) -- (12.5,0);

% \draw[ brown,<->] (0.8,2.8) -- (2.2,2.8);

\node[left] at (-0.7,1.5) {$\bZ_p[R]$};
    \node at (1.5,1.5) {$\widehat{\bZ}_p[R]$};
    \node[right] at (3.7,1.5) {$\bZ_p[\frac{1}{R}]$};
      \node[left,blue] at (0,3) {$\sigma$};
          \node[right,orange] at (3,3) {$s$}; 
  \node[left] at (9.3,1.5) {$\bZ_p[R]$};
    \node[right] at (10.7,1.5) {$\bZ_p[\frac{1}{R}]$};
      \node[left,brown] at (10,3) {$\cN$};
\end{tikzpicture}
\caption{At $R=1$, the $\bZ_p$ theory has a non-invertible defect $N=\sigma s$. The coupling constant $R$ is included explicitly for each theory.  }
\label{Fig:NFG}
\end{figure}

The non-invertible defects can be realized as the combinations of the topological operations $G(\sigma,\tau)$, the gauging $\sigma$ and stacking SPT phase $\tau$, and the duality $F$ from the automorphism group $\MCG(\bP^1 \times \bP^1)$. In general, these operations will change the 2d global variant and the coupling $R$. To realize defects in 2d theories, one needs to find a set of $G(\sigma,\tau)$ and $F$ operations such that their combination keeps the global variant and coupling invariant. 
Note that only the duality operation $F$ will change the coupling $R$. Thus, to guarantee that the coupling $R$ stays the same after acting with $F$, one needs to take $R$ to be the fixed point of it. In our case, the only non-trivial duality operation is $F=s$ with a fixed point at $R=1$. 
We will implicitly take $R=1$ in the remainder of this section.

Next, we need to find some topological manipulation $G(\sigma,\tau)$ which can undo the action of $F=s$ and map the global variant to itself. More precisely, given a distinct 2d global variant labeled by $M$, one needs to find a pair of $F$ and $G(\sigma,\tau)$ satisfying the following conditions,
\begin{equation} \label{Eqn:FMG}
   F^t M G = M, \hspace{2cm} F \in \MCG(\bP^1 \times \bP^1), \quad G(\sigma,\tau) \in \Aut_{\bZ_N}(Q).
\end{equation}
This way one can realize a defect $\cN=FG$ in a theory labeled by $M$. A typical example of this realization is for theory $\bZ_p$ where as shown in Figure \ref{Fig:NFG}, the duality defect is given by $\cN=\sigma s$.
If the set of topological manipulations $G$ contains a gauging operation $\sigma$, then the symmetry realized by $\cN$ is non-invertible, otherwise, the symmetry is invertible.

\begin{table}[!htp] \centering
  \renewcommand{\arraystretch}{1.3}
  \begin{tabular}{c|c|c}
   N & Theory & Defects  \\ \hline \strut
   2 &  $(\bZ_2)_m,(\widehat{\bZ}_2)_m$ & $\tau^m\sigma s \tau^{m}$  \\ \hline \strut
   2&  $(\bZ_2^f)_m$ &$\tau^m\tau s\tau^{m}$  \\   \hline \strut
   p & $\bZ_p, \widehat{\bZ}_p$ & $\sigma s $  \\   \hline \strut
   4 &  $(\bZ_4)_m,(\widehat{\bZ}_4)_m,(\bZ_4^f)_m,(\widehat{\bZ}^f_4)_m$ & 
    $\tau^m\sigma s \tau^{m}$  \\ \hline \strut
   6 & $(\bZ_6)_{\pm \pm m}$ & $\tau^m \sigma s\tau^{m}$  \\ \hline \strut
   6&  $(\bZ_6)_{\pm f m}$ & $\sigma \tau \sigma \tau \sigma s $  
  \end{tabular}
  \caption{Duality defects of $T_N[\bP^1 \times \bP^1]$ at $R=1$.}
  \label{Tab:p1p1N}
\end{table}

In this way, we analyze all possible defects in global variants of $T_N[\bP^1 \times \bP^1]$ for $N=\text{prime}$, $N=4$, and $N=6$. The result is in Table \ref{Tab:p1p1N}. 
Notice that except for the theory $(\bZ^f_2)_m$, the defects realized using $F=s$ involve the gauging operation and are thus non-invertible. In fact, they are duality defects with fusion rule \cite{Kaidi:2022cpf}
\begin{equation} \label{Eqn:fusionDDefect}
   \eta^N=1,  \hspace{0.2 in} \eta \times \cN =\cN,  \hspace{0.2 in} 
% \vphantom{.}
\overline{\cN} \sim \cN 
,\hspace{0.2 in} 
\cN\times \overline{\cN}= \sum \eta(\gamma),
\end{equation}
where $\eta$ is the line generating the $\bZ_N$ symmetry, $\overline{\cN}$ is the orientation reversal of $\cN$ and $\gamma\in H_1(M_1, \bZ_N)$ is a 1-cycle of $\Sigma_2$. 
Note that these are the same as the fusion rules of the Tambara-Yamagami fusion categories $TY(\bZ_N)$.

Besides that for the theory of $(\bZ^f_2)_m$, one can identify a defect composed of stacking SPT phase $\tau$ and duality $s$ at $R=1$. Since it does not involve the gauging operation, it implements an $\bZ_2$ invertible symmetry. 
% from the isometries of the 4-manifold generated by $s$. 
As we will see in the following, this $\bZ_2$ symmetry is anomalous having a mixed anomaly with $\bZ_2^f$. 
After gauging $\bZ_2^f$, the $\bZ_2$ symmetry becomes non-invertible, and the associated duality defect is given by $\tau \sigma \tau s$
\cite{Kaidi:2021xfk}. 
This duality defect can be identified in Figure \ref{Fig:N2f2}. After gauging $\sigma$, the theory becomes $(\bZ_2)_1$ and indeed one can find this duality defect from Table \ref{Tab:p1p1N}.

A typical example of this phenomenon can be seen in the well-known Ising model and its fermionization. The non-invertible duality defect $\mathcal N$ in Ising after fermionization turns out to be the $(-1)^{F_L}$ which is invertible but has a non-trivial 't Hooft anomaly with fermionic parity $(-1)^F$ \cite{Chang:2022hud}. On the other hand, gauging $(-1)^F$ gives back the Ising model and $(-1)^{F_L}$ is mapped back to the non-invertible $\mathcal N$ line correspondingly as a reminiscent of the 't Hooft anomaly.

Since the duality defects found in theories $(\bZ_2)_m$ and $(\widehat{\bZ}_2)_m$ can be related to the anomalous invertible symmetry in $(\bZ_2)^f_m$ by either dualities or topological manipulations, they are non-intrinsic non-invertible \cite{Kaidi:2022cpf, Kaidi:2021xfk, Kaidi:2022uux}. While the duality defects realized in global variants of $T_N[\bP^1\times \bP^1]$ with $N>2$ are not connected to any invertible symmetries, thus they are intrinsic non-invertible.

\paragraph{Mixed anomaly}

Given a choice of maximal isotropic sublattice $L$, the 2d absolute theory can have enhanced 0-form symmetry $G' \subset \MCG(\bP^1 \times \bP^1)$ if for any element $g \in G'$, one has 
\begin{equation}
     g(M_4) = M_4, \qquad g(\cL) = \cL, 
\end{equation}
where $\cL=L\otimes H^1(\Sigma_2,D)$. 
Thus, the partition function $Z_{g(M_4),g(L)}(\Sigma_2,0)$ is invariant under the transformation $g$. 
From the analysis in equation \eqref{Eqn:p1p1duality}, the $\bZ_2$ subgroup generated by reversing orientation is always a symmetry for any absolute theory, while for the other $\bZ_2$ generated by $s$, it depends on the number of M5-branes $N$ and the choice of the maximal isotropic sublattice.

If there is a mixed anomaly between $G'$ and $\bZ_N$, then one can obtain the non-invertible defect by gauging $\bZ_N$ \cite{Kaidi:2021xfk}. 
In our case, we can identify this mixed anomaly from the choice of $L$ and $L^{\perp}$ \cite{Bashmakov:2022jtl}. 
Given a 2d global variant specified by $L\otimes v \subset \cL$ and $\beta \otimes v \in \cL^\perp$ with $\beta \in L^{\perp}$ and $v \in H^1(\Sigma_2,\bZ_N)$ is a $\bZ_N$ cocycle on $\Sigma_2$. 
Although $L$ is invariant under symmetry $g\in G$, the choice $g(\beta \otimes v)$ is not necessarily an element of $\cL^\perp$. In general, it can be written as $g(\beta \otimes v) = \beta \otimes v + \alpha \otimes v $ with $
\alpha \otimes v \in \cL$. There will be a mixed anomaly between $G'$ and $\bZ_N$ if the following condition holds \cite{Bashmakov:2022uek} 
\begin{equation}
    e^{ \frac{i}{2} \langle \alpha\otimes v,\beta\otimes v \rangle} \neq 1.
\end{equation}
As we will see in the following, one can find such mixed anomalies in the 2d absolute theories.

We will derive the mixed anomaly between the duality transformation $s$ and $\bZ_2^f$ in the theory $(\bZ_2^f)_0$ from the partition function. 
For example, consider the absolute theory $(\bZ^f_2)_0$ defined by $L_3$ in equation \eqref{Eqn:LN2} with generator $\alpha=(1,1)$ and $\beta = (0,1)$. In terms of the cycles in $\bP^1 \times \bP^1$, let's denote them by $\alpha = b + f$ and $\beta = b$. 
Obviously, $L_3$ is invariant under tranformation $s\in \MCG(\bP^1 \times \bP^1)$ so as $\alpha$, but the $L^{\perp}$ part $\beta$ is not invariant under $s$, but transform as $s(b) = f$. By direct calculation, one can show 
\begin{equation}
    \begin{aligned}
     Z_{s(L_3)}(\Sigma_2, s(b \otimes v),s(R)) &=  
   \langle \cL_3, f \otimes v | Z \rangle \\
   & = \langle \cL_3, f + (b+ f)\otimes v | Z \rangle\\
   % &=\langle \cL_3, 0| \Phi(\hat \zeta \otimes v) \Phi( \zeta \otimes v+ \hat \zeta \otimes v) | Z \rangle \\
   &=\langle \cL_3, 0| e^{\frac{i}{2}\langle f \otimes v, b \otimes v+ f \otimes v \rangle} \Phi( b \otimes v) | Z \rangle \\
   &
   =e^{\frac{i}{2}\langle f \otimes v, b \otimes v \rangle}
   Z_{L_3}(\Sigma_2, f,\frac{1}{R}) \\
   % &= e^{i \frac{1}{2} \langle v, v\rangle}Z_{L_3}(\Sigma_2,\frac{1}{R}, g) \\
   &= e^{i\pi \text{Arf}(v)} Z_{L_3}(\Sigma_2,\frac{1}{R}, g)
    \end{aligned}
\end{equation}
where we have used that 
$\Arf(v_1+v_2)+\Arf(v_1)+ \Arf(v_2) = \langle v_1, v_2\rangle$ with $v_1,v_2 \in H^1(\Sigma_2,\bZ_N)$. As shown in \cite{Kaidi:2021xfk}, at the fixed point $R=1$, this is the anomaly required to realize the duality defect by gauging $\bZ_2^f$.

\paragraph{Duality defects from SymTFT}

The duality defects constructed above can also be realized in the SymTFT. In general, a subgroup $F \subset \MCG(M_4)$ induces a domain wall in the SymTFT with non-trivial action on the anyons as $L_{\vec{\alpha}} \to L_{F^t\vec{\alpha}}$ 
where $\vec{\alpha}$ represents the charge of $L$. The associated condensation defect is defined by 
\cite{Choi:2022zal, Roumpedakis:2022aik, Gaiotto:2019xmp}
\begin{equation}
    \cC_F(M_2) \; \sim \sum_{\gamma \in H_1(M_2,\bZ_N)} L_{(\mathbb{I}_r -F^T)\vec{\alpha}} (\gamma)
\end{equation}
which is realized geometrically as a surgery defect \cite{Bashmakov:2022uek}. 
A twist defect $T_F(M_2,M_1)$ is obtained by condensing $L_{(\mathbb{I}_r -F^t)\vec{\alpha}}$ on $\cC_F(M_2)$ through $\partial M_2 = M_1$ with Dirichlet boundary condition. 
As studied in \cite{Kaidi:2022cpf}, after gauging $F$ in the SymTFT and shrinking the slab, these twist defects become the $|F|$-ality defect \footnote{Precisely, this is defined when $F$ is a cyclic group.}. 
In our case, $F=\langle s \rangle \subset \MCG(\bP^1 \times \bP^1)$ corresponds to the electro-magnetic duality of $\bZ_N$ gauge theory, which will give rise to the duality defect obtained in this section after gauging $\bZ_2$ and shrinking the interval in the SymTFT.

\subsection{Connected sum of $\bP^1 \times \bP^1$}

We will extend the previous analysis to the connect sum of $\bP^1 \times \bP^1$. The SymTFT in this case is the $\bZ_N \times \bZ_N$ gauge theory, which have very rich structure on the global variants, symmetries and anomalies. Note that $(\bP^1 \times \bP^1) \# (\bP^1 \times \bP^1)$ does not have complex structure and will be treated as a real 4-manifold.

\paragraph{$\bZ_N \times \bZ_N$ gauge theory}

The intersection form of $\#^2(\bP^1 \times \bP^1)$ is 
\begin{equation} \label{Eqn:interF0F0}
Q=    
\begin{pmatrix}
0 & 1 & 0 & 0 \\
1 & 0 & 0 & 0 \\
0 & 0 & 0 & 1 \\
0 & 0 & 1 & 0 \\
\end{pmatrix}.
\end{equation}
Let $b_i$ and $f_i$ with $i=1,2$ be a basis of $H_2(\#^2(\bP^1 \times \bP^1),\bZ)$. 
After compactifying the 7d TQFT on it, the 3d SymTFT is 
\begin{eqnarray}
    S_{3d}
    &=& \frac{2 \pi}{N} \int_{W_3} a_1  \cup \delta  \widehat{a}_1 + a_2  \cup \delta  \widehat{a}_2\;, 
\end{eqnarray}
where $a_i = \int_{b_i} c$ and $\hat a_i = \int_{f_i} c$ are $\bZ_N$ cocycles on $W_3$. 
The defect group 
\begin{equation}
    \mathscr{D} = \bZ_N \times \bZ_N \times \bZ_N \times\bZ_N 
\end{equation}
This is the $\bZ_N \times \bZ_N$ gauge theory that will be the SymTFT for 2d theories with $\bZ_N \times \bZ_N$ symmetries.

% \subsubsection{Orbifold groupoid}

\paragraph{Duality}
The mapping class group of $\MCG(\#^2(\bP^1 \times \bP^1))$ is an infinite group with the following generators  
\begin{equation} \label{Eqn:dualityznzn}
S = \begin{pmatrix}
1 & 0 & 0 & 0 \\
0 & 1 & 0 & 0 \\
0 & 0 & 0 & 1 \\
0 & 0 & 1 & 0 \\
\end{pmatrix},\quad 
T = \begin{pmatrix}
-1 & 0 & 0 & 0 \\
0 & -1 & 0 & 1 \\
1 & 0 & 1 & 0 \\
0 & 0 & 0 & 1 \\
\end{pmatrix},\quad 
D = \begin{pmatrix}
0 & 0 & 1 & 0 \\
0 & 0 & 0 & 1 \\
1 & 0 & 0 & 0 \\
0 & 1 & 0 & 0 \\
\end{pmatrix},\quad 
W = \begin{pmatrix}
-1 & 0 & 0 & 0 \\
0 & -1 & 0 & 0 \\
0 & 0 & 1 & 0 \\
0 & 0 & 0 & 1 \\
\end{pmatrix},
\end{equation}

The element $P$ of mapping class group acts on the 2-cycle (Poincar\'{e} dual to the 2-form in $\Omega^2(M_4)$) 
\be
J=xb_1+yf_1+zb_2+wf_2=\bp x \\ y\\ z\\ w\ep
\ee
as $J\rightarrow PJ$. The volume of 2-cycles are
\be
\ba
V_{b_1}&=J\cdot b_1=y\cr
V_{f_1}&=J\cdot f_1=x\cr
V_{b_2}&=J\cdot b_2=w\cr
V_{f_2}&=J\cdot f_2=z\,.
\ea
\ee
The volume of $\#^2(\bP^1 \times \bP^1)$ is invariant under the action of $P$:
\be
V_{\#^2(\bP^1 \times \bP^1)}=\frac{1}{2}J\cdot J=xy+zw\,.
\ee
We introduce three geometric parameters
\be
R_1=\frac{x}{y}=\frac{V_{f_1}}{V_{b_1}}\ ,\ R_2=\frac{z}{w}=\frac{V_{f_2}}{V_{b_2}}\ ,\ R_3=\frac{y}{z}=\frac{V_{b_1}}{V_{f_2}}\,.
\ee
The $\MCG(\#^2(\bP^1 \times \bP^1))$ generators acting non-trivially on these parameters are
\be
\ba
&S\cdot R_1=R_1\ ,\ S\cdot R_2=\frac{1}{R_2}\ ,\ S\cdot R_3=R_2 R_3\cr
&D\cdot R_1=R_2\ ,\ D\cdot R_2=R_1\ ,\ D\cdot R_3=\frac{1}{R_1 R_2 R_3}\cr
&T\cdot R_1=\frac{R_1 R_2 R_3}{R_2 R_3-1}\ ,\ T\cdot R_2=R_2+R_1 R_2 R_3\ ,\ T\cdot R_3=\frac{1-R_2 R_3}{R_1 R_2 R_3+R_2}\cr
&W\cdot R_1=R_1\ ,\ W\cdot R_2=R_2\ ,\ W\cdot R_3=-R_3\,.
\ea
\ee

Let us discuss the finite subgroups of $\MCG(\#^2(\bP^1 \times \bP^1))$. 
First, let's consider $\bZ_2$ subgroups generated by one of the generators in \eqref{Eqn:dualityznzn}. The fixed points of the coupling constants are 
\begin{align}
 S:\; (R_1,1,R_3),\quad D: \;(R_1,\frac{1}{R_1},\pm 1),\quad T:\; (0,R_2,\frac{1}{2R_2}),\quad
    W:\; (R_1,R_2,0)
\end{align}
which depending on arbitary parameters define the extended loci in the conformal manifold. 
% Note that only smooth 4-manifolds are considered, so we will ignore the cases with zero coupling. 
When the couplings are taken to be
\be \label{Eqn:fixPZnZn}
(R_1,R_2,R_3)=(1,1,\pm 1)\,.
\ee
The symmetry generated by $S$ and $D$ is enhanced to $D_8$ and one can realize more interesting defects at this coupling. 
One can consider more general subgroups of $\MCG(\#^2(\bP^1 \times \bP^1))$ and find more fixed points of these couplings on the conformal manifold.

\paragraph{N=2} We find that the maximal isotropic sublattice are given by
\begin{equation}
    \begin{array}{cccc}
 L_{1} =\{(0,0,0,0) ,& (0,0,0,1) ,& (0,1,0,0) ,& (0,1,0,1)\} \\ L_{2} =\{
 (0,0,0,0) ,& (0,0,0,1) ,& (1,0,0,0) ,& (1,0,0,1) \} \\ L_{3} =\{
 (0,0,0,0) ,& (0,0,0,1) ,& (1,1,0,0) ,& (1,1,0,1) \} \\ L_{4} =\{
 (0,0,0,0) ,& (0,0,1,0) ,& (0,1,0,0) ,& (0,1,1,0) \} \\ L_{5} =\{
 (0,0,0,0) ,& (0,0,1,0) ,& (1,0,0,0) ,& (1,0,1,0) \} \\ L_{6} =\{
 (0,0,0,0) ,& (0,0,1,0) ,& (1,1,0,0) ,& (1,1,1,0) \} \\ L_{7} =\{
 (0,0,0,0) ,& (0,0,1,1) ,& (0,1,0,0) ,& (0,1,1,1) \} \\ L_{8} =\{
 (0,0,0,0) ,& (0,0,1,1) ,& (1,0,0,0) ,& (1,0,1,1) \} \\ L_{9} =\{
 (0,0,0,0) ,& (0,0,1,1) ,& (1,1,0,0) ,& (1,1,1,1) \} \\ L_{10} =\{
 (0,0,0,0) ,& (0,1,0,1) ,& (1,0,1,0) ,& (1,1,1,1) \} \\ L_{11} =\{
 (0,0,0,0) ,& (0,1,0,1) ,& (1,0,1,1) ,& (1,1,1,0) \} \\ L_{12} =\{
 (0,0,0,0) ,& (0,1,1,0) ,& (1,0,0,1) ,& (1,1,1,1) \} \\ L_{13} =\{
 (0,0,0,0) ,& (0,1,1,0) ,& (1,0,1,1) ,& (1,1,0,1) \} \\ L_{14} =\{
 (0,0,0,0) ,& (0,1,1,1) ,& (1,0,0,1) ,& (1,1,1,0) \} \\ L_{15} =\{
 (0,0,0,0) ,& (0,1,1,1) ,& (1,0,1,0) ,& (1,1,0,1) \} 
\end{array}
\end{equation}
Thus, there are 15 absolute theories on the boundary with $\bZ_2 \times \bZ_2$ symmetry. These theories are transformed into each other by different ways of gauging subgroups in $\bZ_2 \times \bZ_2$ forming the orbifold groupoid \cite{Gaiotto:2020iye}. 
For example, there are three ways to gauge a single $\bZ_2$ by gauging the first one, the second one or the diagonal one. One can also gauge the full $\bZ_2 \times \bZ_2$ with or without the SPT phase. In this way, one can obtain 6 bosonic theories. Besides that one can perform fermizations to these theories leading to 9 fermionic theories. In total, there are 15 absolute theories.

For each of these absolute theories, one can stack SPT phase and Arf invariant, it turns out that these two operations generating $D_8$ group. So, there are 8 global variants associated with each absolute theory and totally 120 global variants. 
The counting of global variants can also be understood from the automorphism group of $\Aut_{\bZ_2}(Q)$. 
In this example, we find that $|\Aut_{\bZ_2}(Q)|=720$, which again can be understood as the semi-product of $\Aut(\bZ_2 \times \bZ_2)$ and $\cO_2(Q)$. Since $\Aut(\bZ_2 \times \bZ_2) = S_3$, we find that $|\cO_2(Q)|=120$ that corresponds to global variants of $T_2[\#^2(\bP^1 \times \bP^1)]$.

Similar to the $\bP^1\times \bP^1$ case, we can associate each global variants with a matrix $M \in \cO_2(Q)$. This matrix presentation of a global variant can be obtained from the maximal isotropic sublattice $L$ and the complement $L^{\perp}$. For example, consider $L_1$ and $L_1^{\perp}$, one can obtain 8 the following matrices
\begin{align*}
    &  
M_{L_1}^{(1)}=\left(
\begin{array}{cccc}
 0 & 1 & 0 & 0 \\
 1 & 0 & 0 & 0 \\
 0 & 0 & 0 & 1 \\
 0 & 0 & 1 & 0 \\
\end{array}
\right),
\quad 
M_{L_1}^{(2)}=\left(
\begin{array}{cccc}
 0 & 1 & 0 & 0 \\
 1 & 0 & 0 & 0 \\
 0 & 0 & 0 & 1 \\
 0 & 0 & 1 & 1 \\
\end{array}
\right),
\quad 
M_{L_1}^{(3)}=
\left(
\begin{array}{cccc}
 0 & 1 & 0 & 0 \\
 1 & 0 & 0 & 1 \\
 0 & 0 & 0 & 1 \\
 0 & 1 & 1 & 0 \\
\end{array}
\right), \\
&
M_{L_1}^{(4)}=
\left(
\begin{array}{cccc}
 0 & 1 & 0 & 0 \\
 1 & 0 & 0 & 1 \\
 0 & 0 & 0 & 1 \\
 0 & 1 & 1 & 1 \\
\end{array}
\right),
\quad 
M_{L_1}^{(5)}=\left(
\begin{array}{cccc}
 0 & 1 & 0 & 0 \\
 1 & 1 & 0 & 0 \\
 0 & 0 & 0 & 1 \\
 0 & 0 & 1 & 0 \\
\end{array}
\right),
\quad
M_{L_1}^{(6)}=
\left(
\begin{array}{cccc}
 0 & 1 & 0 & 0 \\
 1 & 1 & 0 & 0 \\
 0 & 0 & 0 & 1 \\
 0 & 0 & 1 & 1 \\
\end{array}
\right),
\\
&
M_{L_1}^{(7)}=
\left(
\begin{array}{cccc}
 0 & 1 & 0 & 0 \\
 1 & 1 & 0 & 1 \\
 0 & 0 & 0 & 1 \\
 0 & 1 & 1 & 0 \\
\end{array}
\right),
\quad 
M_{L_1}^{(8)}=
\left(
\begin{array}{cccc}
 0 & 1 & 0 & 0 \\
 1 & 1 & 0 & 1 \\
 0 & 0 & 0 & 1 \\
 0 & 1 & 1 & 1 \\
\end{array}
\right)\;.
\end{align*}
Each of them is a matrix in $\cO_2(Q)$ associated with a global variants. 
Similarly, one can obtain the matrix representation for other $L$'s.

These 2d theories admit many topological defects. For example, let's consider the global variant defined by the following $\cO_2(Q)$ matrix 
\begin{equation}
    M = \left(
\begin{array}{cccc}
 1 & 0 & 0 & 0 \\
 0 & 1 & 0 & 0 \\
 0 & 0 & 0 & 1 \\
 0 & 0 & 1 & 0 \\
\end{array}
\right)
\end{equation}
From the 1st and 3rd, we can see that they are from the $L_2$ and $B = \{(0,1,0,0),(0,0,1,0)\}$. 
Consider the duality transformation $S$ and $D$. These action can be undone by the following two topological manipulation 
\begin{equation*}
\sigma_4=
 \left(
\begin{array}{cccc}
 1 & 0 & 0 & 0 \\
 0 & 1 & 0 & 0 \\
 0 & 0 & 0 & 1 \\
 0 & 0 & 1 & 0 \\
\end{array}
\right) 
,\qquad 
 \sigma_5=\left(
\begin{array}{cccc}
 0 & 0 & 0 & 1 \\
 0 & 0 & 1 & 0 \\
 0 & 1 & 0 & 0 \\
 1 & 0 & 0 & 0 \\
\end{array}
\right) 
\end{equation*}
which represents gauging of subgroups in $\bZ_2 \times \bZ_2$ with possible stacking of SPT phases. In this way, one can realize topological defects in the theory specified by $M$ at the fixed points found in \eqref{Eqn:fixPZnZn}. 
% , we have two duality defects $\cN_S=S\sigma_4$ and $\cN_D=D\sigma_5$
Since the construction involving gauging, the corresponding symmetry is non-invertible described by the $TY(D_8)$ category \cite{Thorngren:2021yso}.

\paragraph{$N=p>2$}

Let's consider a theory with non-anomalous symmetry $G=\bZ_p \times \bZ_p$ where $p$ is prime larger than $2$. There are three types of topological manipulations include the automorphism of $\bZ_p \times \bZ_p$, the stacking of SPT phase $v_2 \in H^2(\bZ_p \times \bZ_p,U(1)) = \bZ_p$ and gauging subgroups of $\bZ_p \times \bZ_p$. 
With these basic operations, one can find $2(p+1)$ gauging operations \cite{Gaiotto:2020iye}. Take $p=3$ for example,  
there are 4 ways that $\bZ_3$ can embedded in $\bZ_3 \times \bZ_3$. In terms of their generator, $(1,0)$, $(0,1)$, $(1,1)$, $(1,2)$. Besides that, one has 3 ways to gauge $\bZ_3 \times \bZ_3$ with the SPT phase. Taking into account the trivial gauging, there are a total of 8 orbifolding operations leading to a orbifold groupoid.

This result can be confirmed from the study of global variants of $T_3[\#^2(\bP^1 \times \bP^1)]$. The maximal isotropic sublattices in this case are given by 
\begin{equation*}
\begin{aligned}
 L_1=&\{(0,0,0,0) , (0,0,0,1) , (0,1,0,0) , (0,1,0,1) , (0,0,0,2) , (0,1,0,2) , (0,2,0,0) , (0,2,0,1) , (0,2,0,2)\} \\
 L_2=&\{(0,0,0,0) , (0,0,0,1) , (1,0,0,0) , (1,0,0,1) , (0,0,0,2) , (1,0,0,2) , (2,0,0,0) , (2,0,0,1) , (2,0,0,2) \}\\
L_3=& \{(0,0,0,0) , (0,0,1,0) , (0,1,0,0) , (0,1,1,0) , (0,0,2,0) , (0,1,2,0) , (0,2,0,0) , (0,2,1,0) , (0,2,2,0) \}\\
 L_4=&\{(0,0,0,0) , (0,0,1,0) , (1,0,0,0) , (1,0,1,0) , (0,0,2,0) , (1,0,2,0) , (2,0,0,0) , (2,0,1,0) , (2,0,2,0) \}\\
 L_5=&\{(0,0,0,0) , (0,1,0,1) , (0,2,0,2) , (1,0,2,0) , (1,1,2,1) , (1,2,2,2) , (2,0,1,0) , (2,1,1,1) , (2,2,1,2) \}\\
 L_6=&\{(0,0,0,0) , (1,0,1,0) , (0,1,0,2) , (0,2,0,1) , (1,1,1,2) , (1,2,1,1) , (2,0,2,0) , (2,1,2,2) , (2,2,2,1) \}\\
 L_7=&\{(0,0,0,0) , (0,1,1,0) , (0,2,2,0) , (1,0,0,2) , (1,1,1,2) , (1,2,2,2) , (2,0,0,1) , (2,1,1,1) , (2,2,2,1) \}\\
 L_8=&\{(0,0,0,0) , (1,0,0,1) , (0,1,2,0) , (0,2,1,0) , (1,1,2,1) , (1,2,1,1) , (2,0,0,2) , (2,1,2,2) , (2,2,1,2)\} \\
\end{aligned}
\end{equation*}
which defines 8 absolute theories with $\bZ_3 \times \bZ_3$ symmetry on $\Sigma_2$. Considering the possible stacking of SPT phase, there are 3 global variants for each absolute theory. Thus, there are 24 global variants of $T_3[\#^2(\bP^1 \times \bP^1)]$.

The automorphism group has order $|\Aut_{\bZ_3}(\bZ_3 \times \bZ_3)|=1152$. Taking into account the automorphism group $GL(3,\bZ_3)$ with order 48, one has that $|\cO_3(Q)|=24$ which is expected from the physical analyisis. 
Similarly, one can assign each global variant to a $\cO_3(Q)$ matrix. For example, the matrices of global variants defined by $L_1$ and its complement $L_1^{\perp}$ is given by 
\begin{equation*}
M_{L_1}^{(1)}=
\left(
\begin{array}{cccc}
 0 & 1 & 0 & 0 \\
 1 & 0 & 0 & 0 \\
 0 & 0 & 0 & 1 \\
 0 & 0 & 1 & 0 \\
\end{array}
\right),
\quad 
M_{L_1}^{(2)}=
\left(
\begin{array}{cccc}
 0 & 1 & 0 & 0 \\
 1 & 0 & 0 & 2 \\
 0 & 0 & 0 & 1 \\
 0 & 1 & 1 & 0 \\
\end{array}
\right),
\quad 
M_{L_1}^{(3)}=
\left(
\begin{array}{cccc}
 0 & 1 & 0 & 0 \\
 1 & 0 & 0 & 1 \\
 0 & 0 & 0 & 1 \\
 0 & 2 & 1 & 0 \\
\end{array}
\right)
\end{equation*}

Next, consider the global variant defined by $M_{L_1}^{(2)}$. The $S$ duality and $D$ duality will change the global variants. To get back $M_{L_1}^{(2)}$, one can perform the gaugings defined by 
\begin{equation*}
\sigma_5 = \left(
\begin{array}{cccc}
 1 & 1 & 1 & 2 \\
 0 & 1 & 0 & 0 \\
 0 & 2 & 0 & 1 \\
 0 & 1 & 1 & 0 \\
\end{array}
\right),
\qquad
\sigma_7=
\left(
\begin{array}{cccc}
 0 & 2 & 1 & 0 \\
 0 & 0 & 0 & 1 \\
 1 & 0 & 0 & 1 \\
 0 & 1 & 0 & 0 \\
\end{array}
\right)
\end{equation*}
which can undo the duality transformation $S$ and $D$. Again, we construct the non-invertible defects at the fixed point of these self-dual couplings. In general, there are many topological defects and we will study them in future work.

\section{6d $\cN=(2,0)$ SCFTs on Hirzebruch surfaces $\mathbb{F}_l$}
\label{sec:4}

In this section, we will study the compactification of the 6d $\cN=(2,0)$ theories of type $A_{N-1}$ on Hirzebruch surfaces $\mathbb{F}_l$. 
As we will see that it is sufficient to focus on the case $\bF_1$.
Using 3d SymTFT, we determine the global variants of $T_N[\bF_1]$ and possible topological manipulations for various different $N$. Similar to the $\bP^1 \times \bP^1$ case, we identify the duality group $\MCG(\bF_1)$ and the coupling of $T_N[\bF_1]$ from invariant volume of $\bF_1$. Finally, we construct topological defects in each of these global variants.

\subsection{Twisted $\bZ_N$ gauge theory}

Let us denote the divisor classes of Hirzebruch surface $\mb{F}_l$ by $f$ and $b$, and they have the intersection form
\be
Q=\bp f\cdot f & f\cdot b\\ b\cdot f & b\cdot b\ep=\bp 0 & 1\\1 & -l\ep\,.
\ee
After compactification, the 7d TQFT in equation \eqref{eqn:symTFT-7d} becomes 
\begin{eqnarray} \label{Eqn:actionF1}
    S_{3d} &=& \frac{N}{2\pi} \int \hat{a} \wedge da  - \frac{Nl}{4\pi} \int a \wedge d a 
    % &=& -\frac{2\pi}{N} \int \hat{a}  \cup \delta a  + \frac{l\pi}{N} \int a \cup \delta a \nonumber
\end{eqnarray}
where $a = \int_{b} c$ and $\hat{a} = \int_{f} c$. 
% In the last line of the above equation, we have written the $a$ and $\hat{a}$ in terms of $\bZ_N$ cocycle. 
Note that under the gauge transformation of $a\to a + dg$ and $\hat a \to \hat a + d\hat g$, there will be a boundary term
\begin{equation} \label{Eqn:gaugeinvariance}
    \frac{1}{4\pi} \int_{\Sigma_2} (2N \hat g -N l g) da
\end{equation}
which constrains us to consider the transformations satisfying $2N \hat g -N l g \in 2\pi \bZ$ \cite{Kapustin:2014gua}.

From this, the $K$-matrix and its inverse are determined to be
\begin{equation}
    K = \left(\begin{array}{cc}
                0 & N \\
                N & -l N
              \end{array}\right), \quad K^{-1} = \left(\begin{array}{cc}
                \frac{l}{N} & \frac{1}{N} \\
                \frac{1}{N} & 0
              \end{array}\right).
\end{equation}
We see that the defect group is
\begin{eqnarray}
    \mathscr{D}_{\mathbb{F}_l} = \mathbb{Z}_N \times \mathbb{Z}_N,
\end{eqnarray}
The line operators $L_{(e,m)}$ are given similarly by the equation \eqref{Eqn:p1p1L} with $(e,m) \in \bZ_N \times \bZ_N$. 
The topological spin is 
\begin{equation}
    \theta \left( L_{(e,m)} \right) = \exp \left( \frac{2\pi i}{N}\left(em+\frac{le^2}{2}\right) \right)
\end{equation}
and the S-matrix is determined to be
\begin{equation}
    S(\vec{\alpha},\vec{\beta}) = \frac{1}{N} \exp\left[\frac{2\pi i}{N} (\alpha_1 \beta_1 l + \alpha_2 \beta_1 + \alpha_1 \beta_2)\right].
\end{equation}
This matches with the result in reference \cite{Manschot:2011dj} which computed such S-matrices from the point of view of $\mathcal{N}=4$ $SU(N)$ SYM on $\mathbb{F}_l$.

By pushing our 3d defect lines $\gamma$, $\gamma'$ to the 2d boundary $\Sigma$ of our 3-manifold $M_3$, we find the following commutation relation between line operators
\begin{equation}
    L_{\vec{\alpha}}(\gamma) L_{\vec{\beta}}(\gamma') = B(\vec{\alpha},\vec{\beta})^{\gamma \cdot \gamma'} L_{\vec{\beta}}(\gamma')L_{\vec{\alpha}}(\gamma),
\end{equation}
where $\gamma \cdot \gamma'$ denotes the intersection number of the two lines on the 2d boundary or equivalently their linking number in bulk. This allows us to define absolute theories on the boundary $\Sigma$ of the 3-manifold $M_3$ by choosing a maximal commuting subgroup of the defect group or in other words by choosing a polarization. Such a subgroup $L$ is determined by the requirement
\begin{equation}
    B(\vec{\alpha},\vec{\beta}) = 1 \quad \forall ~\vec{\alpha}, \vec{\beta} \in L .
\end{equation}

In fact, it is sufficient to consider the case with $l=1$, since one can always shift the coefficient $-Nl$ in the DW twist to an integer in $\bZ_{2N}$ by $\hat{a} \to \hat a - a$. For even $l$, the DW twist can be turned off, and the action gives the $\bZ_N$ gauge theory. While all the odd $l$ is equivalent to $l=1$ and the 3d TQFT 
% \begin{equation}
%     S = \frac{2\pi}{N} \int \hat a \cup \delta a + \frac{\pi}{N} \int a \cup \delta a 
% \end{equation}
is a twisted $\bZ_N$ gauge theory denoted by $(\bZ_N)_N$. In the following, we will focus on the 4-manifold $\bF_1$. 
For theories $T_N[\bF_l]$, although the local dynamical physics are different, the global variants, symmetries, and anomalies are captured by the SymTFT obtained for $T_N[\bF_1]$.

\subsection{Global variants}

Similarly with the $\bZ_N$ gauge theories, we expect the topological manipulations of $T_N[\bF_1]$ include $\bZ_N$ gauging $\sigma$ and stacking Arf invariant $\xi$ which can be observed from $\Aut_{\bZ_N}(Q)$. Besides, there are also duality transformations from $\MCG(\bF_1)$. For $\bF_1$, we find that $\MCG(\bF_1) = \bZ^2_2$ given by 
\be \label{Eqn:f1duality}
I = \bp 1 & 0 \\ 0 & 1 \ep, \qquad r = \bp 1 & 0 \\ 2 & -1 \ep, \qquad -I=\bp -1 & 0\\0 & -1\ep,\qquad -rI=\bp -1 & 0\\-2 & 1\ep.
% \text{Aut}_{\bZ}(Q)= \left \{\bp 1 & 0 \\ 0 & 1\ep\ ,\ \bp -1 & 0\\-2 & 1\ep\ ,\ \bp -1 & 0\\0 & -1\ep\ ,\ \bp 1 & 0 \\ 2 & -1 \ep\, \right\}.
\ee
The only group element that acts non-trivially on the global variants is $r$.

\paragraph{\underline{$N=2$}}

In this case, we find the following maximal isotropic sublattice by \eqref{Eqn:isotropyM4}
\begin{equation*}
   L=\{(0,0),\quad (1,0)\}\rightarrow \mb{Z}_2
\end{equation*}
So, there is only one absolute theory according to its symmetry denoted by $\bZ_2$. Different from the $\bZ_2$ gauge theory, there is no maximal isotropic sublattice corresponding to theory with $\bZ_2$ gauged because the DW twist $\omega \in H^3(W_3, U(1))$ is a t'Hooft anomaly for $\bZ_2$.

This anomaly can be probed from the braiding between lines in the SymTFT. The theory has four anyons $1$, $s$, $\bar{s}$ and $b=s \times \bar{s}$ with topological spin $\theta(1)=\theta(b)=1$, $\theta(s)=i$ and $\theta(s)=-i$. Notice that these anyons are identical to those in the double semion model. 
Indeed, there is only one type of topological boundary condition found in the double semion model \cite{Ji:2019eqo}. 
The maximal isotropic sublattice corresponds to take Lagrangian algebra $(1,b)$. As one can check that there is non-trivial braiding between either $s$ and $\bar s$ with itself, which implies that $\bZ_2$ is anomalous \cite{Kaidi:2023maf}.

Since there is only one absolute theory, the automorphism group $\Aut_{\bZ_2}(Q)$ is also simple, which is isomorphic to $\bZ_2$ with generator 
\begin{equation}
    \xi = \left(\begin{array}{cc}
        1 & 1 \\
        0 & 1
    \end{array}\right).
\end{equation}
Since $\bF_1$ is not spin, the theory $T_2[\bF_1]$ does not have spin structure on $\Sigma_2$. Different from the $\bP^1 \times \bP^1$ case, we cannot understand this operation as stacking Arf invariant instead stacking some other invertible TQFT related to the anomaly discussed above. Again, we will denote the theory with and without this stacking as $(\bZ_2)_0$ and $(\bZ_2)_1$. 
% Note that $\bF_1$ is not spin, thus 
% This operation gives the stacking of Arf invariant for $T_2[\bF_1]$. 

%
\begin{figure}
\centering
\begin{tikzpicture}[scale=1.5]
\draw node at (0,0) {$(\bZ_2)_0$};
\draw node at (3,0) {$(\bZ_2)_1$};
% \draw [<->,orange] (.3,0) -- (2.7,0);
\draw [<->,blue] (0,.3) arc (120:60:3);
\draw [<->,orange] (0,-.3) arc (-120:-60:3);
 \draw node at (1.5,-0.5) {${\color{orange} r}$};
 \draw node at (1.5,.5) {${\color{blue} \xi}$};
\node[left] at (-0.3,0) {$\left(\begin{matrix} 1 & 0 \\ 0 & 1 \end{matrix}\right)$};
\node[right] at (3.3,0) {$\left(\begin{matrix} 1 & 1 \\ 0 & 1 \end{matrix}\right)$};
\end{tikzpicture}
    \caption{Web of transformations for $T_2[\bF_1]$. The transformations
in orange are the duality transformations.
The transformations in blue are topological manipulations.}
    \label{Fig:F1N2}
\end{figure}

Hence in this case, there are two global variants, labeled by $(\mb{Z}_2)_0$ and $(\mb{Z}_2)_1$, which are transformed into each other with the stacking of the non-trivial phase $\xi$ and the duality $r$. The same as the $\bP^1 \times \bP^1$ case, one can associate a $\Aut_{\bZ_2}(Q)$ matrices to each global variant and perform the operation $\xi$ and $r$ on it using the equations \eqref{Eqn:actG} and \eqref{Eqn:actF}. The result is in Figure \ref{Fig:F1N2}. 

%
% \be
% (\mb{Z}_2)_0\overset{\xi}{\longleftrightarrow}(\mb{Z}_2)_1\,.
% \ee

\paragraph{\underline{$N=p>2$}}

Consider $N$ is a prime number larger than two. 
The maximal isotropic lattices can be obtained from the equation \eqref{Eqn:isotropyM4}. 
For $p=3$, they are  
% \be
% \ba
% L_1&=\{(0,0),\quad (0,1),\quad (0,2)\}\rightarrow\mb{Z}_3\cr
% L_2&=\{(0,0),\quad (1,1),\quad (2,2)\}\rightarrow\mb{Z}^{\rho}_3\,.
% \ea
% \ee
%
\begin{equation*} 
\begin{array}{cccc}
L_1 = \{(0,0),(1,0), (2,0)\} &\to &\bZ_3\\
L_2 = \{(0,0),(1,2),(2,1)\} &\to &\bZ_3^{\rho}
\end{array}
\end{equation*}
%
% when $N=5$, they are 
% %
% \begin{equation*} 
% \begin{array}{cccc}
% L_1 = \{(0,0),(1,0), (2,0),(3,0),(4,0)\} &\to &\bZ_5\\
% L_2 = \{(0,0),(1,2),(3,1),(2, 4), (4, 3)\} &\to &\mb{Z}^{\rho}_5 
% \end{array}
% \end{equation*}
% %
% and when $N=7$, they are 
% %
% \begin{equation*} 
% \begin{array}{cccc}
% L_1 = \{(0,0),(1,0), (2,0),(3,0),(4,0),(5,0),(6,0),(7,0)\} &\to &\bZ_7\\
% L_2 = \{(0,0),(1,2),(3,1),(2, 4), (4, 1),(5,3),(3,6),(6,5)\} &\to &\mb{Z}^{\rho}_7
% \end{array}
% \end{equation*}
% %
In fact, one can show that there are only two maximal isotropic sublattices for any prime $p$. 
% which leads to two absolute theories of $T_p[\bF_1]$ with $\bZ_p$ symmetry

Consider the sublattice generated by a lattice point $(e,m)$ other than $(0,0)$ in $\bZ^2_p$. Thus, the sublattice contains points $(e',m')$ satisfying $(e',m') = k (e,m)$ with $k\in \bZ_p^{\times}$. The inner product between these two points is 
\begin{equation}
    k(2em-m^2) = 0, \hspace{0.3cm} \text{mod}\; p .
\end{equation}
For prime $p$, the only solution is either $m=0$ or $2e=m$ $\text{mod}\; p$. Thus, the isotropic sublattices are generated by points $(1,0)$ and $(1,2)$ which define the following polarization 
\begin{equation} 
L_1 = \langle (1,0) \rangle \;\; \to \;\;\bZ_p,
\hspace{2cm}
L_2 = \langle (1,2) \rangle \;\; \to \;\; \bZ^{\rho}_p 
\end{equation}
%
% $L_1$ and $L_2$ in the above equation. 
Note that physically the above isotropic condition can be understood as the requirement of gauge invariance of the SymTFT discussed in \eqref{Eqn:gaugeinvariance}. 
One can also consider the sublattices generated by two or more linear independent points in $\bZ^2_p$. However, in this case, one always gets the full lattice, which is obviously not isotropic. Thus, one can only find two maximal isotropic lattices when $N$ is prime and larger than two. When  $N=2$, these two polarizations are the same.

\begin{figure}
\centering
\begin{tikzpicture}[scale=1.5]
\draw node at (0,0) {$\bZ_p$};
\draw node at (3,0) {$\bZ^{\rho}_p$};
% \draw [<->,orange] (.3,0) -- (2.7,0);
\draw [<->,blue] (0,.3) arc (120:60:3);
\draw [<->,orange] (0,-.3) arc (-120:-60:3);
 \draw node at (1.5,-0.5) {${\color{orange} r}$};
 \draw node at (1.5,.5) {${\color{blue} \rho}$};
\node[left] at (-0.3,0) {$\left(\begin{matrix} 1 & 0 \\ 0 & 1 \end{matrix}\right)$};
\node[right] at (3.3,0) {$\left(\begin{matrix} 1 & 2 \\ 0 & p-1 \end{matrix}\right)$};
\end{tikzpicture}
    \caption{Web of transformations for $T_p[\bF_1]$. The transformations
in orange are the duality transformations.
The transformations in blue are topological manipulations.}
    \label{Fig:F1N3}
\end{figure}

The automorphism group $\Aut_{\bZ_p}(Q)$ is still $D_{2(p-1)}$. Taking into account the $\Aut(\bZ_p)=\bZ_p^{\times}$, one has $\cO_p(Q)=\bZ_2$ with generator 
\begin{equation}
    \rho = \left(\begin{array}{cc}
        1 & 0 \\
        2 & p-1
    \end{array}\right)
\end{equation}
As one can check $\rho$ switches two polarizations $L_1$ and $L_2$. However, due to the twist, one cannot gauge $\bZ_p$ since the operation $\rho$ cannot be understood as gauging 
\footnote{Note that for $N=p$, we did not find the obstruction to gauging discussed in \cite{Kaidi:2023maf}. For either $L_1$ and $L_2$, one can always find representatives in $B_{1}$ and $B_2$ such that they have trivial braiding and generating $\bZ_p$ symmetry on $\Sigma_2$.}
. Indeed, the gauging operation comes from the electro-magnetic duality in the bulk, but the twisted $\bZ_N$ gauge theory does not have it. 
We will denote one absolute theory defined by $L_1$ as $\bZ_p$ and denote the other one by $\bZ^{\rho}_p$ to emphasize that it can be obtained from the theory $\bZ_p$ by a topological manipulation $\rho$. Note that there is no $\xi$ operation for prime $p$. $\bZ_p$ and $\bZ^{\rho}_p$ are the only two global variants of $T_p[\bF_1]$ related by the operation $\rho$ and duality $r$ as in Figure \ref{Fig:F1N3}.

\paragraph{\underline{$N=4$}}

The maximal isotropic lattices are
\be
\ba
L_1&=\{(0,0),\quad (1,0),\quad (2,0),\quad (3,0)\}\rightarrow\mb{Z}_4\cr
L_2&=\{(0,0),\quad (1,2),\quad (2,0),\quad (3,2)\}\rightarrow\mb{Z}^{\rho}_{4}\,\cr
L_3&=\{(0,0),\quad (0,2),\quad (2,0),\quad (2,2)\}\rightarrow\mb{Z}_2\times\mb{Z}^{\rho}_2.
\ea
\ee
There are two absolute theories with $\bZ_4$ symmetry denoted by $\bZ_4$ and $\mb{Z}^{\rho}_4$ which are defined by the central extension of $\bZ_2$ by $\bZ_2$ in \eqref{Eqn:z4} and \eqref{Eqn:z4hat}, and one aboslute theory with anomalous symmetry $\mb{Z}_2\times\mb{Z}^{\rho}_2$.

The automorphism group $\Aut_{\bZ_4}(Q)$ is $\bZ_2^3$. Taking into account the $\Aut(\bZ_4)=\bZ_2$, one has $\cO_4(Q)=\bZ^2_2$ with generator 
\begin{equation}
    \rho = \left(\begin{array}{cc}
        1 & 0 \\
        2 & 3
    \end{array}\right)
    \qquad 
    \xi = \left(\begin{array}{cc}
        1 & 2 \\
        0 & 1
    \end{array}\right)
\end{equation}
As one can check that $\rho$ switching two polarizations $L_1$ and $L_2$ can be understood as an operation of gauging $\bZ_4$ while $\xi$ is the operation of stacking a non-trivial phase. Thus, there are four global variants of $T_4[\mb{F}_1]$. They transform under duality $r$ and topological manipulation $\xi$ and $\rho$ in Figure \ref{Fig:F1N4}.

\begin{figure}
    \centering
    \begin{tikzpicture}[scale=1.5]
\draw node at (0,0) {$(\mb{Z}_4)_0$};
\draw node at (3,0) {$(\mb{Z}^{\rho}_4)_0$};
\draw node at (0,2) {$( \mb{Z}_4)_1$};
\draw node at (3,2) {$(\mb{Z}^{\rho}_{4})_1$};
\node[left] at (-0.4,0) {$\left(\begin{matrix} 1 & 0 \\ 0 & 1 \end{matrix}\right)$};
\node[right] at (3.4,0) {$\left(\begin{matrix} 1 & 0 \\ 2 & 3 \end{matrix}\right)$};  
\node[left] at (-0.4,2) {$\left(\begin{matrix} 1 & 2 \\ 0 & 1 \end{matrix}\right)$}; 
\node[right] at (3.4,2) {$\left(\begin{matrix} 1 & 2 \\ 2 & 3 \end{matrix}\right)$};
\draw node at (1.5,1) {$(\bZ_2 \times \mb{Z}^{\rho}_2)_{\mu_3}$};
\draw [<->,blue] (.5,0) -- (2.5,0);
\draw [<->,blue] (.5,2) -- (2.5,2);
\draw [<->,blue] (-0.05,0.2) -- (-0.05,1.8);
\draw [<->,blue] (3.05,0.2) -- (3.05,1.8);
\draw [<->,orange] (0.2,2.3) arc (120:60:2.6);
\draw [<->,orange] (0.2,-.3) arc (-120:-60:2.6);
\draw node at (1.5,-.2) {${\color{blue} \rho}$};
\draw node at (1.5,2.2) {${\color{blue} \rho}$};
\draw node at (-0.2,1) {${\color{blue} \xi}$};
\draw node at (3.2,1) {${\color{blue} \xi}$};
\draw node at (1.5,2.8) {${\color{orange} r}$};
\draw node at (1.5,-0.8) {${\color{orange} r}$};
\end{tikzpicture}
    \caption{Web of transformations for $T_4[\bF_1]$. The transformations
in orange are the duality transformations.
The transformations in blue are topological manipulations.}
    \label{Fig:F1N4}
\end{figure}

% the duality symmetry group of $P$-matrices is $\mb{Z}_2^3$, generated by
% \begin{equation}
%     s = \left(\begin{array}{cc}
%         3 & 0 \\
%         0 & 3
%     \end{array}\right)\ ,\ t = \left(\begin{array}{cc}
%         1 & 2 \\
%         0 & 1
%     \end{array}\right)\ ,\ s_4=\left(\begin{array}{cc}
%         1 & 0 \\
%         2 & 3
%     \end{array}\right).
% \end{equation}

% Note that the $s$ generator is just an automorphism group of $\mb{Z}_4$.  There are four global forms of the physical theory, $(\mb{Z}_4)_0$, $(\mb{Z}_4)_1$, $(\mb{Z}^{\rho}_{4})_0$, $(\mb{Z}^{\rho}_{4})_1$, transformed as

% In this case, the gauging operation of $\mb{Z}_4$, $\rho$, exactly coincides with the group element $s_4$. The stacking of SPT phase, $\xi$, coincides with the group element $t$.

Note that the absolute theory defined by $L_3$ is not connected with the other four global variants via topological manipulations because the topological manipulations are $\bZ_4$ preserving operations while the absolute theory has symmetry $\mb{Z}_2\times\mb{Z}^{\rho}_2$ with a mixed anomaly $\mu_3$. This anomaly is the same one found from $\bZ_4$ discrete gauge theory in equation \eqref{Eqn:anomalyz4}. Similarly, one can detect it from the braidings between the line operations \eqref{Eqn:anomalyz4line}.

\paragraph{\underline{$N=6$}}

The maximal isotropic lattices are
\begin{equation*} 
\begin{array}{cccc}
L_1 = \{(0,0),(1,0), (2,0),(3,0),(4,0),(5,0)\} &\to &\bZ_6 = \bZ_3 \times \bZ_2 \\
L_2 = \{(0,0),(1,2),(3,0),(2,4),(4,2),(5,4)\} &\to & \bZ_6^{\rho}=\mb{Z}^{\rho}_3 \times \bZ_2
\end{array}
\end{equation*}
The theory $T_6[\mb{F}_1]$ on the boundary has $\bZ_6$ symmetry. We can also use its subgroups $\bZ_3$ and $\bZ_2$ to denote its global variants. 
The automorphism group $\Aut_{\bZ_6}(Q)$ is $\bZ_2^3$. Taking into account the $\Aut(\bZ_6)=\bZ_2$, one has $\cO_6(Q)=\bZ^2_2$ with generator 
\begin{equation}
    \rho = \left(\begin{array}{cc}
        1 & 0 \\
        2 & 5
    \end{array}\right)
    \qquad 
    \xi = \left(\begin{array}{cc}
        1 & 3 \\
        0 & 1
    \end{array}\right)
\end{equation}
As one can check that $\rho$ switching two polarizations $L_1$ and $L_2$ 
% can be understood as an operation of gauging $\bZ_6$ 
while $\xi$ is the operation of stacking a non-trivial phase. Thus, there are four global variants of $T_6[\mb{F}_1]$. They transform under duality $r$ and topological manipulation $\xi$ and $\rho$ in Figure \ref{Fig:F1N6}.

\begin{figure}
    \centering
    \begin{tikzpicture}[scale=1.5]
\draw node at (0,0) {$(\mb{Z}_6)_0$};
\draw node at (0,2) {$(\mb{Z}_6)_1$};
\draw node at (3,0) {$(\mb{Z}^{\rho}_{6})_0$};
\draw node at (3,2) {$(\mb{Z}^{\rho}_{6})_1$};
\node[left] at (-0.4,0) {$\left(\begin{matrix} 1 & 0 \\ 0 & 1 \end{matrix}\right)$};
\node[left] at (-0.4,2) {$\left(\begin{matrix} 1 & 3 \\ 0 & 1 \end{matrix}\right)$}; 
\node[right] at (3.4,0) {$\left(\begin{matrix} 1 & 3 \\ 2 & 5 \end{matrix}\right)$};
\node[right] at (3.4,2) {$\left(\begin{matrix} 1 & 0 \\ 2 & 5 \end{matrix}\right)$};
\draw [<->,blue] (.5,0) -- (2.5,0);
\draw [<->,blue] (.5,2) -- (2.5,2);
\draw [<->,blue] (-0.05,0.2) -- (-0.05,1.8);
\draw [<->,blue] (3.05,0.2) -- (3.05,1.8);
\draw [<->,orange] (0.2,2.3) arc (120:60:2.6);
\draw [<->,orange] (0.2,.3) arc (120:60:2.6);
\draw node at (1.5,-.2) {${\color{blue} \rho}$};
\draw node at (1.5,2.2) {${\color{blue} \rho}$};
\draw node at (-0.2,1) {${\color{blue} \xi}$};
\draw node at (3.2,1) {${\color{blue} \xi}$};
\draw node at (1.5,2.8) {${\color{orange} r}$};
\draw node at (1.5,0.8) {${\color{orange} r}$};
\end{tikzpicture}
    \caption{Web of transformations for $T_6[\bF_1]$. The transformations
in orange are the duality transformations.
The transformations in blue are topological manipulations.}
    \label{Fig:F1N6}
\end{figure}

\paragraph{\underline{$\text{General}\; N$}}

With the help of twisted $\bZ_N$ gauge theory,
we can study the global variants of $T_N[\bF_1]$ for general $N$. 
The possible topological manipulations are determined by the automorphism group $\cO_N(Q)$.  
These topological manipulations act transitively on the global variants and the number of global variants is $ d(N) = |\cO_N(Q)|$.
By associating each global variant with a $\cO_N(Q)$ matrix, we can determine how they transform under the topological manipulations. 

\begin{table}[]
\centering
\begin{tabular}{|c|c|c|c|c|c|c|c|c|c|c|}
\hline
$N$ & 2 & 3 & 4 & 5 & 6 & 7 & 8 & 9 & 10 & 11\\
\hline
$\text{Aut}_{\bZ_N}(Q)$ & $\mb{Z}_2$ & $\mb{Z}_2^2$ & $\mb{Z}_2^3$ & $D_8$ & $\mb{Z}_2^3$ & $D_{12}$ & $\mb{Z}_2^5$ & $D_{12}$ & $\mb{Z}_2\times D_8$ & $D_{20}$\\
\hline
\end{tabular}
 \caption{The automorphism group $\Aut_{\bZ_N}(Q)$ of $\bF_1$ up to $N=11$.}
  \label{Tab:F1Aut}
\end{table}

The automorphism group $\Aut_{\bZ_N}(Q)$ is important for finding the global variants, and topological manipulations. Besides that, it also gives the 0-form symmetry for the twisted gauge theory $(\bZ_N)_N$. We compute $\Aut_{\bZ_N}(Q)$ for $N=2,3\ldots 11$, and identify them to finite groups in Table \ref{Tab:F1Aut}. Note that, for odd $N$, our results match with the 0-form symmetry for the twisted gauge theory $(\bZ_N)_N$ studied in \cite{Delmastro:2019vnj} while, for even $N$, our approach from the compactification of 6d SCFT gives the 0-form symmetry of the spin $(\bZ_N)_N$ theories.

\subsection{Topological defects}

We discuss the non-invertible symmetries for $T_N[\mb{F}_1]$. Analogous to the cases of $M_4=\mb{P}^1\times\mb{P}^1$, we need to introduce a parameter $R$, where $\MCG(\bF_1)$ acts on. 
Then, we study the non-invertible defect at the fixed point under the duality transformation.

\paragraph{Coupling from geometry}

There is a coupling in the theory $T_N[\mb{F}_1]$. We will determine it from the invariant volume of $\bF_1$. 
The duality group $\MCG(\bF_1)=\bZ_2^2$ is discussed in \eqref{Eqn:f1duality}.  
Let us denote the Kahler class (Poincar\'{e} dual to the Kahler form) as
\be
J=x b+y f\equiv\bp x \\ y\ep\,.
\ee
The volume of the 4-manifold $\mb{F}_1$ is given by
\be
V_{\mb{F}_1}=\frac{1}{2}J^T Q J\,.
\ee
$V_{\mb{F}_1}$ is invariant under the base change $J\rightarrow PJ$, where $P\in GL(2,\mb{Z})$ satisfies
\be
P^T Q P=Q\,.
\ee
Hence we conclude that the action of $P$ on the geometry of $\mb{F}_1$ is exactly given by $J'=PJ$. We introduce the parameter
\be
R=\frac{x}{y}
\ee
that transforms under the elements of $\MCG(\bF_1)$. To see its geometric meaning, we compute the volume of 2-cycles
\be
\ba
V_f&=J\cdot f=x\cr
V_b&=J\cdot b=y-x\cr
V_{b+f}&=V_f+V_b=y\,.
\ea
\ee
Hence $R$ is the ratio of the volume of $f$ over the volume of $b+f$, which are both $S^2$:
\be
R=\frac{V_f}{V_{b+f}}\,.
\ee

Under the duality transformation $r$ from equation \eqref{Eqn:f1duality}, the coupling changes as  
\begin{equation}
    r(R)=\frac{R}{2R-1}.
\end{equation}
The fixed point is hence $R=1$. We will fix the coupling to this value in the following.

\begin{table}[!htp] \centering
  \renewcommand{\arraystretch}{1.3}
  \begin{tabular}{c|c|c}
   N & Theory & Defects  \\ \hline \strut
   2&  $(\bZ_2)_m$ & $\tau r$  \\   \hline \strut
   p & $\bZ_p, \mb{Z}^{\rho}_p$ & $\rho r $  \\   \hline \strut
   4 &  $(\bZ_4)_m,(\mb{Z}^{\rho}_4)_m$ & 
    $\xi^m\rho r \xi^{m}$  \\ \hline \strut
   6 & $(\bZ_6)_{m}$, $(\mb{Z}^{\rho}_6)_m$& $\xi^m \rho r \xi^{m}$  
  \end{tabular}
  \caption{Topological defects of $T_N[\bF_1]$ at $R=1$.}
  \label{Tab:F1N}
\end{table}

In analogy with the $\bP^1 \times \bP^1$ case, to construct the duality defect in a global variant $M$, one needs to search for a combination of the topological operations $G(\rho,\tau)$, the operation $\rho$ and stacking SPT phase $\tau$, and the duality $F$ from the $\MCG(\bF_1)$ such that $FMG=M$, i.e. the global variant keeps the same. Then the duality defect is given by $\cN=FG$ at the fixed point of $R$. In this way, we find all defects in the global variants of $T_N[\bF_1]$ for $N$ is 4, 6 and prime numbers. The result is listed in Table \ref{Tab:F1N}. Note that the physical explanation of the topological manipulation $\rho$ is not clear. We will study the fusion rule of these defects in future work.

\section{6d $\mc{N}=(2,0)$ SCFTs on del Pezzo surfaces}
\label{sec:5}

In this section, we will study the global variants, symmetries, and possible anomalies of theories $T_N[M_4]$ when the 4-manifold is a del Pezzo surface. 
% We will study consider the del Pezzo surfaces in particular the $dP_2$ 
We extend the discussion to del Pezzo surfaces $dP_l$ with higher $l$. The intersection form of $dP_l$ is a rank-$(l+1)$ matrix, of the form
\be
Q_{ij}=\text{diag}(1,-1,\dots,-1)\,.
\ee

We compute the mapping class group $\MCG(dP_l)$ and  $\Aut_{\mb{Z}_N}(Q)$, whose elements correspond to the solutions to the equations (\ref{Eqn:autZ}) and (\ref{Eqn:autZn}). 
We discuss the choices of polarization for some examples of $l,N$. Note that in general
\be
H^2(dP_l,\mb{Z}_N)=\mb{Z}_N^{l+1}\,.
\ee
These group elements one-to-one correspond to the genuine topological line operators, which are labeled by
\be
L_{(c_1,c_2,\dots,c_{l+1})}(\gamma)=\prod_{i=1}^{l+1}\exp\left(\frac{2\pi i}{N}\oint_\gamma c_i a_i\right)\quad (c_i=1,\dots,N)\,.
\ee

Now we discuss the case of $M_4=dP_2$ in detail. The solution of integral $T$-matrix to the equation (\ref{Eqn:autZn})
forms the discrete group $O(1,2;\mb{Z})$. The generators are 
\be
A=\bp 3 & 2 & -2\\2 & 1 & -2\\2 & 2 & -1\ep\ ,\ S=\bp 1 & 0 & 0\\0 & 0 & 1\\0 & 1 & 0\ep\ ,\ T_1=\bp -1 & 0 & 0\\0 & 1 & 0\\0 & 0 & 1\ep\ ,\ T_2=\bp 1 & 0 & 0\\0 & -1 & 0\\0 & 0 & 1\ep
\ee
In particular, they generate all Pythgorean triples $(a,b,c)$ satisfying $a^2=b^2+c^2$ from a given one.

The Picard group generators of $dP_2$ are denoted as $h,e_1,e_2$, which satisfies the intersection relations
\be
h^2=1\ ,\ h\cdot e_i=0\ ,\ e_i\cdot e_j=-\delta_{i,j}\quad (i,j=1,2)\,.
\ee
The $O(1,2;\mb{Z})$ element $P$ act on the Kahler form
\be
J=xh+ye_1+ze_2=\bp x \\y \\z\ep
\ee
as $J\rightarrow PJ$. The volumes of 2-cycles are
\be
\ba
V_{h}&=J\cdot h=x\cr
V_{e_1}&=J\cdot e_1=-y\cr
V_{e_2}&=J\cdot e_2=-z\,.
\ea
\ee
The volume of $dP_2$ is
\be
V_{dP_2}=\frac{1}{2}J\cdot J=\frac{1}{2}(x^2-y^2-z^2)\,.
\ee

Let us introduce two geometric parameters
\be
R_1=-\frac{x}{y}=\frac{V_H}{V_{E_1}}\ ,\ R_2=\frac{y}{z}=\frac{V_{E_1}}{V_{E_2}}\,,
\ee
the $O(1,2;\mb{Z})$ generators acting non-trivially on $R_1$ are $R_2$ are
\be
\ba
&S\cdot R_1=R_1 R_2\ ,\ S\cdot R_2=\frac{1}{R_2}\cr
&A\cdot R_1=-\frac{3R_1 R_2+2R_2-2}{2R_1 R_2+R_2-2}\ ,\ A\cdot R_2=\frac{2R_1 R_2+R_2-2}{2R_1 R_2+2R_2-1}\cr
&T_1\cdot R_1=-R_1\ ,\ T_1\cdot R_2=R_2\cr
&T_2\cdot R_1=-R_1\ ,\ T_2\cdot R_2=-R_2\,.
\ea
\ee

We discuss the fixed point of $(R_1,R_2)$ under the actions of some subgroups of $O(1,2;\mb{Z})$. For the finite subgroup of $O(1,2;\mb{Z})$, the only meaningful one is the $\mb{Z}_2$ subgroup generated by $S$, since if one include $A$, they must generate an infinite subgroup. For the $\mb{Z}_2$ subgroup $\{I,S\}$, the fixed point is given by
\be
(R_1,R_2)=(R_1,1)\,,
\ee
which is an extended loci in the terminology of \cite{Bashmakov:2022uek}.

\paragraph{$N=2$}

The maximal isotropic sublattices are labeled as
\be
\ba
L_1&=\{(0,0,0),\quad (0,1,1)\}\rightarrow\mb{Z}_2^{(1)}\cr
L_2&=\{(0,0,0),\quad (1,0,1)\}\rightarrow\mb{Z}_2^{(2)}\cr L_3&=\{(0,0,0),\quad (1,1,0)\}\rightarrow\mb{Z}_2^{(3)}\,.
\ea
\ee
The $T$ solutions to (\ref{Eqn:autZn}) are
\bea
\left\{I\ ,\ \bp 0 & 1 & 0\\1 & 0 & 0\\0 & 0 & 1\ep\ ,\ \bp 1 & 0 & 0\\0 & 0 & 1\\0 & 1 & 0\ep\,,
\bp 0 & 0 & 1\\1 & 0 & 0\\0 & 1 & 0\ep\ ,\ \bp 0 & 1 & 0\\0 & 0 & 1\\1 & 0 & 0\ep\ ,\ \bp 0 & 0 & 1\\0 & 1 & 0\\1 & 0 & 0\ep
\right\}
\eea
They form an $\Aut_{\mb{Z}_2}(Q)=S_3$ group, with generators
\be
\sigma=\bp 1 & 0 & 0\\0 & 0 & 1\\0 & 1 & 0\ep\ ,\ \tau=\bp 0 & 0 & 1\\0 & 1 & 0\\1 & 0 & 0\ep\ \,,
\ee
satisfying
\be
\sigma^2=\tau^2=I\ ,\ (\sigma\tau)^3=I\,.
\ee

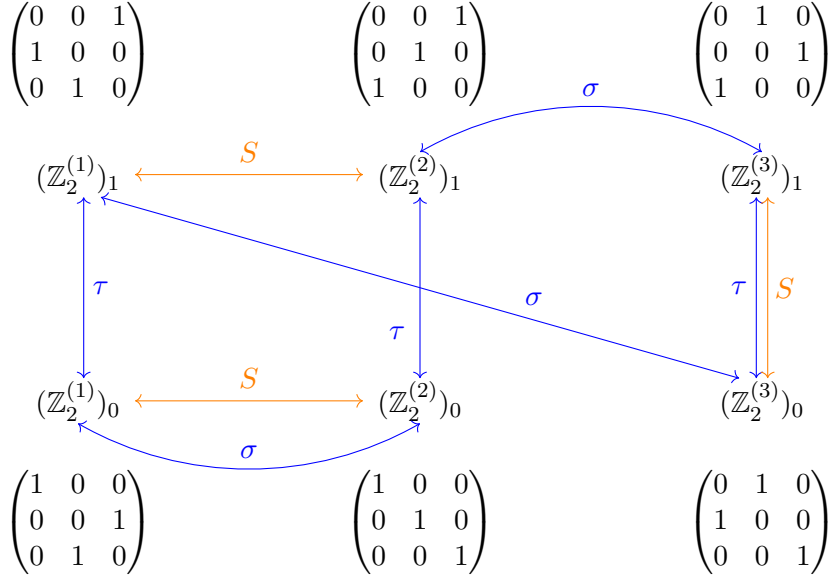
\begin{figure}
    \centering
   \begin{tikzpicture}[scale=1.5]
\draw node at (0,0) {$(\mb{Z}_2^{(1)})_0$};
\draw node at (3,0) {$(\mb{Z}_2^{(2)})_0$};
\draw node at (6,0) {$(\mb{Z}_2^{(3)})_0$};
\draw node at (0,2) {$(\mb{Z}_2^{(1)})_1$};
\draw node at (3,2) {$(\mb{Z}_2^{(2)})_1$};
\draw node at (6,2) {$(\mb{Z}_2^{(3)})_1$};
\draw [<->,orange] (.5,0) -- (2.5,0);
\draw [<->,orange] (.5,2) -- (2.5,2);
% \draw [<->,blue] (3.5,2) -- (5.5,2);
%%
\draw [<->,blue] (0.05,0.2) -- (0.05,1.8);
\draw [<->,blue] (3,0.2) -- (3,1.8);
\draw [<->,blue] (5.95,0.2) -- (5.95,1.8);
\draw [<->,orange] (6.05,0.2) -- (6.05,1.8);
\draw [<->,blue] (0.2,1.8) -- (5.8,0.2);
\draw [<->,blue] (0,-.2) arc (-120:-60:3);
\draw [<->,blue] (3,2.2) arc (120:60:3);
\draw node at (1.5,.2) {${\color{orange} S}$};
\draw node at (1.5,2.2) {${\color{orange} S}$};
\draw node at (1.5,-.45) {${\color{blue} \sigma}$};
\draw node at (4.5,2.75) {${\color{blue} \sigma}$};
\draw node at (0.2,1) {${\color{blue} \tau}$};
\draw node at (5.8,1) {${\color{blue} \tau}$};
\draw node at (6.2,1) {${\color{orange} S}$};
\draw node at (2.8, 0.6) {$\color{blue} \tau$};
\draw node at (4, 0.9) {$\color{blue} \sigma$};
\node[below] at (0,-0.5) {$\left(\begin{matrix} 1 & 0 & 0 \\ 0 & 0 & 1\\0 & 1 & 0\end{matrix}\right)$};
\node[above] at (0,2.5) {$\left(\begin{matrix} 0 & 0 & 1 \\ 1 & 0 & 0\\0 & 1 & 0 \end{matrix}\right)$};
  \node[below] at (3,-0.5) {$\left(\begin{matrix} 1 & 0 & 0 \\ 0 & 1 & 0\\ 0 & 0 & 1 \end{matrix}\right)$};
 \node[above] at (3,2.5) {$\left(\begin{matrix} 0 & 0 & 1 \\ 0 & 1 & 0\\1 & 0 & 0\end{matrix}\right)$};
 \node[below]  at (6,-0.5) {$\left(\begin{matrix} 0 & 1 & 0 \\ 1 & 0 & 0 \\ 0 & 0 & 1 \end{matrix}\right)$};
 \node[above] at (6,2.5) {$\left(\begin{matrix} 0 & 1 & 0 \\ 0 & 0 & 1\\ 1 & 0 & 0\end{matrix}\right)$};
\end{tikzpicture}
    \caption{Web of transformations for $T_2[dP_2]$. The transformations
in orange are the duality transformations.
The transformations in blue are topological manipulations.}
    \label{Fig:N2dP2}
\end{figure}

There are six global forms $(\mb{Z}_2^{(1)})_0$, $(\mb{Z}_2^{(2)})_0$, $(\mb{Z}_2^{(3)})_0$, $(\mb{Z}_2^{(1)})_1$, $(\mb{Z}_2^{(2)})_1$, $(\mb{Z}_2^{(3)})_1$, which are transformed by $S_3$ elements in the same way as the case of $M=\mb{P}^1\times\mb{P}^1$, $N=2$. We plot the transformations in Figure~\ref{Fig:N2dP2}.

We also list the 
topological defects
at the fixed point $(R_1,R_2)=(R_1,1)$, analogous to the case of $T_2[\mb{P}^1\times\mb{P}^1]$:
\be
\begin{tabular}{c|c}
   Theory & Defects  \\ \hline \strut
     $(\mb{Z}_2^{(1)})_m,(\mb{Z}_2^{(2)})_m$ & $\tau^m\sigma S \tau^{m}$  \\ \hline \strut
    $(\mb{Z}_2^{(3)})_m$ &$\tau^m\tau S\tau^{m}$   
  \end{tabular}
  \ee

For $N=3$, there are 48 $T$ solutions to (\ref{Eqn:autZn}), they form a group $S_4\times\mb{Z}_2$, which is generated by
\be
\left\{\bp 1 & 0 & 0\\0 & 1 & 0\\0 & 0 & 2\ep\ ,\ \bp 0 & 1 & 1\\1 & 1 & 2\\2 & 1 & 2\ep\ ,\ \bp 0 & 2 & 2\\2 & 1 & 2\\2 & 2 & 1\ep\ ,\ \bp 2 & 0 & 0\\0 & 2 & 0\\0 & 0 & 2\ep\right\}\,.
\ee
In the above equation, the first three elements generate the $S_4$ factor, while the last diagonal matrix generates the $\mb{Z}_2$ factor.
More generally, we list the number $n$ of the solutions to (\ref{Eqn:autZn}) for different $N$ here:
\be
\begin{tabular}{|c|c|c|c|c|c|c|c|c|c|c|}
\hline
$N$ & 2 & 3 & 4 & 5 & 6 & 7 & 8 & 9 & 10 & 11\\
\hline
$|\Aut_{\mb{Z}_N}(Q)|$ & 6 & 48 & 128 & 240 & 288 & 672 & 2048 & 1296 & 1440 & 2640\\
\hline
\end{tabular}
\ee
From the growth of $|\Aut_{\mb{Z}_N}(Q)|$, we can see that $T_N[dP_2]$ is not simply a $\mb{Z}_N$ gauge theory.
Finally let us briefly discuss the cases of $M_4=dP_l$ for $l>2$. For $M_4=dP_3$, we list the number $n$ of solutions to (\ref{Eqn:autZn}) for some values of $N$:
\be
\begin{tabular}{|c|c|c|c|c|c|c|}
\hline
$N$ & 2 & 3 & 4 & 5 & 6 & 7\\
\hline
$n$ & 48 & 1440 & 12288 & 28800 & 69120 & 235200\\
\hline
\end{tabular}
\ee
For $M_4=dP_4$, when $N=2$, the number of solutions to (\ref{Eqn:autZn}) is 720.
In general, we will leave a detailed dicussion of the physics of $T_N[dP_l]$ $(l\geq 2)$ in the future work.

\section{6d $\mc{N}=(2,0)$ SCFTs on general 4-manifolds} \label{sec:6}

In this section, we consider $T_N[M_4]$ with a general 4-manifold $M_4$, which is allowed to have 1-cycles, 3-cycles, as well as torsional cycles. Using Poincar\'{e} duality and universal coefficient theorem, the general form of homology and cohomology groups are
\be
\ba
H_*(S,\mb{Z})&=(\mb{Z},\mb{Z}^{b_1}\oplus\bigoplus_\alpha\mb{Z}_{l_\alpha},\mb{Z}^{b_2}\oplus\bigoplus_\alpha\mb{Z}_{l_\alpha},\mb{Z}^{b_1},\mb{Z})\cr
H^*(S,\mb{Z})&=(\mb{Z},\mb{Z}^{b_1},\mb{Z}^{b_2}\oplus\bigoplus_\alpha\mb{Z}_{l_\alpha},\mb{Z}^{b_1}\oplus\bigoplus_\alpha\mb{Z}_{l_\alpha},\mb{Z})\,.
\ea
\ee
We denote the free generators of $H^n(M_4,\mb{Z})$ by $v_n^i$, and the torsional generators of $H^n(M_4,\mb{Z})$ by $t_n^\alpha$, then we expand the differential cohomology class $\breve{G}_4$ as
\be
\ba
\label{general-G4}
\br{G}_4&=\sum_{i=1}^{b_1}\br{F}_3^i\star\br{v}_1^i+\sum_{i=1}^{b_2}\br{F}_2^i\star\br{v}_2^i+\sum_{i=1}^{b_1}\br{F}_1^i\star\br{v}_3^i\cr
&+
\sum_\alpha \br{B}_1^\alpha\star \br{t}_3^\alpha+\sum_\alpha\br{B}_2^\alpha\star\br{t}_2^\alpha\,.
\ea
\ee
Plug (\ref{general-G4}) into the SymTFT action of 6d (2,0) theory
\be
S_{\rm 7d}=\frac{N}{4\pi}\int \br{G}_4\star\br{G}_4
\ee
and expand out the terms. The SymTFT action has two parts, the first part involves the primary invariant integrated over $M_4$:
\be
\label{SymTFT-gen1}
S_{3d,1}=\frac{N}{4\pi}\left(\sum_{i,j=1}^{b_2}\int_{M_4}\br{v}_2^i\star\br{v}_2^j\int_{W_3}\br{F}_2^i\star\br{F}_2^j+\sum_{i,j=1}^{b_1}2\int_{M_4}\br{v}_1^i\star\br{v}_3^j\int_{W_3}\br{F}_3^i\star\br{F}_1^j\right)\,.
\ee
The second part involves the secondary invariant integrated over $S$:
\be
\ba
\label{SymTFT-gen2}
S_{3d,2}&=\frac{N}{4\pi}\left(\sum_{i,\alpha}\int_{M_4}\br{v}_2^i\star\br{t}_3^\alpha \int_{W_3}\br{F}_2^i\star\br{B}_1^\alpha+\sum_{i,\alpha}\int_{M_4}\br{v}_3^i\star\br{t}_2^\alpha \int_{W_3}\br{F}_1^i\star\br{B}_2^\alpha\right.\cr
&+\left.\sum_{\alpha,\beta}\int_{M_4}\br t_2^\alpha\star\br{t}_3^\beta \int_{W_3}\br{B}_2^\alpha\star\br{B}_1^\beta\right)\,.
\ea
\ee
After plug in $F_2^i=da^i$, $F_1^i=dc_0^i$ and $F_3^i=db^i$, the terms (\ref{SymTFT-gen1}) become
\be
S_{3d,1}=\frac{N}{4\pi}\left(\sum_{i,j=1}^{b_2}\int_{M_4}\br{v}_2^i\star\br{v}_2^j\int_{W_3}a^i\wedge da^j+\sum_{i,j=1}^{b_1}2\int_{M_4}\br{v}_1^i\star\br{v}_3^j\int_{W_3}c_0^i\wedge db^j\right)
\ee
For the terms (\ref{SymTFT-gen2}), the terms on the first line can be eliminated by redefining $\br{v}_2^i\rightarrow\br{v}_2^i+m_{i\alpha}t_2^\alpha$ and $\br{v}_3^i\rightarrow\br{v}_3^i+n_{i\alpha}t_3^\alpha$ ($m_{i\alpha},n_{i\alpha}\in\mb{Z}$). The remaining terms are
\be
S_{3d,2}=\frac{N}{4\pi}\sum_{\alpha,\beta}\int_{M_4}\br t_2^\alpha\star\br{t}_3^\beta \int_{W_3}\br{B}_2^\alpha\star\br{B}_1^\beta\,.
\ee

\subsection{$T^2\times S^2$}

As an example of $M_4$ with odd-dimensional cycles, we consider the dimensional reduction of 6d (2,0) $A_N$ theory on $M_4=T^2\times S^2$, which results in various 2d theories $T_N[T^2\times S^2]$ with certain amounts of supersymmetries.

The 4-manifold $T^2\times S^2$ has homology and cohomology groups
\be
\ba
H_*(T^2\times S^2,\mb{Z})&=(\mb{Z},\mb{Z}^2,\mb{Z}^2,\mb{Z}^2,\mb{Z})\cr
H^*(T^2\times S^2,\mb{Z})&=(\mb{Z},\mb{Z}^2,\mb{Z}^2,\mb{Z}^2,\mb{Z})\,.
\ea
\ee
We denote the generators of $H^i(T^2\times S^2,\mb{Z})$ by $\omega_i,\hat{\omega}_i$ ($i=1,2,3$). We can thus expand the 3-form $c$ as
\be
c=b\wedge\omega_1+\hat{b}\wedge\hat{\omega}_1+a\wedge\omega_2+\hat{a}\wedge\hat{\omega}_2+c_0\wedge\omega_3+\hat{c}_0\wedge\hat{\omega}_3\,.
\ee
After integrating the 7d topological action (\ref{eqn:symTFT-7d}) over $T^2\times S^2$, we obtain the 3d topological action
\be
\ba
S_{3d}&=\frac{N}{4\pi}\int_{M_3}a\wedge d\hat{a}+\hat{a}\wedge da+b\wedge d\hat{c}_0-\hat{c}_0\wedge db-\hat{b}\wedge dc_0+c_0\wedge d\hat{b}\cr
&=\frac{2\pi}{N}\int_{M_3}a\delta\hat{a}+b\delta\hat{c}_0-\hat{b}\delta c_0\,.
\ea
\ee
On the second line, the gauge fields all become $\mb{Z}_N$-valued cochains. $a$ and $\hat{a}$ are gauge fields for $\mb{Z}_N$ 0-form symmetries, analogous to the case of $T_N[\mathbb P^1\times \mathbb P^1]$. $b$ and $\hat{b}$ are gauge fields for $\mb{Z}_N$ 1-form symmetries. $c_0$ and $\hat{c}_0$ are scalars, which can be thought as background gauge fields for $\mb{Z}_N$ $(-1)$-form symmetries. Note that the two $\mb{Z}_N$ 1-form symmetries are mutually local, which leads to the prediction that the 2d theory may have a $\Gamma^{(1)}=\mb{Z}_N^2$ 1-form symmetry.

\subsection{Enriques surface}

An Enriques surface $S$ is a complex surface with torsional homology and cohomology groups
\be
\ba
H_*(S,\mb{Z})&=(\mb{Z},\mb{Z}_2,\mb{Z}^{10}\oplus\mb{Z}_2,0,\mb{Z})\cr
H^*(S,\mb{Z})&=(\mb{Z},0,\mb{Z}^{10}\oplus\mb{Z}_2,\mb{Z}_2,\mb{Z})\,.
\ea
\ee

The surface has topological invariants
\be
\chi(S)=12\ ,\ \sigma(S)=-8\,.
\ee
Hence the central charge of the 2d theory $T_G[S]$ is
\be
\ba
c_R&=6r_G\cr
c_L&=12r_G\,.
\ea
\ee

Let us expand the differential cohomology class $\br{G}_4$ as
\be
\label{Enriques-G4}
\br{G}_4=\sum_{i=1}^{10}\br{F}_2^i\star\br{v}_2^i+\br{B}_1\star \br{t}_3+\br{B}_2\star\br{t}_2\,.
\ee
$\br{F}_2^i$ are field strengths of gauge fields $a_i$ $(i=1,\dots,10)$, the background gauge fields of $U(1)$ 0-form global symmetries. $\br{B}_1$ and $\br{B}_2$ are background gauge fields of $\mb{Z}_2$ 0-form and 1-form global symmetries. Geometrically, $\br{v}_2^i$ corresponds to the free part $\mb{Z}^{10}$ of $H^2(S,\mb{Z})$. $\br{t}_3$ and $\br{t}_2$ corresponds to the $\mb{Z}_2$ torsion of $H^3(S,\mb{Z})$ and $H^2(S,\mb{Z})$ respectively. In the Poincar\'{e} dual language, $\br{t}_3$ corresponds to the $\mb{Z}_2$ torsional 1-cycle $[\Sigma_1]$, and $\br{t}_2$ corresponds to the $\mb{Z}_2$ torsional 2-cycle $[\Sigma_2]$ on $S$.

Plug (\ref{Enriques-G4}) into the SymTFT action of 6d (2,0) theory
\be
S_{3d}=\frac{N}{4\pi}\int \br{G}_4\star\br{G}_4
\ee
and expand out the terms. The SymTFT action has two parts, the first part involves the primary invariant integrated over $S$:
\be
S_{3d,1}=\frac{N}{4\pi}\sum_{i,j=1}^{10}\int_{S}\br{v}_2^i\star\br{v}_2^j\int_{W_3}\br{F}_2^i\star\br{F}_2^j\,.
\ee
$\int_{S}\br{v}_2^i\star\br{v}_2^j$ can be computed with the intersection form on the Enriques surface, with is the unimodular matrix $I_{1,9}$ with signature $(1,9)$.
\be
\ba
\label{Enriques-int}
\int_{S}\br{v}_2^i\star\br{v}_2^j&=Q_{ij}\cr
Q&=\diag(1,-1,-1,\dots,-1)\,.
\ea
\ee
Hence after reducing to the 3d 't Hooft anomaly polynomial, we have
\be
S_{3d,1}=\frac{N}{4\pi}\int_{W_3} Q_{ij}a^i\wedge da^j\,,
\ee
where we contracted $i,j$ indices.

The second part involves the secondary invariant integrated over $S$:
\be
S_{3d,2}=\frac{N}{2\pi}\left(\sum_{i=1}^{10}\int_{S}\br{v}_2^i\star\br{t}_3\int_{W_3}\br{F}_2^i\star\br{B}_1+\int_{S}\br{t}_2\star\br{t}_3\int_{W_3}\br{B}_2\star\br{B}_1\right)\,.
\ee

Now let us evaluate the coefficients in the SymTFT action. First, we can set
\be
\int_{S}\br{v}_2^i\star\br{t}_3=0
\ee
by redefining $\br{v}_2^i\rightarrow\br{v}_2^i+m_i\br{t}_2$ ($m_i\in\mb{Z}$). This shifting do not affect the computation of (\ref{Enriques-int}). Finally, the last term $\int_{S}\br{t}_2\star\br{t}_3$ is computed via the linking pairing\footnote{The linking pairing must equal to $\frac{1}{2}$, because it is non-degenerate.}
\be
\int_{S}\br{t}_2\star\br{t}_3=\text{link}([\Sigma_1],[\Sigma_2])=\frac{1}{2}\,.
\ee
The final SymTFT expression is
\be
S_{3d}=\frac{N}{4\pi}\left(\sum_{i,j=1}^{10}Q_{ij}\int_{W_3}\br{F}_2^i\star\br{F}_2^j+\int_{W_3}\br{B}_2\star\br{B}_1\right)\,.
\ee

It is qualitatively different for even $N$ and odd $N$. For odd $N$, there is a mixed 't Hooft anomaly between a $\mb{Z}_2$ 1-form symmetry (with background gauge field $B_2$) and a $\mb{Z}_2$ 0-form symmetry (with background gauge field $B_1$). For even $N$, the mixed 't Hooft anomaly term is absent.

\section{Conclusion and outlook} \label{sec:7}

In this paper we studied the symmetry TFT and duality defects of 2d CFTs obtained from compactification of the 6d $(2,0)$ theory of $A_{N-1}$-type on 4-manifolds $M_4$. We focused mainly on the case of $M_4 = \bP^1 \times \bP^1$ while also working out details of $\bF_1$ and del Pezzo surfaces as well as more general surfaces. We find that such compactifications give rise to a rich multitude of duality networks and interesting defect fusion categories including non-invertible defects. A main message is that the combination of geometric dualities coming from $M_4$ and topological transformations on the field theory side together produce the full structure of $0$-form symmetries. While in some cases the global variants and their connections on the 2d side were known previously from field theory constructions, we find that our geometric approach allows for a much more efficient screening. We find invertible defects by studying maximal isotropic sublattices of $H^2(M_4,\bZ_N)$ and their complements giving rise to Abelian fusion categories in 2d. The corresponding fusion category in the bulk SymTFT is then the quantum double of the one on the boundary.  Moreover, we find non-invertible defects which are realized at fixed points of coupling constants under the discrete automorphisms of $M_4$. Here, coupling constants correspond to ratios of volumes of 2-cycles of $M_4$ and the discrete automorphisms act as generalized T-dualities on these. 

In the case of del Pezzo surfaces, the structure seems to be more intricate and it is not immediately clear how to choose topological boundary conditions for the SymTFT which does not seem to be the quantum double of any known fusion category. However, the geometric method of choosing maximal isotropic sublattices does give rise to Abelian fusion categories on the boundary which can then be identified with possible topological boundary conditions. To get a more complete picture of the ultimate structure of the 2d TDLs one needs a thorough analysis of the SymTFT at the Lagrangian and field theoretical level. We leave this to future work.

Another interesting direction is to study SymTFTs arising from reductions for 6d (1,0) SCFTs. Reductions of such theories on a 4-manifold would give rise to 2d theories with half of the amount of supersymmetry and have richer physics. 
Besides that the compactification of 6d (2,0) SCFTs on $T^2\times S^2$ is expected to give various 2d SQCDs. It would be compelling to investigate the TDL structures of these theories in the far infra-red region, and compare them to our general SymTFT analysis employed in this paper. We also plan to explore these topics in future studies.

\subsection*{Acknowledgments}
We would like to thank Chi-Ming Chang, Michele Del Zotto, Jiahua Tian and Zheyan Wan for valuable discussions.
BH is supported by the Young Thousand Talents grant of China as well as by the NSFC grant 12250610187. BH would like to thank the Max-Planck Institute for Mathematics and the Theoretical Physics Department of Utrecht University, where part of this work was completed, for hospitality. YNW is  supported by National Science Foundation of China under Grant No. 12175004, by Peking University under startup Grant No. 7100603667, and by Young Elite Scientists Sponsorship Program by CAST (2022QNRC001). 
The work of WC is supported by the fellowship of China Postdoctoral Science Foundation NO.2022M720507 and in part by the Beijing Postdoctoral Research Foundation.
WC would also like to thank the Tsinghua Sanya International Mathematics Forum (TSIMF) for hospitality where part of this work was completed.

\bibliographystyle{JHEP}
\bibliography{main}

\end{document}